\documentclass[letterpaper,twoside,12pt]{article}
\usepackage{graphicx,natbib,amsmath,amssymb}
\title{Channeling of protons through carbon nanotubes}

\author{Du\v{s}ko Borka, Srdjan Petrovi\'{c} and Neboj\v{s}a Ne\v{s}kovi\'{c} \\
Laboratory of Physics (010), Vin\v{c}a Institute of Nuclear Sciences, \\
P. O. Box 522, 11001 Belgrade, Serbia}

\date{}

\begin{document}

\maketitle


\tableofcontents
\markright{CONTENTS}

\vfil\newpage

\section*{Abstract}
\markright{Abstract}
\addcontentsline{toc}{section}{Abstract}

This book contains a thorough theoretical consideration of the
process of proton channeling through carbon nanotubes. We begin with
a very brief summary of the theoretical and experimental results of
studying ion channeling through nanotubes. Then, the process of ion
channeling is described briefly. After that, the crystal rainbow
effect is introduced. We describe how it was discovered, and present
the theory of crystal rainbows, as the proper theory of ion
channeling in crystals and nanotubes. We continue with a description
of the effect of zero-degree focusing of protons channeled through
nanotubes. It is shown that the evolution of the angular
distribution of channeled protons with the nanotube length can be
divided in the cycles defined by the rainbow effect. Further, we
analyze the angular distributions and rainbows in proton channeling
through nanotubes. This is done using the theory of crystal
rainbows. The angular distributions are generated by the computer
simulation method, and the corresponding rainbow patterns are
obtained in a precise analysis of the mapping of the impact
parameter plane to the transmission angle plane. We demonstrate that
the rainbows enable the full explanation of the angular
distributions. The analysis demonstrates that the angular
distributions contain the information on the transverse lattice
structure of the bundle. In this study we investigate the rainbows
in transmission of protons of the kinetic energy of 1 GeV through a
straight very short bundle of (10, 10) single-wall carbon nanotube,
a bent short bundle of (10, 10) nanotubes, and the straight long
(11, 9) nanotubes.

We also investigate how the effect of dynamic polarization of the
carbon atoms valence electrons influences the angular and spatial
distributions of protons transmitted through short nanotubes in
vacuum and embedded in dielectric media. It is demonstrated that
this effect can induce the additional rainbow maxima in the angular
distributions. We have established that the changing of the spatial
distribution with the proton kinetic energy, based on the changing
of the dynamic polarization effect, may be used to probe the atoms
or molecules intercalated in the nanotubes. In these study we
investigate the rainbows in transmission of 0.223-2.49 MeV protons
through the short (11, 9) nanotubes in vacuum and embedded in
SiO$_2$, Al$_2$O$_3$ and Ni. Besides, we report on the donut effect
in transmission of 0.223 MeV protons through a short (11, 9)
nanotube, which occurs when the initial ion velocity vector is not
parallel to the nantoube axis. In addition, we explore the
channeling star effect in 1 GeV proton channeling through bundles of
nanotubes, which appears when the proton beam divergence angle is
larger than the critical angle for channeling.

\section{Introduction}
\markright{Introduction}

Carbon nanotubes were discovered by Iijima in 1991 \citep{iiji91}.
One can describe them as the sheets of carbon atoms rolled up into
the cylinders with the atoms lying at the hexagonal crystal lattice
sites. The diameters of nanotubes are of the order of a nanometer
and their lengths can be more then a hundred micrometers. Nanotubes
can be the single-wall and multi-wall ones, depending on the number
of cylinders they include. They have remarkable geometrical and
physical properties \citep{sait01}. As a result, nanotubes have
begun to play an important role in the field of nanotechnologies.
For example, they are expected to become the basic elements for
creating nanoelectronic devices \citep{yao99}.

\begin{figure}[ht!]
\centering
\includegraphics[width=0.6\textwidth]{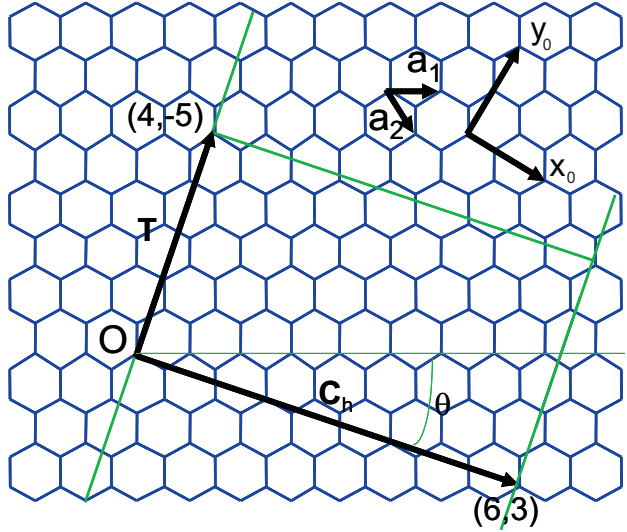}
\caption{The hexagonal crystal lattice: $x_0$ and $y_0$ are the axes of the local reference frame, and $\vec a_1$ and $\vec a_2$ the unit vectors of the lattice. When carbon atoms are placed at the lattice sites, a sheet of carbon atoms is obtained. When rolled up into a cylinder, the sheet of carbon atoms becomes a carbon nanotube: $\vec C_h$ is the chiral vector of the nanotube, $\vec \Theta_h$ the chiral angle of the nanotube, and $\vec T$ the translational vector of the nanotube, which is perpendicular to $\vec C_h$ and extends from its origin (point O) to the first lattice point. Rectangle defined by $\vec C_h$ and $\vec T$ is a unit cell of the nanotube. In the displayed case $\vec C_h$ = (6,3) and $\vec T$ = (4,-5).}
\label{fig1_1}
\end{figure}

\begin{figure}[ht!]
\centering
\includegraphics[width=0.5\textwidth]{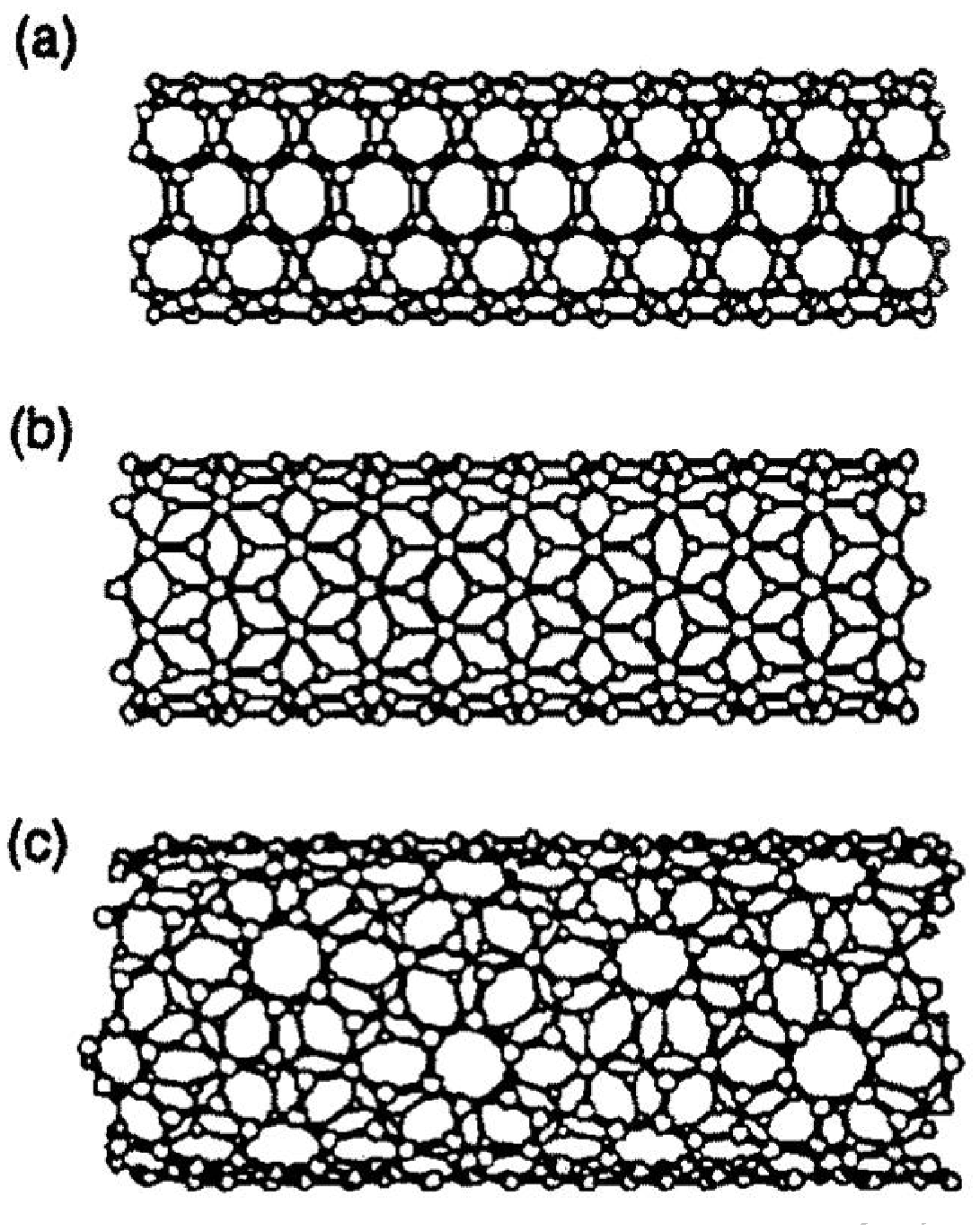}
\caption{The classification of carbon nanotubes: (a) an armchair nanotube, (b) a zigzag nanotube, and (c) a chiral nanotubes.}
\label{fig1_2}
\end{figure}

The structure of a carbon nanotube is described by vector $\vec C_h
= m \vec a_1 + n \vec a_2 \equiv (m,n)$, where $\vec a_1$ and $\vec
a_2$ are the unit vectors of the hexagonal crystal lattice,   and
the integers satisfying inequalities $0 \le \left| n \right| \le m$
\citep{sait01}. This vector is called the chiral vector of the
nanotube. Its magnitude is $C_h = (m^2 + mn + n^2)^{\frac{1}{2}}
a_h$, where $a_h$ is the magnitude of $\vec a_1$ and $\vec a_2$. The
angle of $\vec C_h$ relative to $\vec a_1$, which is called the
chiral angle of the nanotube, is given by expression $\cos \Theta_h
= (m + n/2) / (m^2 + mn + n^2)^{\frac{1}{2}}$ and inequalities $0
\le \left| \Theta_h \right| \le \pi /6$. This is shown in Fig.
\ref{fig1_1}. The sheet of carbon atoms is rolled up into the
cylinder in such a way to transform the segment corresponding to
$\vec C_h$ into a circle. Hence, the nanotube diameter equals $\vec
C_h / \pi$. The nanotube is achiral or chiral. If it is achiral, the
nanotube consists of the atomic strings parallel to its axis. In
particular, if $n$ = 0, when $\Theta_h$ = 0, or $m$ = $n$, when
$\Theta_h$ = $\pi / 6$, the nanotube is achiral. In accordance with
the shape of its transverse cross-section, in the former case the
nanotube is called the zigzag one, and in the latter case the
armchair one. If it is chiral, the nanotube consists of the atomic
strings that spiral around its axis. This is illustrated in Fig.
\ref{fig1_2}, which is taken from Ref. \citet{sait01}.

Soon after the discovery of carbon nanotubes, Klimov and Letokhov
\citep{klim96} predicted that they could be used to channel
positively charged particles. After that, a number of theoretical
groups have studied ion channeling through nanotubes
\citep{gevo98,zhev98,biry02,gree03,zhev03,artr05,bell05,kras05,mour05,bork05,nesk05,petr05a,petr05b,bork06a,bork06b,bork07,maty07,misk07,mour07,bork08a,bork08b,maty08,petr08a,petr08b,zhen08a,zhen08b,petr09}.
The main objective of most of those studies was to investigate the
possibility of guiding ion beams with nanotubes. Biryukov and
Bellucci \citep{biry02} looked for the nanotube diameter optimal for
channeling high energy ion beams. Krasheninnikov and Nordlund
\citep{kras05} studied the channeling of low energy Ar$^+$ ions
through achiral and chiral nanotubes. Petrovi\'{c} et al.
\citep{petr05a} demonstrated that the rainbow effect could play an
important role in ion channeling through nanotubes.

The experimental studying of ion channeling through carbon nanotubes
is in the initial phase. The most challenging task in such
experiments still is to solve the problems of ordering,
straightening and holding nanotubes. The first experimental data on
ion channeling through nanotubes were reported by Zhu et al.
\citep{zhu05}. They were obtained with He$^+$ ions of the kinetic
energy of 2 MeV and an array of the well-ordered multi-wall
nanotubes grown in a porous anodic aluminum oxide (Al$_2$O$_3$)
membrane. The authors measured and compared the yields of ions
transmitted through the bare Al$_2$O$_3$ sample and the Al$_2$O$_3$
sample including the nanotubes. The first experiment with electrons
and nanotubes was performed by Chai et al. \citep{chai07}. They used
the 300 keV electrons and studied their transport through the
multi-wall nanotubes of the length in the range of 0.7-3.0 $\mu$m
embedded in a carbon fiber coating. The misalignment of the
nanotubes was below 1$^{\circ}$. Berdinsky et al. \citep{berd08}
succeeded in growing the single-wall nanotubes in the etched ion
tracks in a silicon oxide (SiO$_2$) layer on a silicon substrate.
This result has opened a possibility for conducting precise
measurements of ion channeling through nanotubes in a wide ion
energy range.

This chapter is devoted to our theoretical studies of proton
channeling through carbon nanotubes. We shall first describe briefly
the process of ion channeling. Then, the crystal rainbow effect will
be introduced, and the theory of crystal rainbows will be presented.
We shall continue with a description of the effect of zero-degree
focusing of protons channeled through nanotubes. Further, the
angular distributions and rainbows in proton channeling through
nanotubes will be analyzed. We shall focus on the rainbows in
transmission of 1 GeV protons through a straight very short bundle
of (10, 10) single-wall carbon nanotube, a bent short bundle of (10,
10) nanotubes, and the straight long (11, 9) nanotubes. Also, the
influence of the effect of dynamic polarization of the carbon atoms
valence electrons on the angular and spatial distributions of
protons transmitted through short nanotubes in vacuum and embedded
in dielectric media will be investigated. We shall concentrate on
the rainbows in transmission of 0.223-2.49 MeV protons through the
short (11, 9) nanotubes in vacuum and embedded in SiO$_2$,
Al$_2$O$_3$ and Ni. Besides, we shall report on the donut effect in
transmission of 0.223 MeV protons through a short (11, 9) nanotube,
which occurs when the initial ion velocity vector is not parallel to
the nantoube axis. In addition, we shall explore the channeling star
effect in 1 GeV proton channeling through bundles of nanotubes,
which appears when the proton beam divergence angle is larger than
the critical angle for channeling.

\section{Ion channeling}
\markright{Ion channeling}

An axial crystal channel is a part of a crystal in between its
neighboring atomic strings being parallel to one of its
crystallographic axes. The process of ion motion through the axial
crystal channel, in which the angle of its velocity vector relative
to the channel axis remains small, is called axial ion channeling
\citep{robi63,lind65,gemm74}. This process is explained by the ion
repulsions from the atomic strings defining the channel, which are
the results of the series of its correlated collisions with the
atoms of the strings. Analogously, a planar crystal channel is a
part of the crystal in between its neighboring atomic planes being
parallel to one of its crystallographic planes. The process of ion
motion through the planar crystal channel is called planar ion
channeling \citep{robi63,lind65,gemm74}.

Theoretical studies of ion channeling tbrough crystals have been
going on along two major lines. The first line was founded by
Lindhard \citep{lind65} and the second one by Barrett
\citep{barr71}. The Lindhard's approach was analytical and it was
developed by the methods of statistical mechanics. He included the
continuum approximation, i.e., neglected the longitudinal
correlations between the positions of the atoms of one atomic
string, did not take into account the transverse correlations
between the positions of the atomic strings, and included the
assumption of statistical equilibrium in the transverse position
plane. The Barrett's approach was via the ion-atom scattering theory
and it was numerical. He developed a very realistic computer code
for the three-dimensional following of ion trajectories through
crystal channels, in which the three approximations used by Lindhard
were avoided. The Barrett's approach is more complex, but a number
of calculations have shown that it is much more accurate than the
Lindhard's approach.

The results that will be presented here have been obtained by a
realistic computer code based on the numerical solution of the ion
equations of motion through crystal channels. The continuum
approximation is included. Consequently, one has to solve only the
ion equations of motion in the transverse position plane. The
transverse correlations between the positions of the atomic strings
are taken into account.

Let us take that the $z$ axis of the reference frame coincides with
the channel axis and that its origin lies in the entrance plane of
the crystal. The $x$ and $y$ axes of the reference frame are the
vertical and horizontal axes, respectively. It is assumed that the
interaction of the ion and crystal can be treated classically
\citep{lind65,gemm74,barr71}. The interaction of the ion and a
crystal's atom is described by an approximation of the Thomas-Fermi
interaction potential or any other appropriate interaction potential
-- $V(r')$, where $r'$ is the distance between the ion and crystal's
atom. As a result, we obtain the ion-crystal continuum interaction
potential,

\begin{equation}
U(x,y) = \sum\limits_{j = 1}^J {U_j(x,y)},
\label{equ01}
\end{equation}

\noindent where

\begin{equation}
U_j(x,y) = \frac{1}{d}\int\limits_{-\infty}^{+\infty} {V(\rho_j^2 + z^2)^{\frac{1}{2}}} dz,
\label{equ02}
\end{equation}

\noindent is the continuum interaction potential of the ion and the
$j^{th}$ atomic string,
						
\begin{equation}
\rho_j = \left[ {(x - x_j)^2 + (y - y_j)^2} \right]^{1/2},
\label{equ03}
\end{equation}

\noindent is the distance between the ion and this atomic string,
and $x$ and $y$ are the transverse components of the proton position
vector; $J$ is the number of atomic strings and $d$ the distance
between the atoms of an atomic string.

The thermal vibrations of the crystal's atoms can be taken into
account. This is done by averaging $U(x,y)$ over the transverse
displacements of the crystal's atoms from their equilibrium
positions \citep{appl67,nesk86}. As a result, one has to substitute
$U_j(x,y)$ in Eq. (\ref{equ02}) with

\begin{equation}
U_j^{th}(x,y) = U_j(x,y) + \frac{\sigma_{th}^2}{2} \left[ {\partial_{xx} U_j(x,y) + \partial_{yy} U_j(x,y)} \right],
\label{equ04}
\end{equation}

\noindent where $\sigma_{th}$ is the one-dimensional thermal
vibration amplitude of the crystal's atoms. Thus, $U^{th}(x,y)$ is
obtained instead of $U(x,y)$.

We neglect the nuclear ion energy loss, i.e., the energy loss
resulting from its collisions with the crystal's nuclei. But, the
electronic ion energy loss and dispersion of its transmission angle,
which are caused by its collisions with the crystal's electrons, can
be included. For the specific ion energy loss we use expression

\begin{equation}
-\frac{dE}{dz} = \frac{4 \pi Z_1^2 e^4}{m_e v^2} {n_e} \left( \ln \frac{2m_e \gamma^2 v(z)^2}{\hbar \omega_e} -\beta ^2 \right),
\label{equ05}
\end{equation}

\noindent where $Z_1$ is the atomic number of the ion, $e$ the
elementary charge, $m_e$ the electron mass, $v(z)$ the magnitude of
the ion velocity vector, $\beta = v(z) / c$, $c$ the speed of light,
$\gamma^2 = (1 - \beta^2)^{-1}$, $n_e = (\partial_{xx} +
\partial_{yy})U^{th}(x,y) / 4 \pi$ the average of the density of the
crystal's electron gas along the $z$ axis, $\hbar$ the reduced
Planck constant, and $\omega_e = (4 \pi e^2 n_e /
m_e)^{\frac{1}{2}}$ the angular frequency of the ion induced
oscillations of the crystal's electron gas \citep{gemm74, petr02}.
For the specific change of the dispersion of the ion transmission
angle we use expression

\begin{equation}
\frac{d \Omega_e^2}{dz} = \frac{m_e}{m^2 v^2(z)} \left( -\frac{dE_e}{dz} \right),
\label{equ06}
\end{equation}

\noindent where $m = m_0 \gamma$ is the relativistic ion mass and
$m_0$ the ion rest mass \citep{gemm74,petr02}. This dispersion is a
measure of the uncertainty of the ion transmission angle. The
dispersions of the vertical and horizontal components of the ion
transmission angle, $\Theta_x$ and $\Theta_y$, are $d\Omega_{ex}^2 =
d\Omega_e^2 / 2$ and $d\Omega_{ex}^2 = d\Omega_e^2 / 2$,
respectively.

The ion equations of motion in the transverse position plane are

\begin{equation}
m \frac{d^2x}{dt^2} = - \partial_x U^{th}(x,y)
\label{equ07}
\end{equation}

\noindent and

\begin{equation}
m \frac{d^2y}{dt^2} = - \partial_y U^{th}(x,y),
\label{equ08}
\end{equation}

\noindent where $t$ denotes the time. They are solved by the
Runge-Kutta method of the fourth order \citep{pres93}. In the
calculation the time step is chosen in such a way to obtain the
fixed longitudinal distance step with the desired accuracy.

At the end of each time step of the calculation the magnitude of the
ion velocity vector is corrected using the value of the ion energy
loss during the step. The vertical and horizontal components of the
ion transmission angle at the end of each time step are chosen
within the Gaussian distribution functions defined by $\Theta_x$ and
$d\Omega_{ex}^2$ and by $\Theta_y$ and $d\Omega_{ey}^2$,
respectively. The (relativistic) ion mass at the end of each time
step is calculated using the corrected value of the magnitude of the
ion velocity vector.

The initial direction of the ion velocity vector is defined by the
angles its vertical and horizontal components make with the $x$ and
$y$ axes, $\Theta_{x0}$ and $\Theta_{y0}$, respectively. However, if
this direction is uncertain, with the dispersions of $\Theta_{x0}$
and $\Theta_{y0}$ being $\Omega_{dx}^2 = \Omega_d^2 / 2$ and
$\Omega_{dy}^2 = \Omega_d^2 / 2$, respectively, where $\Omega_d$ is
the divergence angle of the initial ion beam, the vertical and
horizontal components of the initial ion velocity vector are chosen
via the Gaussian distribution functions defined by $\Theta_{x0}$ and
$\Omega_{dx}^2$ and by $\Theta_{y0}$ and $\Omega_{dy}^2$,
respectively.

\section{Crystal rainbows}
\markright{Crystal rainbows}

\begin{figure}[h!]
\centering
\includegraphics[width=\textwidth]{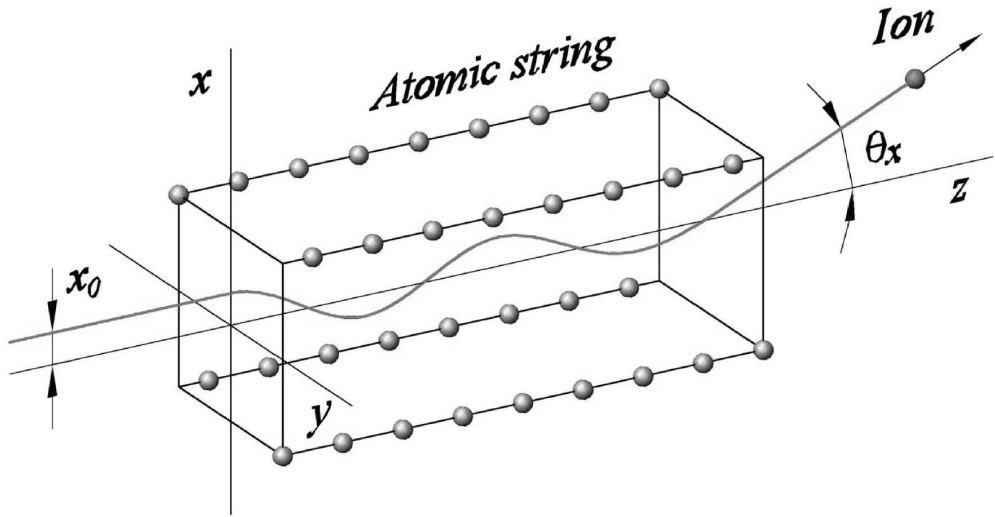}
\caption{A scheme of the process of ion transmission through an axial crystal channel. The $z$ axis of the reference frame coincides with the channel axis and its origin lies in the entrance plane of the crystal. The $x$ and $y$ axes are the vertical and horizontal axes, respectively. In this example the channel is defined by four atomic strings of the crystal, the initial ion velocity vector lies in the $xz$ plane and it is parallel to the $z$ axis. The vertical components of the ion impact parameter vector and transmission angle are designated by $x$ and $\Theta_x$, respectively. The horizontal components of these variables, $y$ and $\Theta_y$, respectively, are equal to zero.}
\label{fig3_1}
\end{figure}

Let us now examine the process of ion transmission through an axial
crystal channel. The scheme of the process is given in Figure
\ref{fig3_1}. It is well-known that the motion of an ion close to
the channel axis can be treated as an oscillatory motion around the
axis. If the ion impact parameter vector and the angle between its
initial velocity vector and the channel axis are fixed, the number
of oscillations the ion makes before leaving the crystal depends on
its initial kinetic energy. If this energy is sufficiently high for
the ion to make less than about a quarter of an oscillation around
the channel axis, its trajectory can be approximated by a straight
line. If this is true for the majority of ions, one says that the
crystal is very thin. If, however, the majority of ions make between
about a quarter of an oscillation and about one oscillation, the
crystal is thin. The crystal is thick if the majority of ions make
more than about one oscillation. Finally, if the majority of ions
make much more than one oscillation, the crystal is very thick.

\subsection{Discovery of crystal rainbows}

Ne\v{s}kovi\'{c} \citep{nesk86} showed theoretically that in the ion
transmission through an axial channel of a very thin crystal the
rainbow occurred, i.e., the ion differential transmission cross
section could be singular. His approach was via the ion-molecule
scattering theory. In the created model the continuum approximation
was included implicitly, the correlations between the positions of
the atomic strings of the crystal were included, and the assumption
of statistical equilibrium in the transverse position plane was
avoided. The effect was observed experimentally for the first time
by Krause et al. \citep{krau86}. It was named the crystal rainbow
effect \citep{nesk93}.

The first measurement of crystal rainbows was performed with 7 MeV
protons and the $<$100$>$ and $<$110$>$ Si very thin crystals
\citep{krau86}. The obtained angular distributions of channeled
protons were reproduced using the model of Ne\v{s}kovi\'{c}
\citep{nesk86} with the Lindhard's proton-crystal continuum
interaction potential \citep{lind65}. Krause et al. have also
performed the high resolution measurements of the angular
distributions of 2-9 MeV protons and 6-30 MeV C$^{4+}$, C$^{5+}$ and
C$^{6+}$ ions transmitted through the $<$100$>$ Si thin crystal
\citep{krau94}. The interpretation of the obtained results was done
using the computer code of Barrett \citep{barr71} with the
Moli\`{e}re's ion-crystal continuum interaction potential. Those
measurements demonstrated clearly the possibilities of using crystal
rainbows for exploring the properties of thin crystals, e.g., for
precise local measurement of their thickness, with the accuracy of a
few atomic layers, and for precise differential measurement of the
electron density and ion energy loss within their channels
\citep{nesk03}.

As it has been said in part \S 1 of this chapter, Petrovi\'{c} et
al. \citep{petr05a} showed theoretically that the rainbow effect can
also appear with carbon nanotubes. This will be discussed in detail
in part \S 5 of this chapter.

\subsection{Theory of crystal rainbows}

Krause et al. \citep{krau86} and Mileti\'{c} et al.
\citep{mile96,mile97} investigated the periodicity of evolution of
the angular distribution of channeled ions with the reduced crystal
thickness, which is

\begin{equation}
\Lambda = \frac{f(q,m) L}{v_0},
\label{equ09}
\end{equation}

\noindent where $q = Z_1e$, $m$ and $v_0$ are the ion charge, mass
and magnitude of the initial ion velocity vector, respectively, $L$
is the crystal thickness, and $f(q,m)$ is the average frequency of
the ion motion close to the channel axis. They found that the
evolution of the angular distribution of channeled ions could be
divided into cycles. The first cycle lasts for $\Lambda$ between 0
and 0.5, the second cycle for $\Lambda$ between 0.5 and 1, and so
on. These cycles were named the rainbow cycles. We note that the
crystal is very thin if $\Lambda$ is smaller than $\sim$0.25, it is
thin if $\Lambda$ is between $\sim$0.25 and $\sim$1, it is thick if
$\Lambda$ is larger than $\sim$1, and it is very thick if $\Lambda$
is much larger than $\sim$1. For $\Lambda = (n - 1)/2 + 1/4$, where
$n$ = 1, 2, ... is the order of the rainbow cycle, a large part of
the ion beam is focused around the origin in the transverse position
plane. This is an effect of spatial focusing, which is called the
effect superfocusing of channeled ions \citet{demk04,nesk09}. For
$\Lambda = n/2$, with $n$ = 1, 2, ..., a large part of the ion beam
is parallel to the channel axis, i.e., it is focused around the
origin in the transmission angle plane. This is an effect of angular
focusing, which is called the effect of zero-degree focusing of
channeled ions \citep{mile96,mile97}.

In order to describe the ion transmission through axial channels of
crystals that are not necessarily very thin, Petrovi\'{c} et al.
\citep{petr00} generalized the model of Ne\v{s}kovi\'{c}
\citep{nesk86} and formulated the theory of crystal rainbows. Let us
describe this theory briefly. We take that the $z$ axis of the
reference frame coincides with the channel axis and its origin lies
in the entrance plane of the crystal. The $x$ and $y$ axes of the
reference frame are the vertical and horizontal axes, respectively.
This is shown in Fig. \ref{fig3_1}. It is assumed that the
interaction of the ion and crystal can be treated classically
\citep{lind65}. The interaction of the ion and a crystal's atom is
described by the Lindhard's \citep{lind65} or Moli\`{e}re's
approximation \citep{moli47} of the Thomas-Fermi interaction
potential, or any other appropriate interaction potential. We apply
the continuum approximation, i.e., the atomic strings are treated as
if they are continual, rather than discrete \citep{lind65}. As a
result, we obtain the Lindhard's or Moli\`{e}re's ion-crystal
continuum interaction potential, or another appropriate continuum
interaction potential. The thermal vibrations of the crystal's atoms
can be taken into account as it is described in part 2 of this
chapter \citep{appl67}. We neglect the nuclear ion energy loss,
i.e., the energy loss resulting from its collisions with the
crystal's nuclei. But, the electronic ion energy loss and dispersion
of its transmission angle, which are caused by its collisions with
the crystal's electrons, can be included as it is described in part
2 of this chapter \citep{gemm74, petr02}.

The transverse components of the initial ion position vector are
equal to the components of its impact parameter vector, $x_0$ and
$y_0$, while the transverse components of the initial ion velocity
vector are $v_{x0}$ and $v_{y0}$. In order to obtain the components
of the final ion position and velocity vectors, one must solve the
ion equations of motion. Since we apply the continuum approximation
and the electronic ion energy loss during the whole transmission
process remains small, we assume that the longitudinal ion motion is
uniform and solve only the ion equations of motion in the transverse
position plane. In order to compensate the inaccuracies resulting
from this assumption, the magnitude of the ion velocity vector at
the end of each step of the calculation is corrected using the value
of the ion energy loss during the step. We also correct the
direction of the ion velocity vector at the end of each step of the
calculation using the value of the change of the dispersion of its
transmission angle during the step. This is how the ion energy loss
and change of the dispersion of its transmission angle are included
in the calculations. The relativistic effect is taken into account
by changing the ion mass at the end of each step of the calculation
using the corrected value of the magnitude of the ion velocity
vector. The obtained transverse components of the final ion position
vector are $x(x_0,y_0,\Lambda)$ and $y(x_0,y_0,\Lambda)$, and the
transverse components of the final ion velocity vector are
$v_x(x_0,y_0,\Lambda)$ and $v_y(x_0,y_0,\Lambda)$. Since the ion
transmission angle during the whole transmission process remains
small, the components of the final ion transmission angle are
$\Theta_x(x_0,y_0,\Lambda) = v_x/v$ and $\Theta_y(x_0,y_0,\Lambda) =
v_y/v$, where $v = v(x_0,y_0,\Lambda)$ is the magnitude of the final
ion velocity vector. The spatial and angular distributions of
transmitted ions, in the exit plane of the crystal, are generated by
the computer simulation method. The spatial and angular
distributions of ions in the entrance plane of the crystal are
determined by the spatial and angular characteristics of the chosen
initial ion beam. If the ion, during its motion along the channel,
enters any of the cylindrical regions around the atomic strings of
the radius equal to the screening radius of a crystal's atom, it is
excluded from the calculations.

Since the ion transmission angle during the whole transmission
process remains small, the ion differential transmission cross
section is

\begin{equation}
\sigma(x_0,y_0,\Lambda) = \frac{1}{\left| J_\Theta (x_0,y_0,\Lambda ) \right|},
\label{equ10}
\end{equation}

\noindent where

\begin{equation}
{J_\Theta}(x_0,y_0,\Lambda) = \partial_{x_0} \Theta_x \partial_{y_0} \Theta_y - \partial_{y_0} \Theta_x \partial_{x_0} \Theta_y
\label{equ11}
\end{equation}

is the Jacobian of the components of its transmission angle, which
equals the ratio of the infinitesimal surfaces in the transmission
angle plane and impact parameter plane. It describes the mapping of
the impact parameter plane to the transmission angle plane. Hence,
equation $J_\Theta(x_0,y_0,\Lambda) = 0$ gives the angular rainbow
lines in the impact parameter plane, along which the mapping is
singular. The images of these lines determined by functions
$\Theta_x(x_0,y_0,\Lambda)$ and $\Theta_y(x_0,y_0,\Lambda)$ are the
rainbow lines in the transmission angle plane.

On the other hand, the mapping of the impact parameter plane to the
transverse position plane is described by the Jacobian of the
transverse components of the ion position vector. This Jacobian,
which equals the ratio of the infinitesimal surfaces in the
transverse position plane and impact parameter plane, reads

\begin{equation}
J_\rho(x_0,y_0,\Lambda) = \partial_{x_0} x \partial_{y_0} y - \partial_{y_0} x \partial_{x_0} y.
\label{equ12}
\end{equation}

\noindent Thus, equation $J_\rho(x_0,y_0,\Lambda) = 0$ gives the
spatial rainbow lines in the impact parameter plane, along which the
mapping is singular. The images of these lines determined by
functions $x(x_0,y_0,\Lambda)$ and $y(x_0,y_0,\Lambda)$ are the
rainbow lines in the transverse position plane.

The rainbow lines in the transverse position plane and transmission
angle plane separate the bright and dark regions in these planes.
Their shapes are classified by catastrophe theory \citep{thom75,
nesk87, nesk93}. Hence, one says that the ion beam dynamics in the
channel has the catastrophic character.

\section{Zero-degree focusing of protons channeled through bundles of nanotubes}
\markright{Zero-degree focusing of protons...}

The first calculations of the effect of zero-degree focusing of
channeled ions, which is defined in part \S 3.2 of this chapter,
were performed by Mileti\'{c} et al. \citep{mile96, mile97}. In the
former study the projectiles were 25 MeV C$^{6+}$ ions and the
target was the $<$100$>$ Si crystal of the thickness between 0 and
2.7 $\mu$m, while in the latter study the projectiles were
Ne$^{10+}$ ions of the kinetic energies between 1 and 37 MeV while
the target was the 1.0 $\mu$m thick $<$100$>$ Si crystal. The effect
has not yet been observed experimentally.

We shall describe here the effect of zero-degree focusing of 1 GeV
protons channeled through the (10, 10) single-wall carbon nanotubes
\citep{nesk05,petr05b}. Such a nantoube is achiral, i.e., its atomic
strings are parallel to its axis. It is assumed that the nanotubes
form a bundle whose transverse cross-section can be descirbed via a
(two-dimensional) hexagonal or rhombic superlattice with one
nanotube per primitive cell \citep{thes96}. We choose the $z$ axis
of the reference frame to coincide with the bundle axis and its
origin to lie in the entrance plane of the bundle. The arrangement
of the nanotubes is such that their axes intersect the $x$ and $y$
axes of the reference frame, which are the vertical and horizontal
axes, respectively \citep{petr05a}. We take into account the
contributions of the nanotubes lying on the two nearest rhombic
coordination lines, relative to the center of the primitive cell of
the (rhombic) superlattice.

The calculations are performed using the theory of crystal rainbows,
which is described in part \S 3.2 of this chapter. The interaction
of the proton and a nanotube atom is described by the Moli\`{e}re's
approximation \citep{moli47} of the Thomas-Fermi interaction
potential, which reads

\begin{equation}
V(r') = \frac{Z_1 Z_2 e^2}{r'} \left[ {0.35 \exp(-0.3r'/a) + 0.55 \exp(-1.2r'/a) + 0.10 \exp(-6.0r'/a)} \right]
\label{equ13}
\end{equation}

\noindent where $Z_1$ = 1 and $Z_2$ = 6 are the atomic numbers of
the proton and nanotube atom, respectively, $e$ is the elementary
charge, $r'$ is the distance between the proton and nanotube atom,
$b = 0.3/a$, $a = \left[ {9 \pi^2 / (128 Z_2)} \right]^{\frac{1}{3}}
a_0$ is the screening radius of the nanotube atom, and $a_0$ is the
Bohr radius. It has been proven that this expression provides
excellent agreement with experimental results in the field of ion
channeling \citep{krau94}. The application of the continuum
approximation gives the Moli\`{e}re's proton-bundle continuum
interaction potential, which reads

\begin{equation}
U(x,y) = \sum\limits_{i = 1}^I {\sum\limits_{j = 1}^J U_{ij}(x,y)},
\label{equ14}
\end{equation}

\noindent where

\begin{equation}
U_{ij}(x,y) = \frac{2 Z_1 Z_2 e^2}{d} \left[ 0.35K_0(0.3 \rho_{ij}/a) + 0.55K_0(1.2 \rho_{ij}/a) + 0.10K_0(6.0 \rho_{ij}/a) \right]
\label{equ15}
\end{equation}

\noindent is the Moli\`{e}re's continuum interaction potential of
the proton and the jth atomic string of the $i^{th}$ nanotube within
the bundle,

\begin{equation}
\rho_{ij} = \left[ (x - x_{ij})^2 + (y - y_{ij})^2 \right]^{\frac{1}{2}}
\label{equ16}
\end{equation}

\noindent is the distance between the proton and this atomic string,
and $x$ and $y$ are the transverse components of the proton position
vector; $I$ = 16 is the number of nanotubes within the bundle, $J$ =
40 is the number of atomic strings of a nanotube, $d$ = 0.24 nm is
the distance between the atoms of an atomic string, and $K_0$
denotes the modified Bessel function of the second kind and $0^{th}$
order. This means that the number of atomic strings within the
bundle is $I \times J$ = 640. The thermal vibrations of the nanotube
atoms are taken into account. This is done by substituting
$U_{ij}(x,y)$ in Eq. (\ref{equ14}) with

\begin{equation}
U_{ij}^{th}(x,y) = U_{ij}(x,y) + \frac{\sigma_{th}^2}{2} \left[ \partial_{xx} U_{ij}(x,y) + \partial_{yy} U_{ij}(x,y) \right],
\label{equ17}
\end{equation}

\noindent where $\sigma_{th}$ = 5.3 pm is the one-dimensional
thermal vibration amplitude of the nanotube atoms \citep{hone00}.
Thus, $U^{th}(x,y)$ is obtained instead of $U(x,y)$. The electronic
proton energy loss and dispersion of its transmission angle, caused
by its collisions with the nanotube electrons, are neglected. The
nanotube radius is 0.67 nm \citep{sait01} and the distance between
the axes of two neighboring nanotubes is 1.70 nm \citep{thes96}. The
components of the proton impact parameter vector are chosen
uniformly within the primitive cell of the (rhombic) superlattice.
The initial proton velocity vectors are all taken to be parallel to
the bundle axis.

The nanotube walls define two separate regions in the transverse
position plane: inside the nanotubes and in between them
\citep{petr05a}. Accordingly, the bundle contains two types of
channels: the circular one, whose center coincides with the center
of the region inside each nanotube, and the triangular one, whose
center coincides with the center of the region in between each three
neighboring nanotubes. It is clear that in this case one has to
define two average frequencies of the proton motion close to the
channel axis, corresponding to the protons moving close to the
centers of the circular and triangular channels. The two frequencies
can be determined from the second order terms of the Taylor
expansions of the proton-bundle continuum interaction potential in
the vicinities of the centers of the two types of channels.
Consequently, there are two reduced bundle lengths, $\Lambda_1$ and
$\Lambda_2$, corresponding to the proton motions close to the axes
of the circular and triangular channels, respectively.

\begin{figure}
\centering
\includegraphics[width=0.45\textwidth]{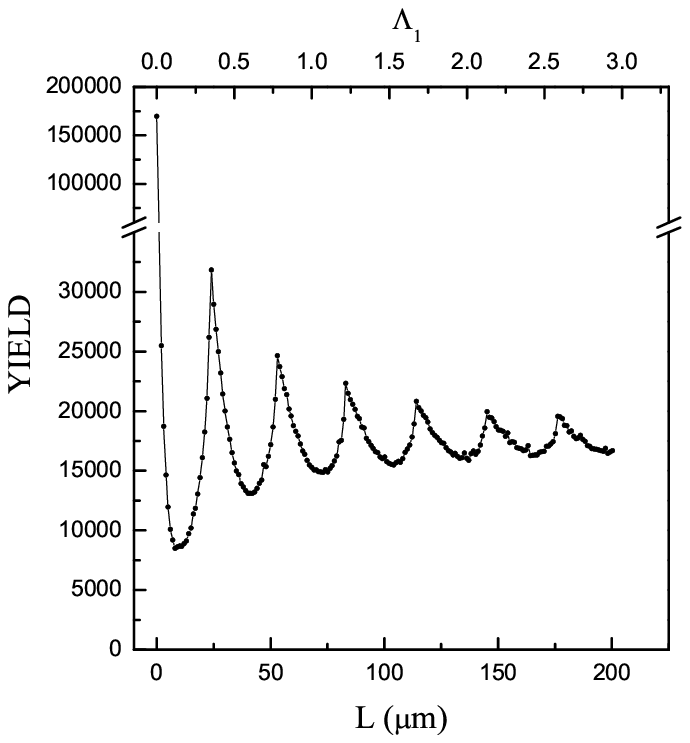}
\hspace*{1cm}
\includegraphics[width=0.45\textwidth]{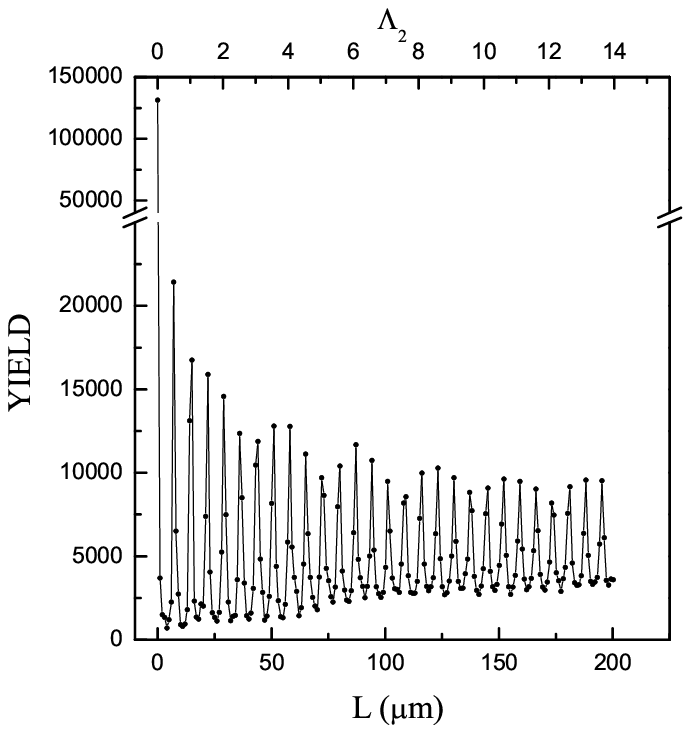}
\caption{(a) The zero-degree yield of 1 GeV protons channeled through the circular channels of the bundle of (10, 10) single-wall carbon nanotubes as a function of the bundle length, $L$. (b) The zero-degree yield of 1 GeV protons channeled through the triangular channels of the bundle of (10, 10) single-wall carbon nanotubes as a function of the bundle length, $L$.}
\label{fig4_1}
\end{figure}

Figure \ref{fig4_1}(a) shows the dependence of the zero-degree yield
of protons propagating along the circular channels of the bundle on
the bundle length, $L$, in the range of 0-200 $\mu$m; the
$\Lambda_1$ axis is shown too. For the region in the transmission
angle plane in the vicinity of its origin we take the region in
which the proton transmission angle, $\Theta = (\Theta_x^2 +
\Theta_y^2)^{\frac{1}{2}}$, where $\Theta_x$ and $\Theta_y$ are its
vertical and horizontal components, respectively, is smaller than
0.0109 mrad. The initial number of protons is 174~976. The six
maxima of the dependence correspond to the ends of the first six
rainbow cycles, where the proton beam channeled through the circular
channels is quasi-parallel \citep{mile96,mile97}. The positions of
these maxima should be close to $\Lambda_1$ = 0.5, 1, 1.5, ....
However, this is not true. For example, the position of the first
maximum of the dependence is $\Lambda_1$ = 0.35. The large
deviations of the values of these position from their expected
values are explained by the strong anharmonicity of the
proton-bundle continuum interaction potential in the vicinities of
the centers of the circular channels \citep{mile97}.

The dependence of the zero-degree yield of protons propagating along
the triangular channels of the bundle on $L$ in the range of 0-200
$\mu$m is given in Fig. \ref{fig4_1}(b); the $\Lambda_2$ axis is
shown too. The initial number of protons is 136~658. The dependence
has 27 maxima, and they correspond to the ends of the first 27
rainbow cycles, where the proton beam channeled through the
triangular channels is quasi-parallel \citep{mile96,mile97}. The
positions of these maxima are close to $\Lambda_2$ = 0.5, 1, 1.5,
.... For example, the position of the first maximum of the
dependence is $\Lambda_2$ = 0.49. The small deviations of the values
of these position from their expected values are attributed to the
weak anharmonicity of the proton-bundle continuum interaction
potential in the vicinities of the centers of the triangular
channels \citep{mile97}. It should be noted that the total
zero-degree yield of channeled protons, i.e., of protons channeled
through both the circular and triangular channels, as a function of
$L$ is the sum of the zero-degree yields shown in Fig.
\ref{fig4_1}(a) and Fig. \ref{fig4_1}(b).

We have also investigated the effect of zero-degree focusing of 1
GeV protons channeled through the (10, 0) single-wall carbon
nanotubes \citep{bork08c}. The study has been performed in the same
way as in the case of (10, 10) nanotubes. The number of nanotubes
within the bundle is $I$ = 16, the number of atomic strings of a
nanotube is $J$ = 20, the distance between the atoms of an atomic
string is $d$ = 0.21 nm. The nanotube radius is 0.39 nm
\citep{sait01}, and the distance between the axes of two neighboring
nanotubes is 1.10 nm \citep{zhan03}. In this case the dependence of
the zero-degree yield of protons channeled through the circular
channels of the bundle on $L$ in the range of 0-200 $\mu$m contains
26 maxima. The positions of these maxima are close to $\Lambda_1$ =
0.5, 1, 1.5, .... The average distance between the maxima is $\Delta
\Lambda_1$ = 0.49. This means that the anharmonicity of the
proton-bundle continuum interaction potential in the vicinities of
the centers of the circular channels is weak \citep{mile97}. In this
case the dependence of the zero-degree yield of protons channeled
through the triangular channels of the bundle on $L$ in the range of
0-200 $\mu$m contains 37 maxima. The positions of these maxima are
very close to $\Lambda_2$ = 0.5, 1, 1.5, .... The average distance
between the maxima is $\Delta \Lambda_2$ = 0.50. Thus, one can say
that the anharmonicity of the proton-bundle continuum interaction
potential in the vicinities of the centers of the triangular
channels is very weak \citep{mile97}.

The comparison of the zero-degree yields of protons channeled
through the circular and triangular channels of the bundles as
functions of $L$ in the cases of (10, 10) and (10, 0) nanotubes
shows that:

\begin{list}{-}{}
\item (i) when the circular channels are considered, the average
    frequency of the proton motion close to the channel axis is much
    lower in the former than in the latter case,
\item (ii) when the triangular channels are considered, the average
    frequency of the proton motion close to the channel axis is
    lower in the former than in the latter case, and
\item (iii) when the circular and triangular channels are compared
    to each other, the ratio of the average frequencies of the
    proton motions close to the channel axes in the circular and
    triangular channels is much smaller in the former than in the
    latter case.
\end{list}

Thus, one can conclude that the measurements of the effect of
zero-degree focusing of channeled protons can give the information
on the transverse lattice structures of the nanotubes and of the
bundles.

\section{Angular distributions and rainbows in proton channeling through nanotubes}
\markright{Angular distributions and rainbows...}

We shall present in this part of the chapter the angular
distributions and rainbows in channeling 1 GeV protons through a
straight very short bundle of (10, 10) single-wall carbon nanotubes
\citep{petr05a}, a bent short bundle of (10, 10) nanotubes
\citep{nesk05}, and the straight long (11, 9) nanotubes
\citep{petr08b}.

\subsection{Rainbows with a straight very short bundle of na\-notubes}

The system we investigate here is a 1 GeV proton moving through a
bundle of (10, 10) single-wall carbon nanotubes that is described in
part 4 of this chapter. The bundle length is 1 $\mu$m
\citep{petr05a}. The reduced bundle lengths corresponding to the
proton channeling inside the nanotubes and in between them, i.e.,
through the circular and triangular channels of the bundle, are
$\Lambda_1$ = 0.015 and $\Lambda_2$ = 0.070, respectively. These
values tell us that in both cases the majority of protons make
before leaving the bundle less than a quarter of an oscillation
around the channel axis ($\Lambda_1,\Lambda_2 <$ 0.25). Therefore,
one can say that the bundle we investigate is very short.

\begin{figure}[ht!]
\centering
\includegraphics[width=0.65\textwidth]{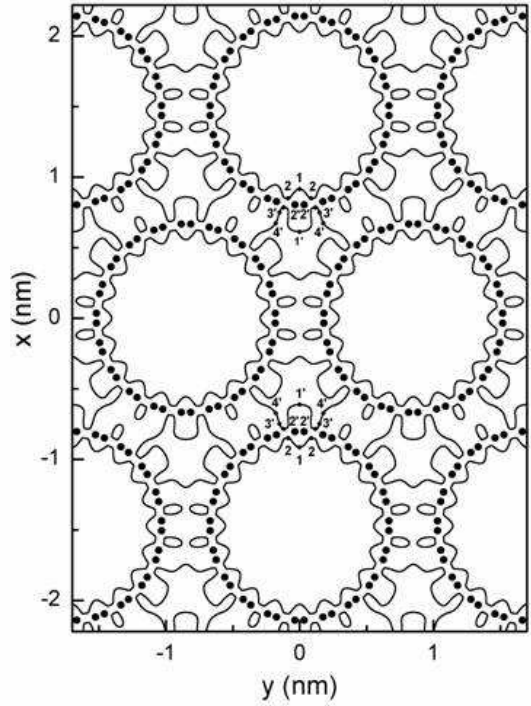}
\caption{The rainbow lines in the impact parameter plane for 1 GeV protons transmitted through the 1 $\mu$m long (10, 10) single-wall carbon nanotubes.}
\label{fig5_1_1}
\end{figure}

The calculations are performed using the theory of crystal rainbows,
which is described in part \S 3.2 of this chapter. The interaction
of the proton and bundle is described by Eqs.
(\ref{equ14})-(\ref{equ16}). We take into account the effect of
thermal vibrations of the nanotube atoms, and this is done by Eq.
(\ref{equ17}). However, the electronic proton energy loss and
dispersion of its transmission angle are neglected. This is
justified by the fact that the bundle is very short. If the bundle
were long ($\Lambda_1,\Lambda_2 >$ 1), these effect would cause the
smearing out of the angular distributions of channeled protons
\citep{petr02}.

\begin{figure}
\centering
\includegraphics[width=0.48\textwidth]{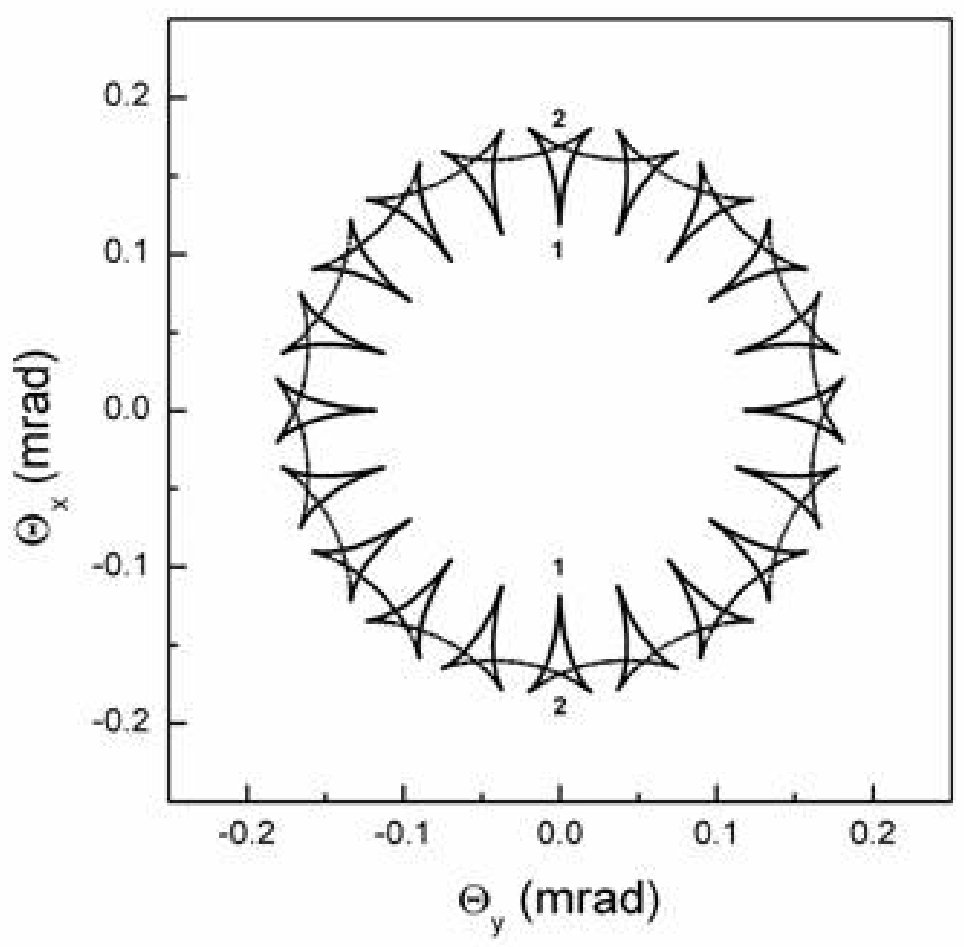}
\includegraphics[width=0.48\textwidth]{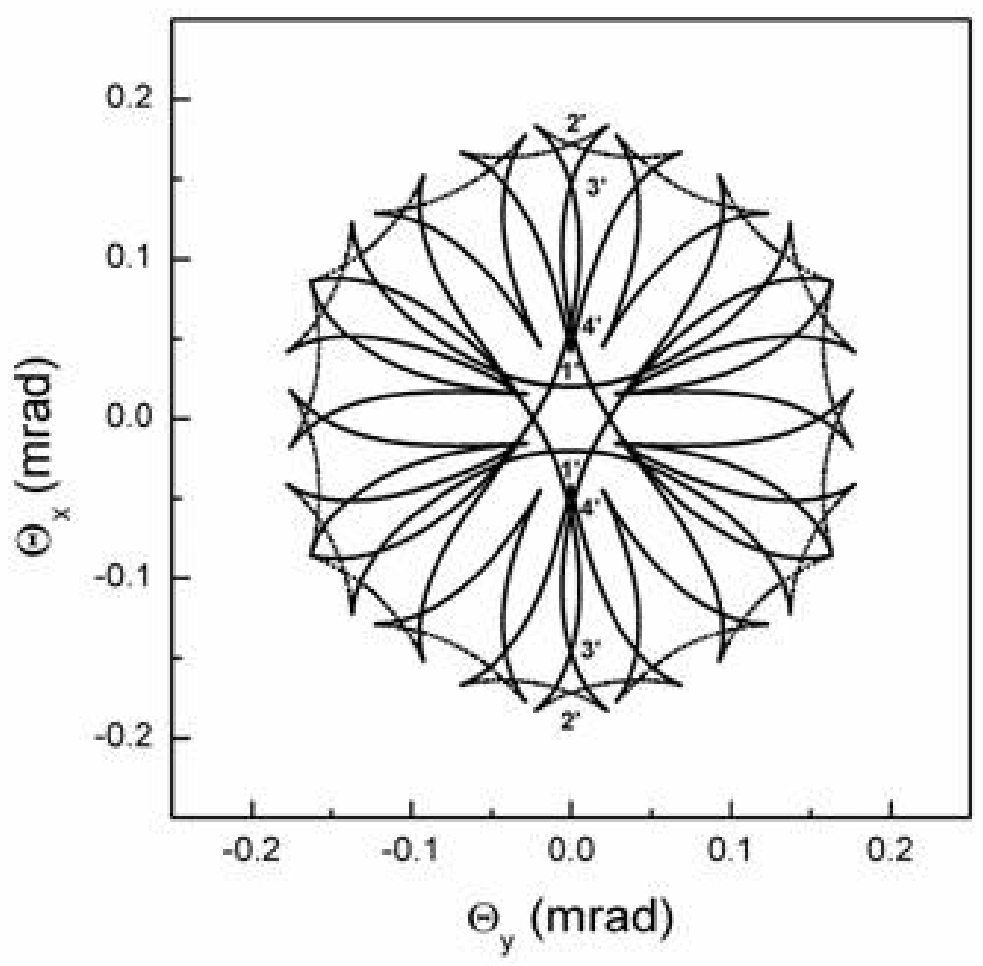}
\caption{The rainbow lines in the transmission angle plane corresponding to the rainbow lines in the impact parameter plane (shown in Fig. 1) lying (a) inside each nanotube, and (b) in between each four neighboring nanotubes.}
\label{fig5_1_2}
\end{figure}

\begin{figure}
\centering
\includegraphics[width=0.40\textwidth]{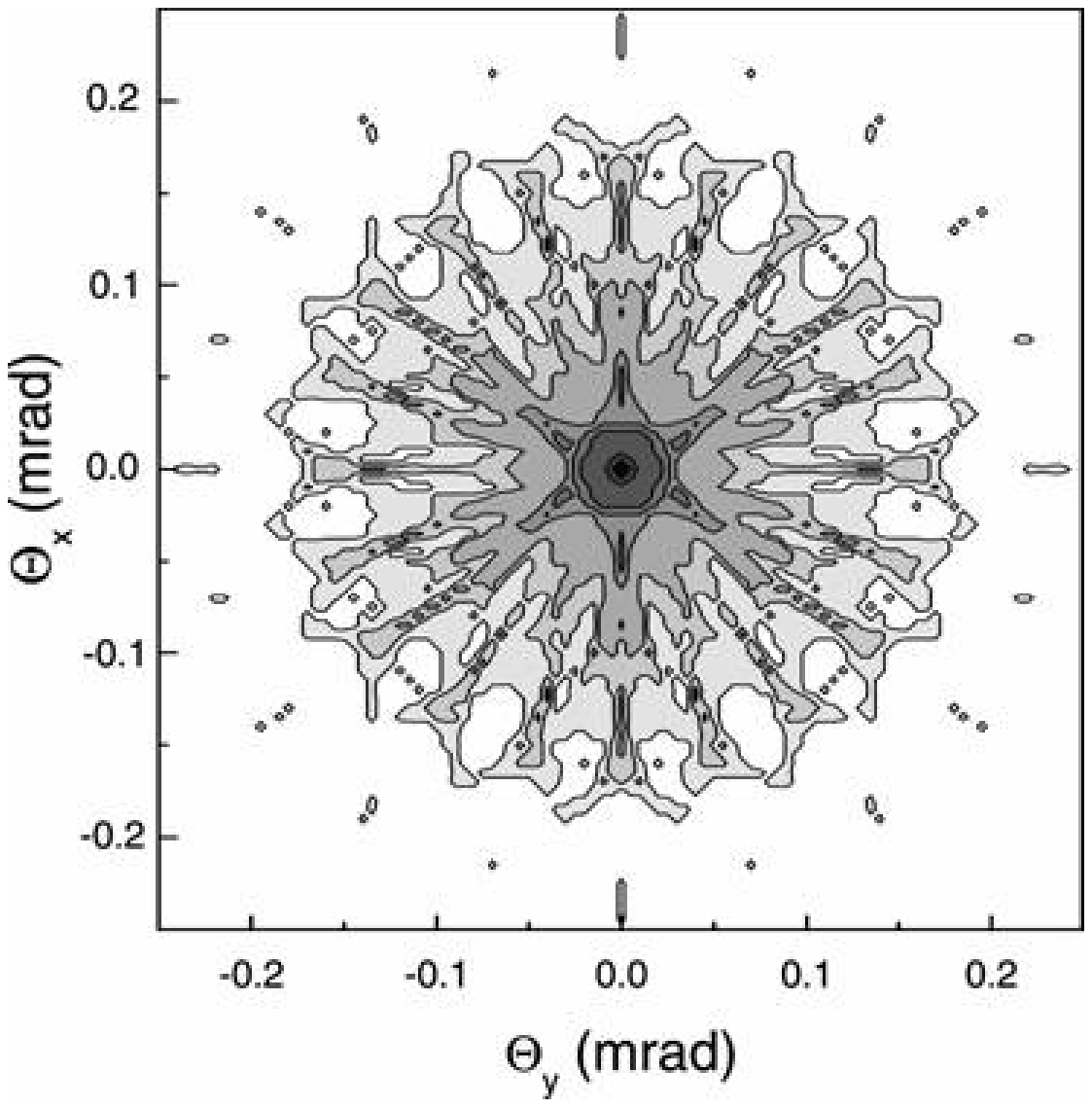}
\hspace*{0.3cm}
\includegraphics[width=0.55\textwidth]{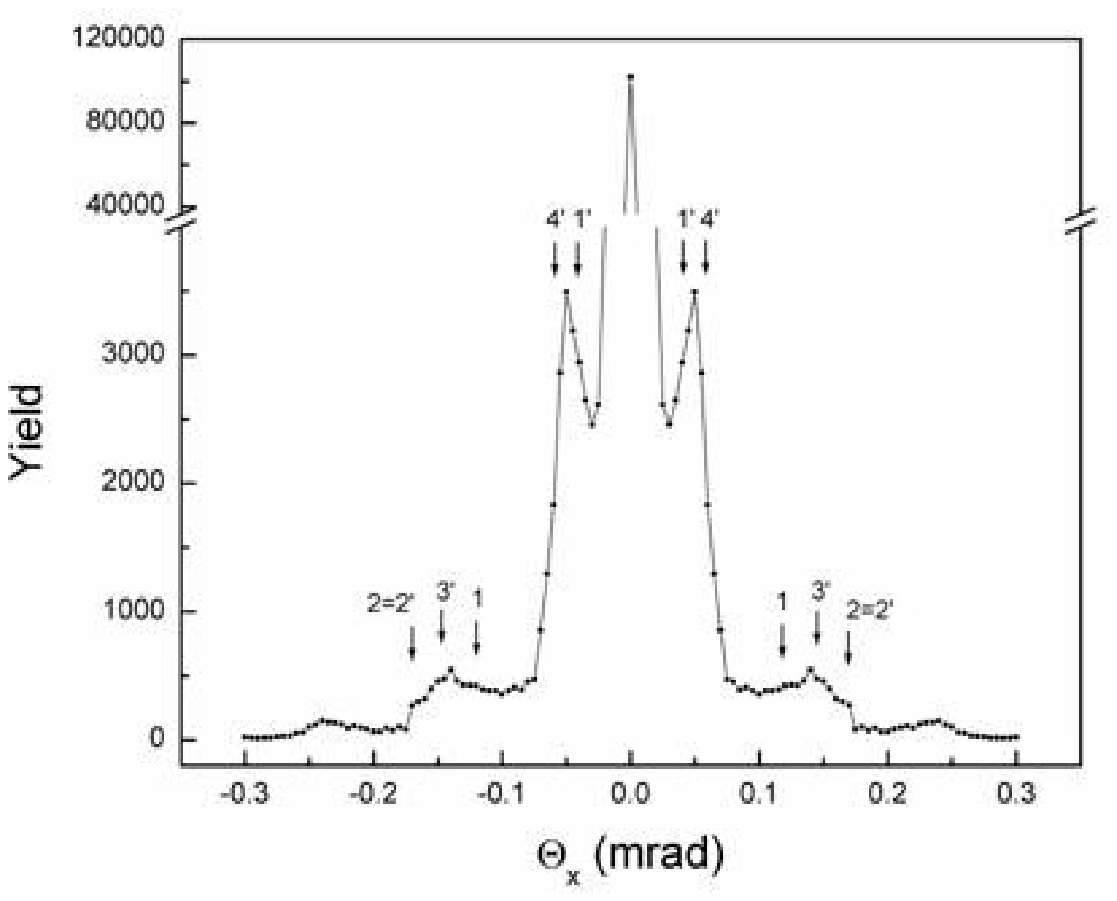}
\caption{(a) The angular distribution of 1 GeV protons transmitted through the 1 $\mu$m long bundle of (10, 10) single-wall carbon nanotubes. The areas in which the yields of transmitted protons are larger than 0.13, 0.26, 0.39, 1.3, 2.6, 3.9, 13, 26, 39 $\%$ of the maximal yield are designated by the increasing tones of gray color. (b) The corresponding yield of transmitted protons along the line $\Theta_y$ = 0.}
\label{fig5_1_3}
\end{figure}

Figure \ref{fig5_1_1} shows the rainbow lines in the impact
parameter plane in the case under consideration. One can see that
inside each nanotube, i.e., inside each circular channel of the
bundle, there is one (closed) rainbow line, while in between each
three neighboring nanotubes, i.e., inside each triangular channel of
the bundle, there are one larger and four smaller (closed) rainbow
lines.

The rainbow line in the transmission angle plane that is the image
of the rainbow line in the impact parameter plane lying inside each
nanotube is shown in Figure \ref{fig5_1_2}(a). It consists of 20
connected cusped triangular lines lying along the lines $\varphi =
\tan^{-1}(\Theta_y / \Theta_x) = 2(n + 1) \pi / 20$, $n$ = 0-19,
which correspond to the parts of the rainbow line in the impact
parameter plane in front of the 20 pairs of atomic strings defining
the nanotube (see Fig. \ref{fig5_1_1}). Points 1 and 2 are the
intersection points of the rainbow line in the transmission angle
plane with the line $\Theta_y$ = 0. Points 1 are the apices of the
cusps and points 2 are the intersections of the parts of the rainbow
line. The corresponding points in the impact parameter plane are
also designated by 1 and 2.

Figure \ref{fig5_1_2}(b) shows the rainbow lines in the transmission
angle plane that are the images of the rainbow lines in the impact
parameter plane lying in between each four neighboring nanotubes.
The analysis shows that the rainbow pattern consists of two cusped
equilateral triangular rainbow lines in the central region of the
transmission angle plane with the cusps lying along the lines
$\varphi = 2n \pi / 3$ and $\varphi = (2n + 1) \pi / 3$, $n$ = 0-2,
each connected with three pairs of cusped triangular rainbow lines
lying along the same lines, and eight cusped triangular rainbow
lines lying in between the six pairs of triangular lines. The two
equilateral triangular lines each connected with the three pairs of
triangular lines are the images of the two larger rainbow lines
while the eight triangular lines are the images of the eight smaller
rainbow lines in the impact parameter plane (see Fig.
\ref{fig5_1_1}). Points 1', 2', 3' and 4' are the intersection
points of the rainbow lines in the scattering angle plane with line
$\Theta_y$ = 0. Points 1' are the apices of the cusps, and points
2', 3' and 4' are the intersections of the parts of the larger
rainbow lines. The corresponding points in the impact parameter
plane are also designated by 1', 2', 3' and 4'.

Figure \ref{fig5_1_3}(a) shows the angular distribution of
transmitted protons. The number of transmitted protons is 2~142~538.
The areas in which the yields of transmitted protons are larger then
0.13, 0.26 and 0.39 $\%$, 1.3, 2.6 and 3.9 $\%$, and 13, 26 and 39
$\%$ of the maximal yield are designated by the increasing tones of
gray color. At the very low level of the yield, corresponding to the
boundary yields of 0.13, 0.26 and 0.39 $\%$ of the maximal yield,
there are 20 triangular forms in the peripheral region of the
transmission angle plane, with the maxima lying on the lines
$\varphi = 2(n + 1) \pi / 20$, $n$ = 0-19. Further, at the low level
of the yield, corresponding to the boundary yields of 1.3, 2.6 and
3.9 $\%$ of the maximal yield, there is a hexagonal structure in the
central region of the transmission angle plane, with the maxima
lying on the lines $\varphi = n \pi / 3$, $n$ = 0-5. Finally, at the
high level of the yield, corresponding to the boundary yields of 13,
26 and 39 $\%$ of the maximal yield, there is a pronounced maximum
at the origin of the transmission angle plane. The analysis shows
that the first part of the angular distribution is generated by the
protons with the impact parameters close to the atomic strings
defining the nanotubes - the 20 triangular forms correspond to the
20 pairs of atomic strings defining the nanotubes. The second part
of the angular distribution is generated by the protons with the
impact parameters in between the nanotubes but not close to the
centers of the triangular channels. The third part of the angular
distribution is generated to a larger extent by the protons with the
impact parameters close to the centers of the circular channels and
to a smaller extent by the protons with the impact parameters close
to the centers of the triangular channels. It should be noted that
most of the protons that generate the third part of the angular
distribution interact with the nanotubes very weakly - they move
through the space inside the nanotubes virtually as through a drift
space. Thus, we can say that the angular distribution contains the
information on the transverse lattice structure of the bundle. Its
first part (the peripheral region of the transmission angle plane)
provides the information on the individual nanotubes while its
second part (the central region of the transmission angle plane)
provides the information on the way they are connected to each
other.

The comparison of Figs. \ref{fig5_1_2}(a) and \ref{fig5_1_2}(b) with
Fig. \ref{fig5_1_3}(a) shows clearly that the shape of the rainbow
pattern determines the shape of the angular distribution of
transmitted protons. Also, each maximum of the angular distribution,
except the maximum lying at the origin of the transmission angle
plane, can be attributed to one of the above mentioned
characteristic rainbow points in the transmission angle plane. Thus,
one can conclude that the rainbow pattern enables the full
explanation of the angular distribution.

Figure \ref{fig5_1_3}(b) gives the low and very low levels of the
yield of transmitted protons along line $\Theta_y$ = 0. The arrows
indicate the above mentioned characteristic rainbow points in the
transmission angle plane. It is evident that the two maxima at the
low level of the yield can be explained by points 1' and 4' in the
transmission angle plane, and the two shoulders at the very low
level of the yield by points 1, 2, 2' and 3'. This means that the
characteristic rainbow points in the transmission angle plane can
provide the information on the continuum potential of the bundle at
the corresponding points in the impact parameter plane, and, hence,
on the average electron density in the bundle at these points (see
Fig. \ref{fig5_1_1}). The two maxima at the low level of the yield
can be used to measure the average electron density at points 1' and
4', lying in between the nanotubes, while the two shoulders at the
very low level of the yield can be used to measure the average
electron density at points 1 and 2, lying in the nanotube, and at
points 2' and 3', lying in between the nanotubes. The obtained data
can help one compare various theoretical approaches and determine
the electron structure of the bundle.

We have also performed the analyses of the angular distributions of
1 GeV protons transmitted through the 1 $\mu$m long bundles of (10,
0) and (5, 5) single-wall carbon nanotubes. These nanotubes are
achiral too. In the former case each nanotube consists of 20 atomic
strings and in the latter case of 10 pairs of atomic strings. The
angular distribution in the (10, 0) case is similar to the one in
the (10, 10) case while the angular distribution in the (5, 5) case
is very different from the one in the (10, 10) case. In both cases
it is easy to establish the correspondence between the parts of the
angular distributions and the transverse lattice structures of the
bundles. It should be also noted that the evolution of the angular
distribution in the (10, 0) case \citep{bork05} with the proton
energy or the bundle length is different from the evolution in the
(10, 10) case \citep{petr05b}, enabling one to distinguish between
the two types of bundles. Thus, the obtained results can lead to a
new method of characterization of achiral carbon nanotubes, based on
the rainbow effect. This method would be complementary to the
existing method of characterization of nanotubes by electrons
impinging on them transversely rather than longitudinally, which is
based on the diffraction effect (see, e.g., \citet{luca02}).

\subsection{Rainbows with a bent short bundle of nanotubes}

Again, the system we explore is a 1 GeV proton moving through a
bundle of (10, 10) single-wall carbon nanotubes that is described in
part \S 4 of this chapter. However, now, the bundle is bent along
the $y$ axis. The bundle length is 7 $\mu$m and its bending angle is
varied in the range of 0-1.5 mrad \citep{nesk05}. In this case the
critical angle for channeling is $\Psi_C$ = 0.314 mrad. If the angle
of the proton velocity vector relative to the channel axis is above
$\Psi_C$, the proton is dechanneled. The reduced bundle lengths
corresponding to the proton channeling through the circular and
triangular channels of the bundle are $\Lambda_1$ = 0.103 and
$\Lambda_2$ = 0.488, respectively. Thus, in the circular channels
the majority of protons make before leaving the bundle less than a
quarter of an oscillation around the channel axis ($\Lambda_1 <$
0.25), and in the triangular channel they make close to half an
oscillation ($\Lambda_2 \approx$ 0.5). Hence, for the protons in the
circular channels the bundle is very short while for those in the
triangular channels it is short ($\Lambda_1 <$ 1).

The calculations are performed using the theory of crystal rainbows,
which is described in part \S 3.2 of this chapter. The interaction
of the proton and bundle is described by Eqs.
(\ref{equ14})-(\ref{equ16}). We take into account the effect of
thermal vibrations of the nanotube atoms, and this is done by Eq.
(\ref{equ17}). However, the electronic proton energy loss and
dispersion of its transmission angle are neglected. This is
justified by the above given values of $\Lambda_1$ and $\Lambda_2$.

The bundle length has been chosen to correspond to the first maximum
of the zero-degree yield of protons propagating along the triangular
channels of the bundle as a function of the bundle length, which is
given in Fig. \ref{fig4_1}(b). Thus, the part of the proton beam
transmitted through the triangular channels is quasi-parallel, and
we can investigate in an easier manner the behavior of the proton
beam transmitted through the circular channels.

\begin{figure}[ht!]
\centering
\includegraphics[width=0.45\textwidth]{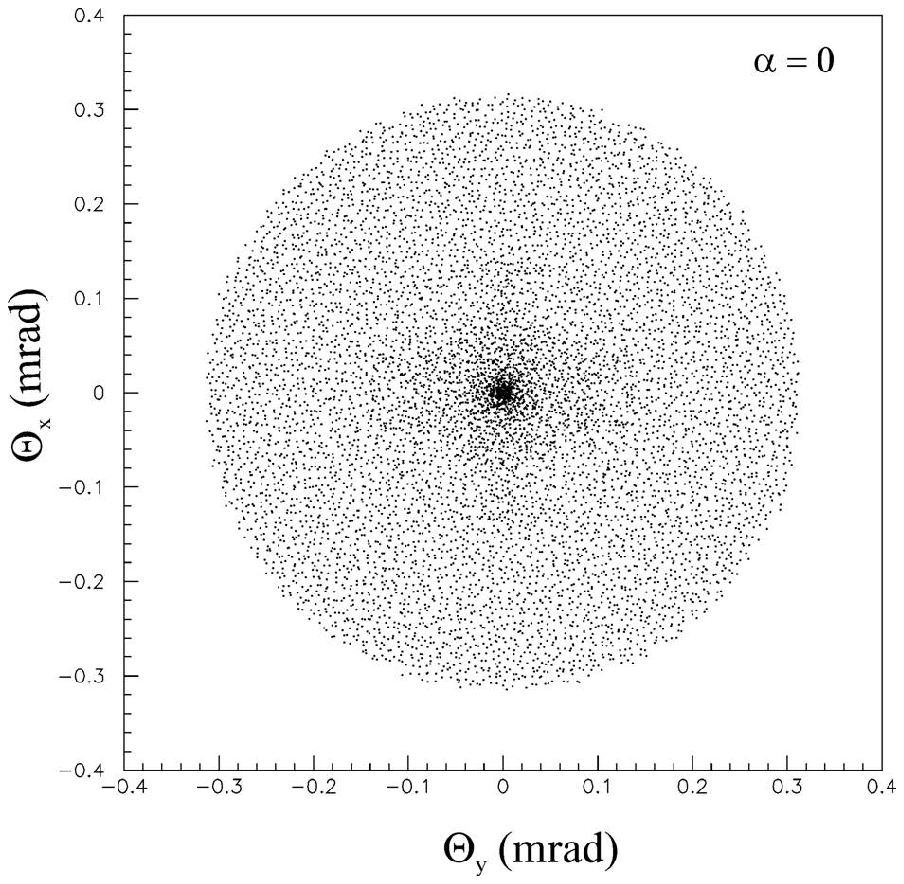}
\hspace*{1cm}
\includegraphics[width=0.45\textwidth]{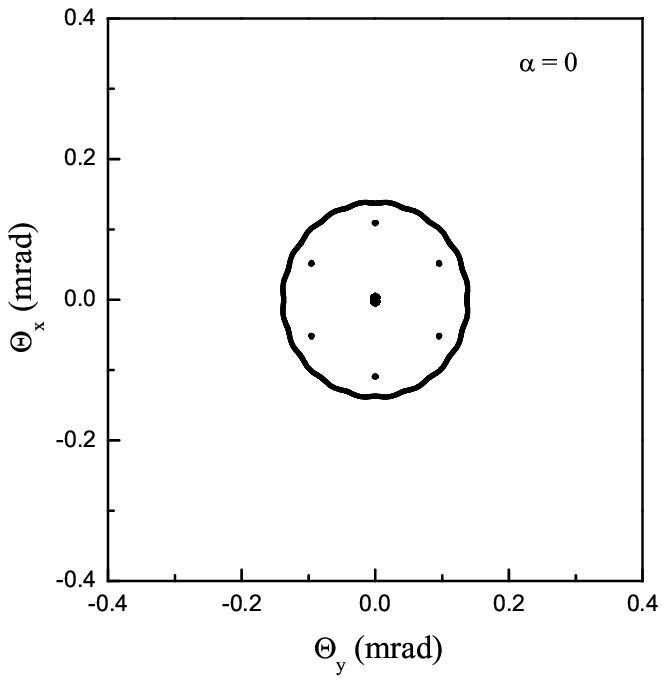}
\caption{(a) The angular distribution of 1 GeV protons transmitted through the straight bundle of (10, 10) single-wall carbon nanotubes of the length of $L$ = 7 $\mu$m. (b) The corresponding rainbow pattern in the transmission angle plane.}
\label{fig5_2_1}
\end{figure}

Figure 5.4(a) shows the angular distribution of protons transmitted
through the straight bundle, i.e., for its bending angle $\alpha$ =
0. The sizes of a bin along the $\Theta_x$ and $\Theta_y$ axes are
equal to 0.00667 mrad while the initial number of protons is
866~976. The angular distribution contains (i) a pronounced maximum
at the origin of the transmission angle plane, (ii) six
non-pronounced maxima lying along the lines defined by $\varphi =
\tan^{-1}(\Theta_y / \Theta_x) = n \pi / 3$, $n$ = 0-5, and (iii) a
non-pronounced circular part. The rainbow pattern in the
transmission angle plane obtained for $\alpha$ = 0 is given in Fig.
\ref{fig5_2_1}(b). It contains (i) two six-cusp lines very close to
the origin of the transmission angle plane, corresponding to two
complex lines in the impact parameter plane close to the centers of
the two triangular channels making each rhombic channel of the
bundle, (ii) six points lying along the lines defined by $\varphi =
n \pi / 3$, $n$ = 0-5, corresponding to two times three points in
the impact parameter plane within the two triangular channels making
each rhombic channel, and (iii) a circular line, corresponding to a
circular line in the impact parameter plane within each circular
channel. The comparison of the results given in Figs.
\ref{fig5_2_1}(a) and 5.4(b) shows that the angular distribution can
be explained as follows: (i) the pronounced maximum at the origin of
the transmission angle plane is generated by the protons with the
impact parameter vectors close to the centers of each circular
channel and each triangular channel, (ii) the six non-pronounced
maxima by the protons with the impact parameter vectors close to the
six rainbow points (within each rhombic channel), and (iii) the
non-pronounced circular part by the protons with the impact
parameter vectors close to the circular rainbow line (within each
circular channel).

\begin{figure}[b!]
\centering
\includegraphics[width=0.45\textwidth]{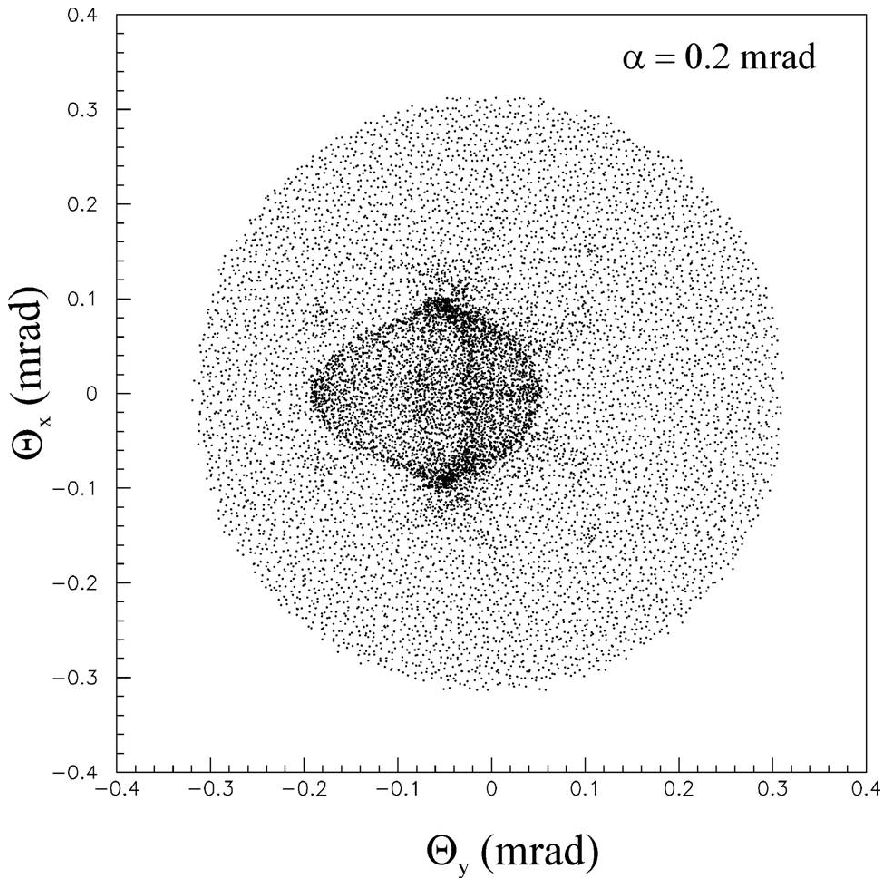}
\hspace*{1cm}
\includegraphics[width=0.45\textwidth]{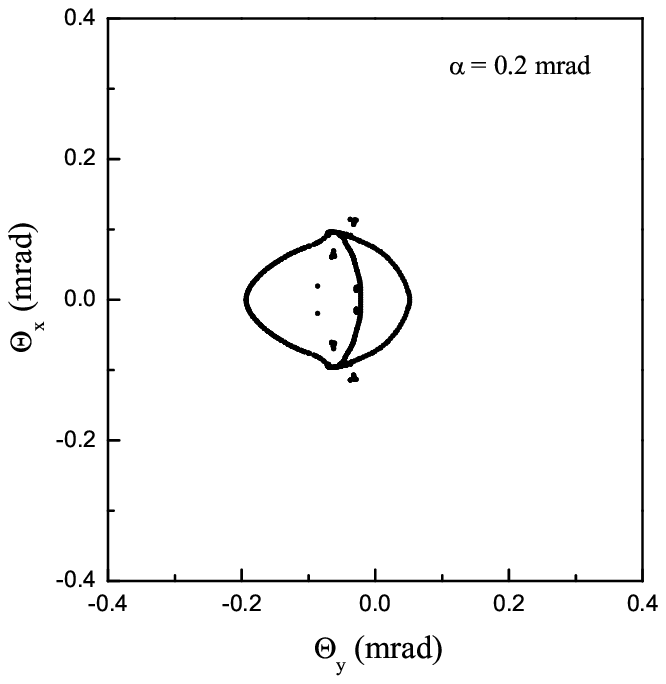}
\caption{(a) The angular distribution of 1 GeV protons transmitted through the bent bundle of (10, 10) single-wall carbon nanotubes of the length of $L$ = 7 $\mu$m for the bending angle $\alpha$ = 0.2 mrad. (b) The correspoding rainbow pattern in the transmission angle plane.}
\label{fig5_2_2}
\end{figure}

Figure \ref{fig5_2_2}(a) shows the angular distribution of protons
transmitted through the bundle obtained for $\alpha$ = 0.2 mrad. The
transmission angle plane is parallel to the exit plane of the
bundle. The sizes of a bin along the $\Theta_x$ and $\Theta_y$ axes
and the initial number of protons are the same as in the case of
$\alpha$ = 0. One can see easily that the angular distribution lies
in the region $\Theta_y >$ -0.2 mrad, demonstrating that the proton
beam is bent by the bundle effectively. The centrifugal force made
the average bending angle of the proton beam smaller than 0.2 mrad.
The angular distribution contains (i) a part having the shape of an
acorn with two maxima at points ($\pm$0.096 mrad, -0.054 mrad), and
(ii) eight additional maxima close and in between the two maxima.
The rainbow pattern in the transmission angle plane for $\alpha$ =
0.2 mrad is given in Fig. \ref{fig5_2_2}(b). It contains (i) an
acorn-like line with two joining points of its branches,
corresponding to a complex line in the impact parameter plane within
each circular channel of the bundle, and (ii) eight groups of
points, corresponding to two times four groups of points in the
impact parameter plane within the two triangular channels making
each rhombic channel. The comparison of the results given in Figs.
\ref{fig5_2_2}(a) and \ref{fig5_2_2}(b) shows that the angular
distribution can be explained as follows: (i) the acorn-like part
with the two maxima is generated by the protons with the impact
parameter vectors close to the acorn-like rainbow line with the two
joining points (within each circular channel), and (ii) the eight
additional maxima by the protons with the impact parameter vectors
close to the eight groups of rainbow points (within each rhombic
channel).

\begin{figure}[ht!]
\centering
\includegraphics[width=0.6\textwidth]{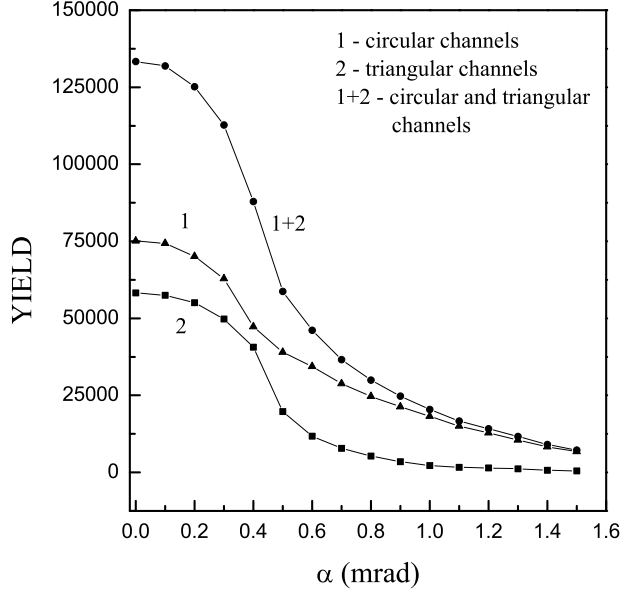}
\caption{The yields of protons transmitted through the circular channels, the triangular channels, and both the circular and triangular channels of the bent bundle of (10, 10) single-wall carbon nanotubes of the lenght of $L$ = 7 $\mu$m as functions of the bending angle, $\alpha$.}
\label{fig5_2_3}
\end{figure}

Figure \ref{fig5_2_3} gives the dependences of the yields of protons
transmitted through the circular channel and the triangular channels
of the bundle on $\alpha$ in the range of 0-1.5 mrad. The initial
numbers of protons are 77~866 for the former dependence and 60~770
for the latter dependence. The inflection points of the dependences
appear at $\alpha$ = 0.35 and 0.45 mrad, respectively. These values
ought to be compared to the value of $\Psi_C$, which is 0.314 mrad.
However, the former yield decreases with $\alpha$ after the
inflection point considerably slower than the latter yield. They
fall below 10 $\%$ of the yields for the straight bundle at $\alpha$
= 1.45 and 0.75 mrad, respectively. This figure also gives the total
yield of protons transmitted through the bundle, both through the
circular and triangular channels, as a function of $\alpha$. Its
inflection point appears at $\alpha$ = 0.45 mrad, and it falls below
10 $\%$ of the yield for the straight bundle at $\alpha$ = 1.25
mrad.

\subsection{Rainbows with long nanotubes}

Let us now consider a 1 GeV proton moving through a long (11, 9)
single-wall carbon nanotube \citep{petr08b}. The nanotube length,
$L$, is varied from 10 to 500 $\mu$m. Such a nanotube is chiral,
i.e. its atomic strings spiral around its axis. The $z$ axis of the
reference frame coincides with the nanotube axis and its origin lies
in the entrance plane of the nanotube. The $x$ and $y$ axes of the
reference frame are the vertical and horizontal axes, respectively.

In the calculations we employ the theory of crystal rainbows, which
is decribed in part \S 3.2 of the chapter. The interaction of the
proton and a nanotube atom is described by the Moli\`{e}re's
approximation \citep{moli47} of the Thomas-Fermi interaction
potential, which is given by Eq. (\ref{equ13}). Since the nanotube
is chiral, the interaction potential of the proton and nanotube is
obtained by the azimuthal averaging of the Moli\`{e}re's
proton-nanotube continuum interaction potential, which is given by
Eqs. (\ref{equ14})-(\ref{equ16}), with the thermal vibrations of the
nanotube atoms taken into account, via Eq. (\ref{equ17}); in these
equations $I$ = 1, $J$ = 40, and $\sigma_{th}$ = 5.3 pm
\citep{thes96}. This interaction potential reads

\begin{equation}
U^{th}(x,y) = \frac{16 \pi Z_1 Z_2 e^2 R}{3 \sqrt 3 a_{CC}^2}\sum\limits_{i = 1}^3 {\left( {\alpha_i + \frac{\sigma_{th}^2 \beta_i^2}{2a^2}} \right)} K_0(\beta_i R/a) I_0(\beta_i \rho/a),
\label{equ18}
\end{equation}

\noindent where $a_{CC}$ = 0.14 nm is the bond length of the carbon
atoms \citep{sait01}, R = 0.69 nm is the nanotube radius
\citep{sait01}, ($\alpha_i$) = (0.35, 0.55, 0.10) and ($\beta_i$) =
(0.1, 1.2, 6.0) are the fitting parameters, $\rho = (x^2 +
y^2)^{\frac{1}{2}}$, $x$ and $y$ are the transverse components of
the proton position vector, and $I_0$ denotes the modified Bessel
function of the first kind and $0^{th}$ order \citet{artr05}. It is
evident that $U^{th}(x,y)$ is cylindrically symmetric.

The electronic proton energy loss and dispersion of its transmission
angle, caused by its collisions with the nanotube electrons, are not
taken into account. The components of the proton impact parameter
vector are chosen randomly within the circle around the origin of
radius $R-a$. The initial proton velocity vectors are all taken to
be parallel to the nanotube axis.

Since $U^{th}(x,y)$ is cylindrically symmetric, the problem we
consider is in fact one-dimensional. This means that the rainbow
lines in the impact parameter plane and transmission angle plane, if
they exist, will show up as the circles. Consequently, it is
sufficient to analyze the mapping of the $x$ axis in the impact
parameter plane to the $\Theta$ axis in the transmission angle
plane, i.e., the $\Theta_x(x)$ deflection function. The abscissas
and ordinates of the extrema of this deflection function determine
the radii of the rainbow lines in the impact parameter plane and
transmission angle plane, respectively.

\begin{figure}[ht!]
\centering
\includegraphics[width=0.45\textwidth]{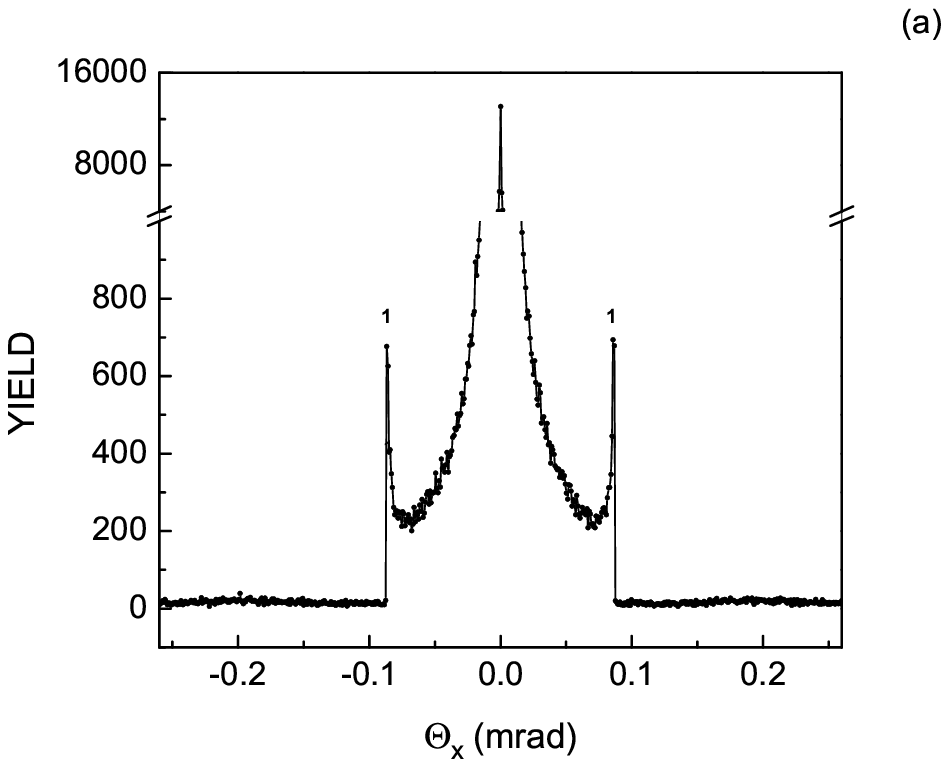}
\vspace*{0.5cm} \hspace*{0.8cm}
\includegraphics[width=0.45\textwidth]{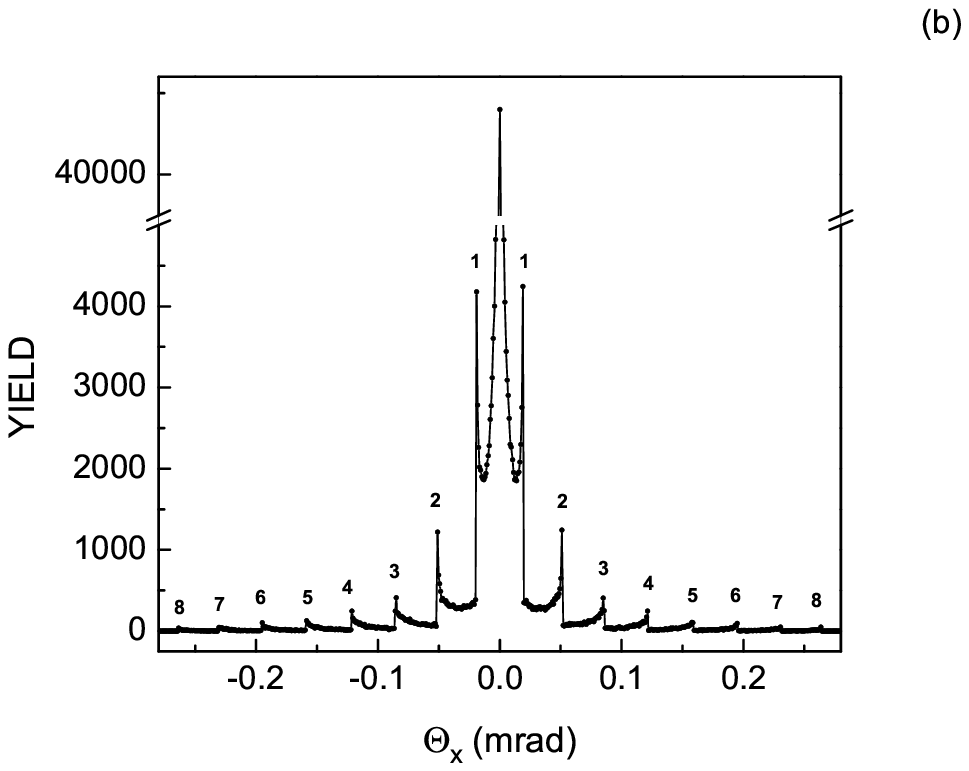}
\includegraphics[width=0.45\textwidth]{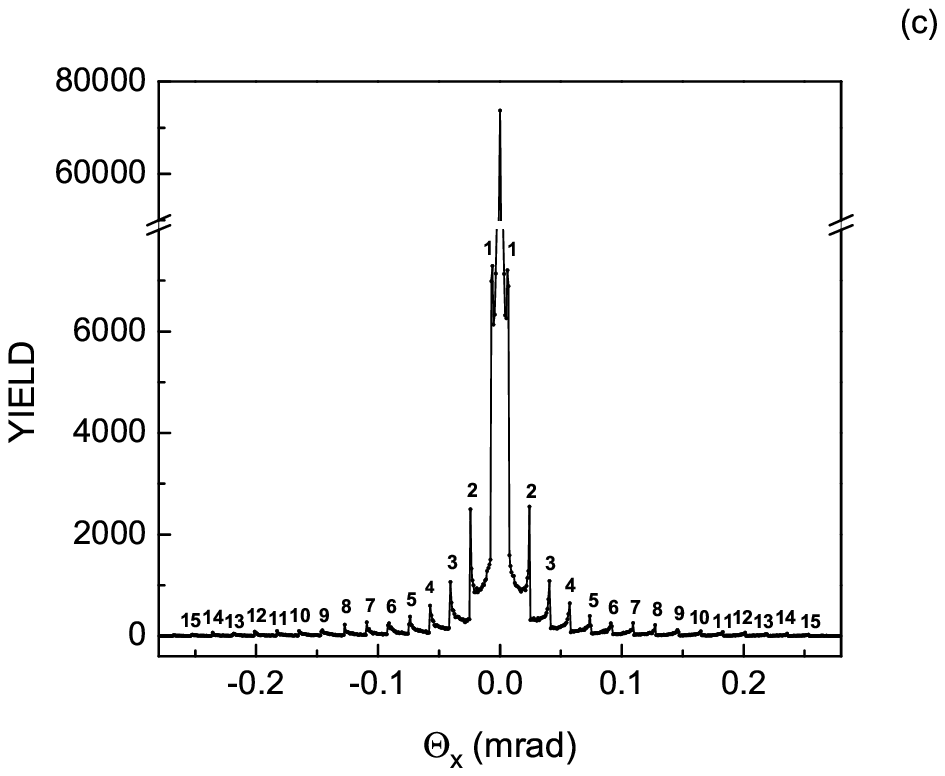}
\hspace*{0.8cm}
\includegraphics[width=0.45\textwidth]{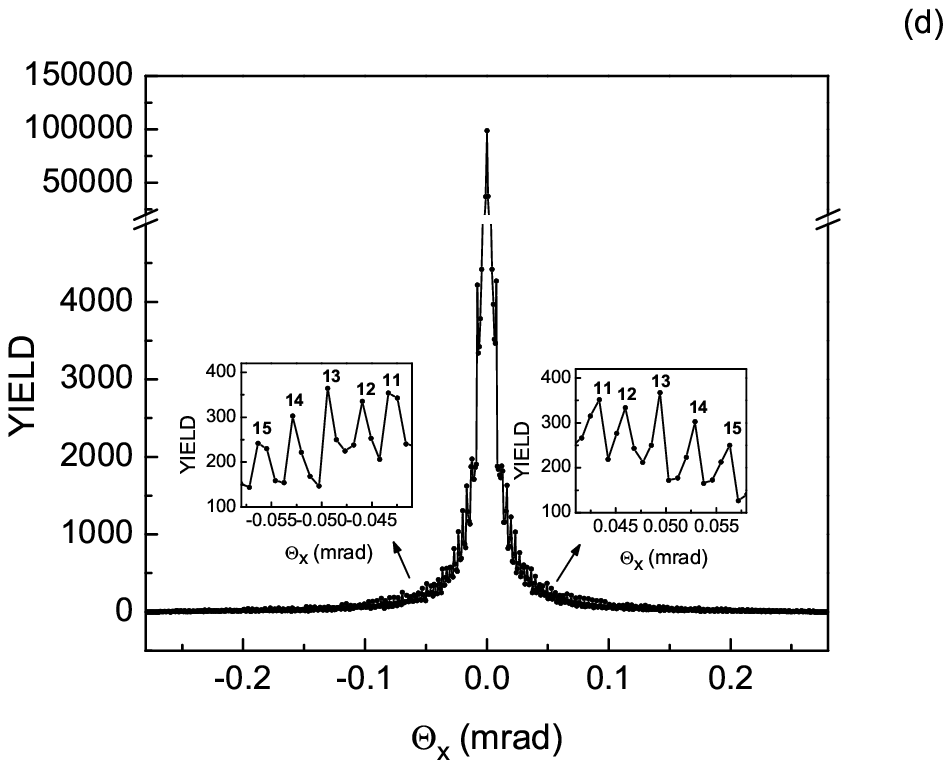}
\caption{The angular distributions along the $\Theta_x$ axis of 1 GeV protons channeled through the (11, 9) SWCNTs of the length of (a) 10 $\mu$m, (b) 50 $\mu$m, (c) 100 $\mu$m and (d) 500 $\mu$m.}
\label{fig5_3_1}
\end{figure}

Figures \ref{fig5_3_1}(a)-(d) show the angular distributions of
channeled protons along the $\Theta$ axis for $L$ = 10, 50, 100 and
500 $\mu$m, respectively. The size of a bin along the $\Theta$ axis
is 0.866 $\mu$rad and the initial number of protons is 16~656~140.
These values enabled us to attain a high resolution of the angular
distributions in a reasonable computational time. The angular
distributions obtained for $L$ = 10, 50 and 100 $\mu$m contain a
central maximum and a number of symmetric pairs of maxima
characterized by a sharp decrease of the proton yield on the large
angle side; the numbers of symmetric pairs of sharp maxima are one,
eight and 15, respectively. The angular distribution obtained for
$L$ = 500 $\mu$m contains a central maximum and a large number of
symmetric pairs of sharp maxima close to each other. The analysis
shows that in the region where $\left| \Theta_x \right| \le$ 0.2
mrad one can identify 55 pairs of sharp maxima. Figure
\ref{fig5_3_1}(d) also shows that in the region where 0.041 mrad
$\le \left| \Theta_x \right| \le$ 0.058 mrad there are five pairs of
sharp maxima; they are designated by 11-15. In the region in which
$\left| \Theta_x \right| >$ 0.2 mrad the resolution of the angular
distribution is not sufficiently high to distinguish easily between
the adjacent pairs of sharp maxima.

\begin{figure}[ht!]
\centering
\includegraphics[width=0.45\textwidth]{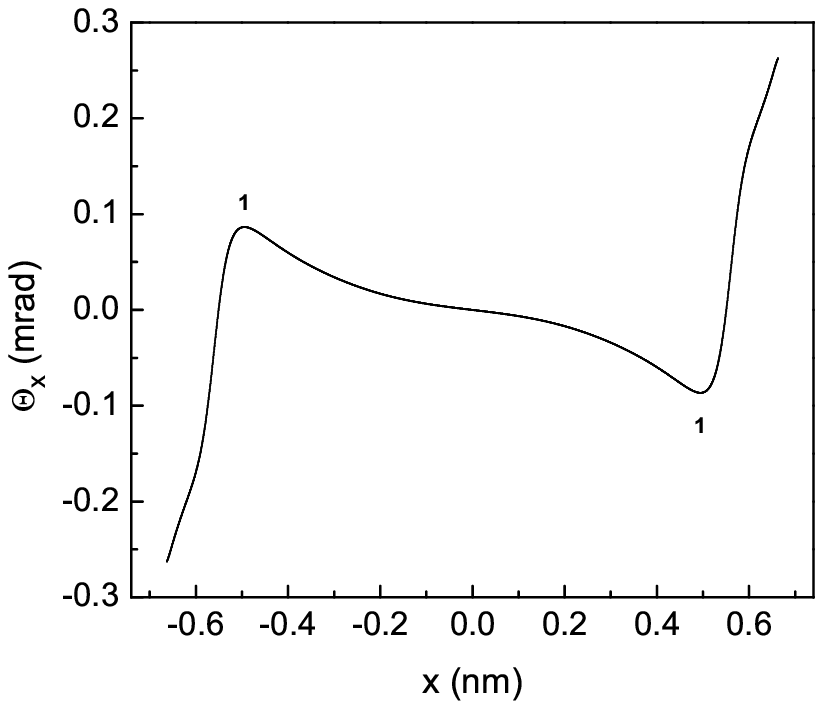}
\vspace*{0.5cm} \hspace*{0.8cm}
\includegraphics[width=0.45\textwidth]{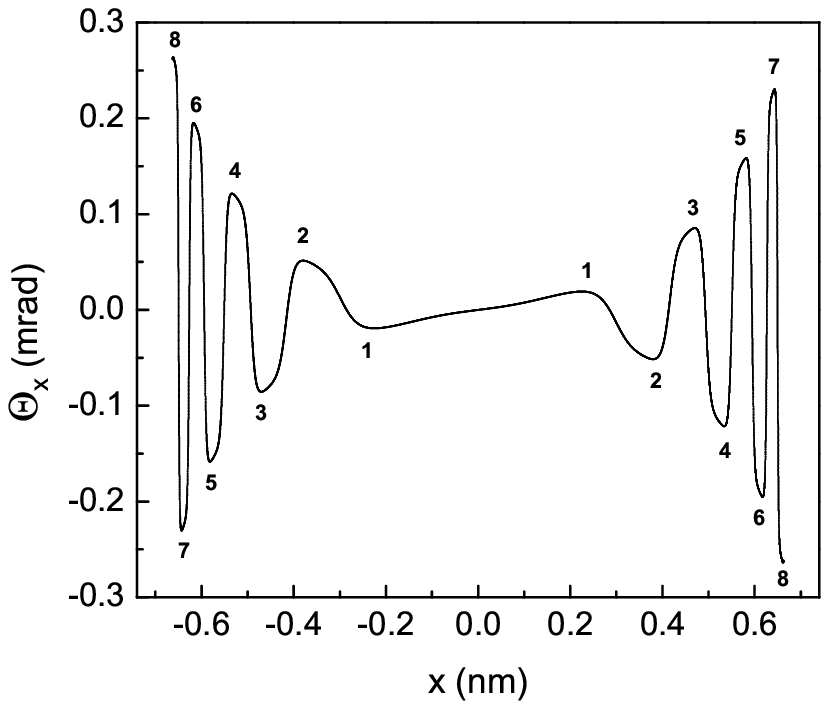}
\includegraphics[width=0.45\textwidth]{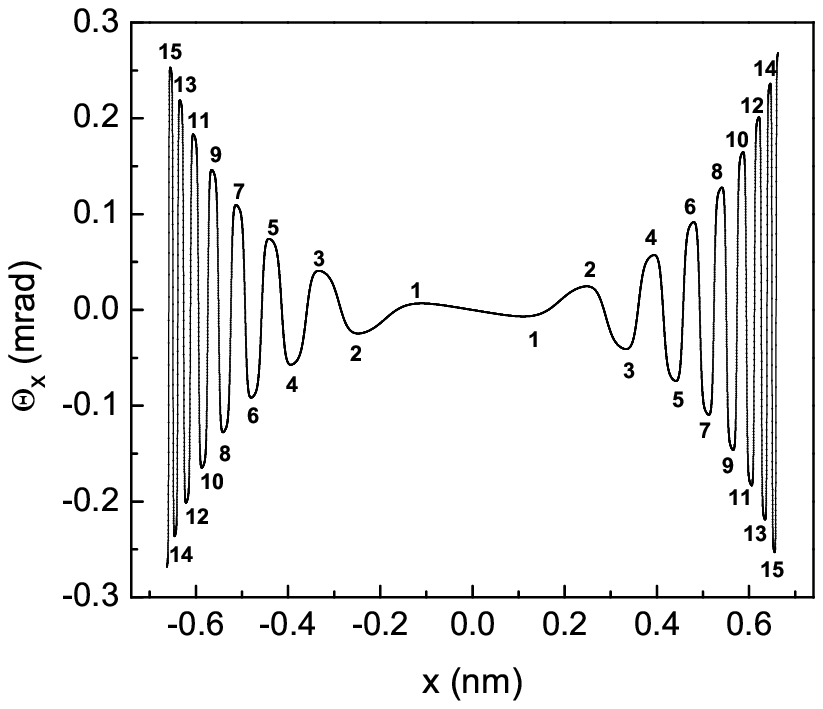}
\hspace*{0.8cm}
\includegraphics[width=0.45\textwidth]{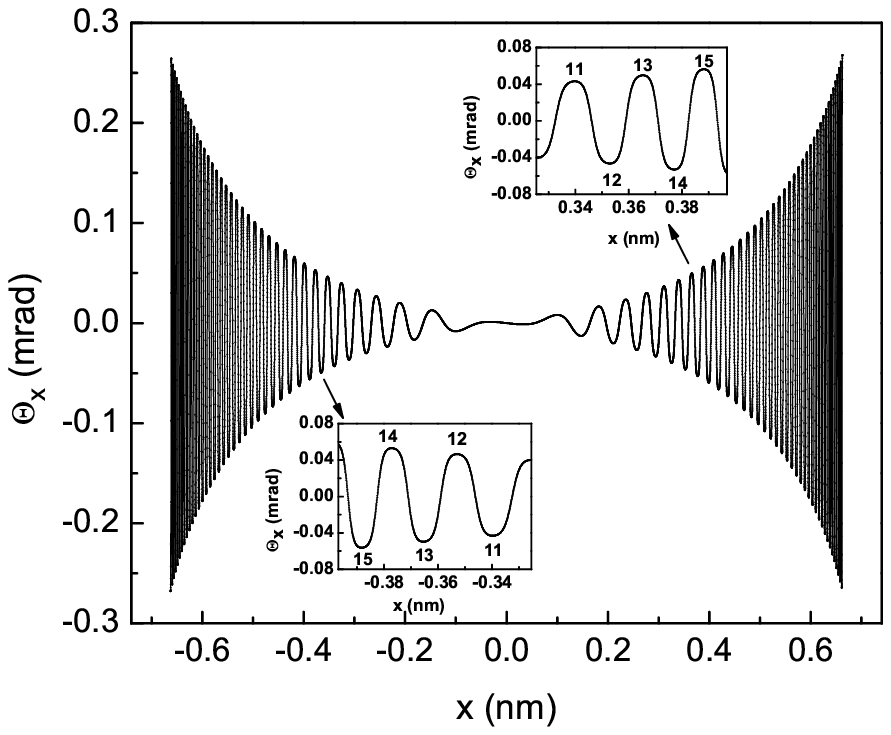}
\caption{The deflection functions $\Theta_x$ of the $x$ axis in the impact parameter plane for 1 GeV protons channeled through the (11, 9) SWCNTs of the length of (a) 10 $\mu$m, (b) 50 $\mu$m, (c) 100 $\mu$m and (d) 500 $\mu$m.}
\label{fig5_3_2}
\end{figure}

The $\Theta_x(x)$ deflection functions obtained for $L$ = 10, 50,
100 and 500 µm are given in Figs. \ref{fig5_3_2}(a)-(d),
respectively. The deflection functions for $L$ = 10, 50 and 100
$\mu$m contain one, eight and 15 symmetric pairs of extrema,
respectively; each pair includes a minimum and a maximum. The
analysis shows that the deflection function obtained for $L$ = 500
$\mu$m contains 76 symmetric pairs of extrema. Figure
\ref{fig5_3_2}(d) also gives five pairs of extrema, designated by
11-15, that correspond to the five sharp maxima shown in Fig.
\ref{fig5_3_1}(d). Each symmetric pair of extrema defines a circular
rainbow line in the impact parameter plane and a circular rainbow
line in the transmission angle plane. One can observe that each of
the deflection functions for $L$ = 50, 100 and 500 $\mu$m has two
envelopes, one of them connecting its extrema designated by odd
numbers and the other connecting its extrema designated by even
numbers, and that it  oscillates between these envelopes.

\begin{figure}[ht!]
\centering
\includegraphics[width=0.65\textwidth]{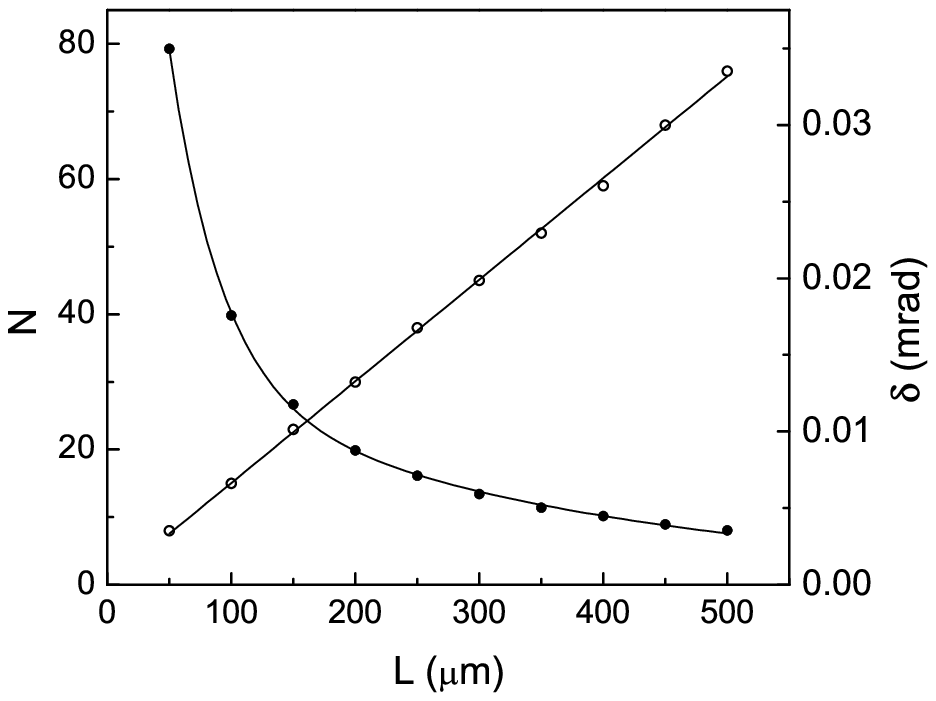}
\caption{The dependences of the number of circular rainbows in the angular distributions of 1 GeV protons channeled through the (11, 9) single-wall carbon nanotubes -- $N$, open circles, and the average distance between them -- $\delta$, closed circles, on the nanotube length -- $L$. The full lines represent the fitting curves.}
\label{fig5_3_3}
\end{figure}

The comparison of Figs. \ref{fig5_3_1}(a)-(c) and Figs.
\ref{fig5_3_2}(a)-(c) shows that the abscissas of the pairs of sharp
maxima of each angular distribution of channeled protons along the
$\Theta_x$ axis coincide with the ordinates of the corresponding
pairs of extrema of the accompanying $\Theta_x(x)$ deflection
function. One can draw the same conclusion for the pairs of sharp
maxima of the angular distribution given in Fig. \ref{fig5_3_1}(d)
and the corresponding pairs of extrema of the deflection function
given in Fig. \ref{fig5_3_2}(d), in spite of the fact that in the
region where $\left| \Theta_x \right| >$ 0.2 mrad it is not possible
to identify the additional 21 pairs of sharp maxima of the angular
distribution. Hence, it is clear that the sharp maxima of the
angular distributions can be attributed to the rainbow effect. The
bright and dark sides of the rainbows are the small and large angle
sides of the corresponding sharp maxima.

Figures \ref{fig5_3_2}(a)-(d) show clearly that the number of
rainbows generated by the transmitted protons increases and the
average distance between them decreases as $L$ increases. These two
dependences, for $L$ between 50 and 500 $\mu$m, are given in Fig.
\ref{fig5_3_3}. The numbers and positions of the rainbows are
determined from the accompanying $\Theta_x(x)$ deflection functions.
The analysis shows that the former dependence can be approximated
excellently by fitting function  , where $a_1$ = 0.15 is the fitting
parameter. The latter dependence can be approximated excellently by
fitting function $f_2 = a_2 \exp(-L/b_2) + a_3 \exp(-L/b_3)$, where
$a_2$ = 0.072 mrad, $b_2$ = 42.6 $\mu$m, $a_3$ = 0.015 mrad and
$b_3$ = 340.5 $\mu$m are the fitting parameters. These fitting
functions are also shown in Fig. \ref{fig5_3_3}. When the nanotube
becomes sufficiently long for the average distance between the
rainbows to become smaller than the resolution of the angular
distribution, one cannot distinguish between the adjacent rainbows.
This means that the rainbows disappear and the angular distribution
becomes a bell-shaped one. The analysis shows that, in parallel, the
spatial distribution of channeled protons in the exit plane of the
nanotube also becomes a bell-shaped one. Hence, one can say that,
when the nanotube becomes sufficiently long, the angular and spatial
distributions equilibrate, and, as one would expect, this does not
happen in accordance with the ergodic hypothesis \citet{barr73}. We
call this route to equilibration, which is characterized by the
linear increase of the number of rainbows and the exponential
decrease of the distance between them as $L$ increases, the rainbow
route to equilibration.

Recently, a theoretical analysis of the channeling of 1 GeV protons
through the long (10, 10) single-wall carbon nanotubes has been
reported \citet{petr09}. The angular distributions of transmitted
protons were generated for $L$ between 10 and 100 $\mu$m. The
obtained results show that when $L <$ 30 $\mu$m the transverse
lattice structure of the nanotube can be deduced from the angular
distribution. When $L \ge$ 40 $\mu$m the angular distribution
contains the concentric circular ridges whose number increases
linearly and the distance between them decreases exponentially as
$L$ increases, in a similar way as in the above described case of
(11, 9) nanotubes. Consequently, these ridges can be attributed to
the rainbow effect.

The above presented results demonstrate that one can employ the
angular distributions of protons channeled through long carbon
nanotubes for their characterization. Namely, it seems that each
type of nanotube for each value of $L$ has a characteristic pattern
of concentric circular ridges, and that this can be checked
experimentally. However, an easier way to check this experimentally
would be to vary the proton kinetic energy rather than $L$. This
method of characterization of nanotubes would be complementary to
the method proposed in part \S 5.1 of this chapter.

\section{Dynamic polarization effect in proton channeling through short nanotubes}
\markright{Dynamic polarization effect...}

In this part of the chapter we shall investigate how the angular and
spatial distributions of protons channeled through a short (11, 9)
single-wall carbon nanotube in vacuum or embedded in a dielectric
medium is influenced by the effect of dynamic polarization of the
nanotube atoms valence electrons \citep{bork06a,bork08b,bork08a}.
The magnaitude of the proton velocity vector, $v$, is varaied
between 3 and 8 a.u., corresponding to the proton kinetic energy
between 0.223 and 1.59 MeV, respectively. It is expected that the
dynamic polarization effect, which is induced by the proton, is
pronounced in this range of its kinetic energy. The nanotube length,
$L$, is varied between 0.1 and 0.8 $\mu$m. We shall also explore the
influence of the angle of the initial proton velocity vector
relative to the nanotube axis on the angular distribution of protons
channeled through a short (11, 9) nanotube in vacuum for $v$ = 3
a.u. and $L$ = 0.2 $\mu$m \citep{bork10}.

The $z$ axis of the reference frame coincides with the nanotube axis
and its origin lies in the entrance plane of the nanotube. The $x$
and $y$ axes of the reference frame are the vertical and horizontal
axes, respectively. The calculations are performed using the theory
of crystal rainbows, which is described in part \S 3.2 of the
chapter. The thermal vibrations of the nanotube atoms are not taken
into account. The electronic proton energy loss and dispersion of
its transmission angle, caused by its collisions with the nanotube
electrons, are neglected too. The components of the proton impact
parameter vector are chosen randomly within the circle around the
origin of radius $R-a$, where $R$ is the nanotube radius, $a =
\left[ 9 \pi^2/(128 Z_2) \right]^{\frac{1}{3}}a_0$ the screening
radius of the nanotube atom, and $a_0$ the Bohr radius. The initial
proton velocity vectors are all taken to be parallel to the nanotube
axis.

\subsection{Rainbows with short nanotubes in vacuum}

The system under consideration here is a proton moving through a
short (11, 9) single-wall carbon nanotube in vacuum for $v$ = 3 a.u.
and $L$ = 0.1-0.3 $\mu$m \citep{bork06a}; the corresponding proton
kinetic energy is 0.223 MeV. This nanotube is chiral.

The interaction of the proton and a nanotube atom is described by
the Doyle-Turner interaction potential \citep{doyl68}. Since the
nanotube is chiral, the static part of the interaction potential of
the proton and nanotube is obtained by the azimuthal averaging of
the Doyle-Turner proton-nanotube continuum interaction potential
\citep{lind65,zhev98,zhev00}; the number of atomic strings of the
nanotube is $J$ = 40. This interaction potential, in atomic units,
is

\begin{equation}
U_s(x,y) = \frac{32\pi Z_1 Z_2 R}{3\sqrt 3 a_{CC}^2}\sum\limits_{i = 1}^4 {\alpha_i} \delta_i^2I_0(\delta_i^2R\rho )\exp [-\delta_i^2(\rho ^2 + R^2)],
\label{equ19}
\end{equation}

\noindent where ($\gamma_i$) = (0.115, 0.188, 0.072, 0.020) and
($\delta_i$) = (0.547, 0.989, 1.982, 5.656) are the fitting
parameters (in a.u.), $\rho = (x^2 + y^2)^{\frac{1}{2}}$, $x$ and
$y$ are the transverse components of the proton position vector, and
$I_0$ denotes the modified Bessel function of the first kind and
$0^{th}$ order. It is clear that $U_s(x,y)$ is repulsive and
cylindrically symmetric.

The dynamic polarization effect is treated by a two-dimensional
hydrodynamic model of the nanotube atoms valence electrons based on
a jellium-like description of the ion cores on the nanotube walls
\citep{mowb04a,mowb04b,bork06a}. This model includes the axial and
azimuthal averaging as in the case of $U_S(x,y)$. As a result, the
dynamic part of the interaction potential of the proton and nanotube
is

\begin{equation}
{U_d}\left( {x,y} \right) = \frac{Z_1^2}{\pi} \sum\limits_{l = - \infty}^\infty  {P\int\limits_0^\infty {K_l^2(kR)\frac{{4\pi {n_0}R\left( {k^2 + {{l^2} \mathord{\left/
 {\vphantom {{l^2} {R^2}}} \right.
 \kern-\nulldelimiterspace} {R^2}}} \right)}}{{{{(kv)}^2} - \omega_l^2(k)}}I_l^2(k\rho)dk} },
\label{equ20}
\end{equation}			

\noindent where

\begin{equation}
\omega_l^2(k) = (k^2 + l^2/R^2)\left[ {4\pi n_0RI_l (kR)K_l(kR) + v_s^2} \right]
\label{equ21}
\end{equation}

\noindent is the square of the angular frequency of the proton
induced oscillations of the nanotube electron gas of angular mode
$l$ and longitudinal wave number $k$, $n_0$ = 0.428 the ground state
density of the nanotube electron gas, $v_s = (\pi
n_0)^{\frac{1}{2}}$ is the velocity of propagation of the density
perturbations of the nanotube electron gas, $0 \le \rho < R$, $I_l$
and $K_l$ denote the modified Bessel function of the first and
second kinds and $l^{th}$ order, respectively, and $P$ designates
that only the principal part of the integral is taken into account.
One can see that $U_d(x,y)$, which is also called the image
interaction potential of the proton and nanotube, is attractive and
cyllindrically symmetric. Thus, the total interaction potential of
the proton and nanotube is

\begin{equation}
U(x,y) = U_s(x,y) + U_d(x,y).
\label{equ22}
\end{equation}

\begin{figure}[ht!]
\centering
\includegraphics[width=0.7\textwidth]{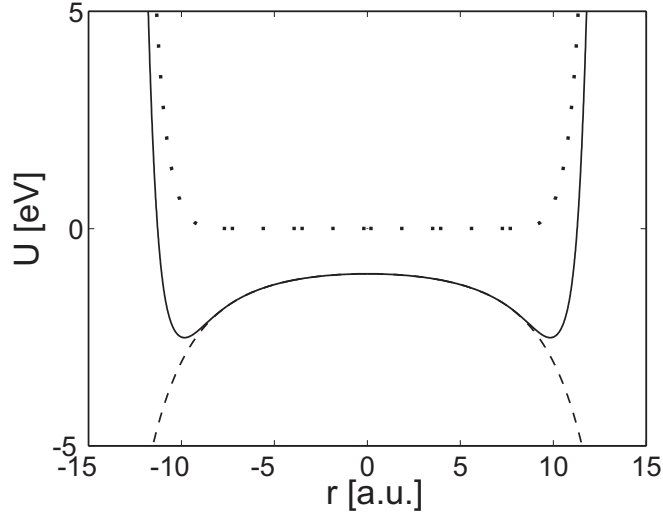}
\caption{The total interaction potential of the proton and nanotube, $U(x,y)$, along the $x$ axis in the interval ($-R,R$) for $v$ = 3 a.u - solid line. The static and dynamic parts of the interaction potential are also shown - dotted and dashed lines, respectively.}
\label{fig6_1_1}
\end{figure}

\noindent Figure \ref{fig6_1_1} shows the total interaction
potential of the proton and nanotube, $U(x,y)$, along the $x$ axis
in interval ($-R,R$) for $v$ = 3 a.u. \citep{bork06a,petr08a}. The
static and dynamic parts of the interaction potential are also shown
in the figure. One can see that due to the inclusion of the image
interaction potential the total interaction potential of the proton
and nanotube has two inflection points. They appear for $x$ =
$\pm$8.1 a.u. If the nanotube we consider is short, the
corresponding $\Theta_x(x)$ deflection function will have a pair of
symmetric extrema appearing for the values of $x$ close to $\pm$8.1
a.u.

\begin{figure}[ht!]
\centering
\includegraphics[width=0.45\textwidth]{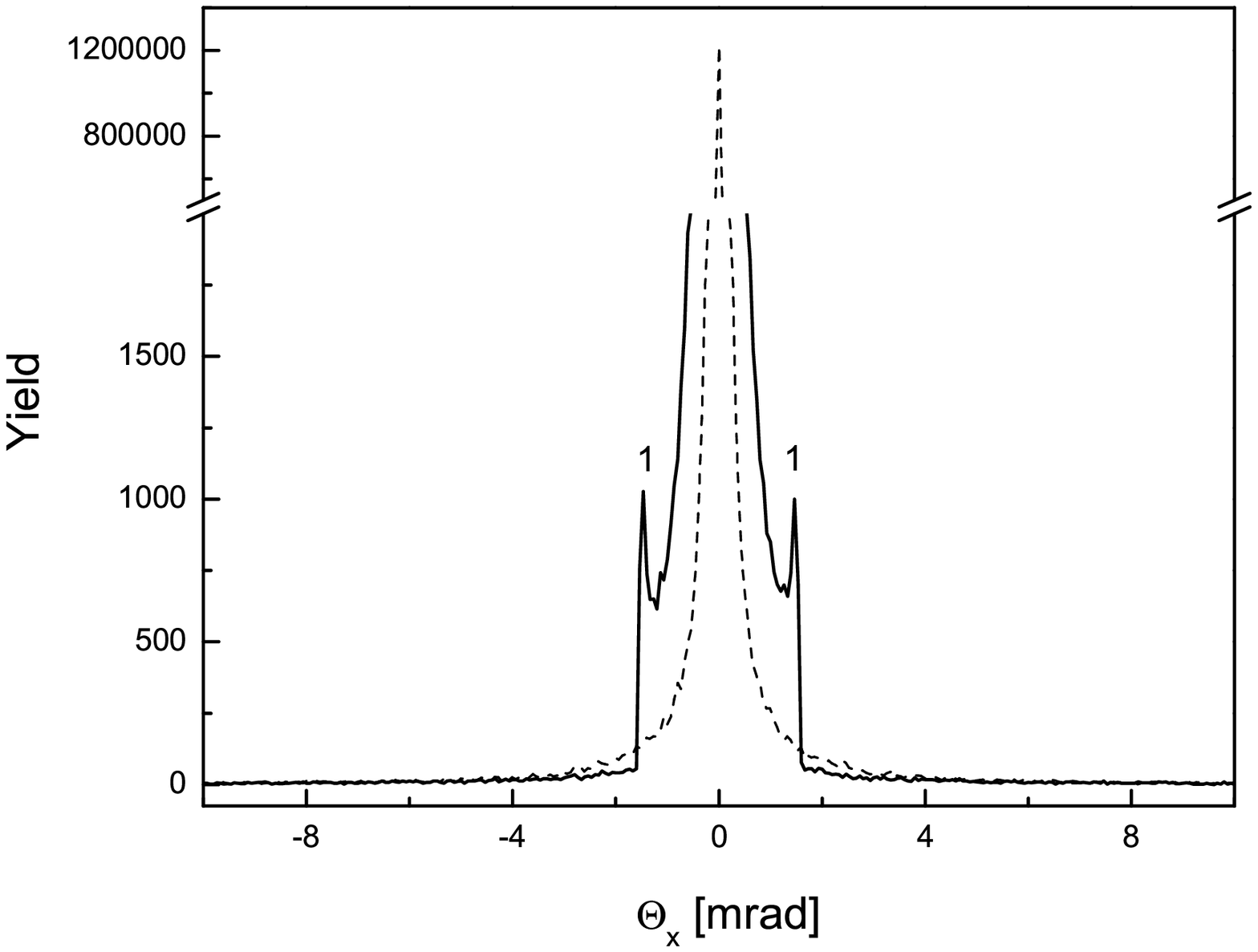}
\hspace*{1cm}
\includegraphics[width=0.45\textwidth]{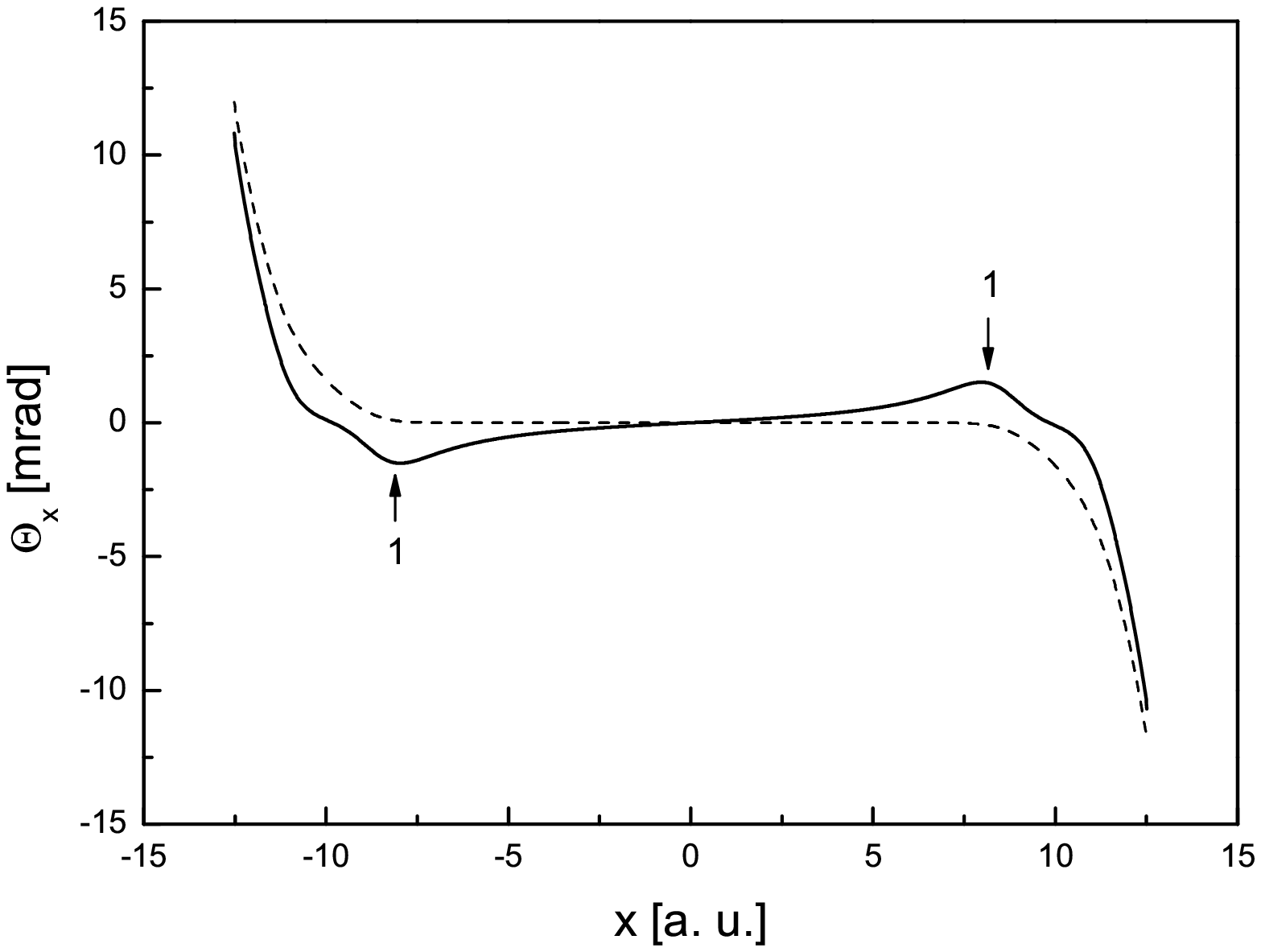}
\caption{(a) The angular distributions along the $\Theta_x$ axis of protons transmitted through a (11, 9) single-wall carbon nanotubes for $v$ = 3 a.u. and $L$ = 0.1 $\mu$m with and without the dynamic polarization effect included -- solid and dashed lines, respectively. (b) The corresponding $\Theta_x(x)$ deflections functions.}
\label{fig6_1_2}
\end{figure}

The angular distribution of transmitted protons along the $\Theta_x$
axis for $v$ = 3 a.u. and $L$ = 0.1 $\mu$m is given in Fig.
\ref{fig6_1_2}(a) \citep{bork06a}. This figure also shows the
corresponding angular distribution when the image interaction
potential of the proton and nanotube is not included. The size of a
bin along the $\Theta_x$ axis is equal to 0.066 mrad and the initial
number of protons is 3~141~929. One can see that a central maximum
and a pair of symmetric peripheral maxima appear in the angular
distribution when the image interaction potential is included; the
peripheral maxima are labeled by 1. Figure \ref{fig6_1_2}(b) shows
the $\Theta_x(x)$ deflection functions corresponding to the angular
distributions given in Fig. \ref{fig6_1_2}(a). It is evident that
the deflection function without the image interaction potential
taken into account has no extrema. On the other hand, the deflection
function with the image interaction potential taken into account
exhibits a pair of symmetric extrema; they are also labeled by 1.
The analysis shows that the two peripheral maxima of the angular
distribution are connected to the two extrema of the deflection
function. Therefore, we conclude that they are the rainbow maxima,
and that they occur as a consequence of the inclusion of the dynamic
polarization effect. Besides, one should note that the extrema of
the deflection function appear for $x$ = $\pm$8.0 a.u. Since these
positions are close to the positions of the inflection points of the
total interaction potential, one should conclude that the nanotube
in question is short.

We have established that for $v$ = 3 a.u. and $L$ = 0.2 $\mu$m the
angular distribution of transmitted protons along the $\Theta_x$
axis contains three pairs of rainbow maxima, and that for $v$ = 3
a.u. and $L$ = 0.3 $\mu$m the angular distribution includes five
pairs of rainbow maxima \citep{bork06a}. All these maxima are
attributed to the image interaction potential of the proton and
nanotube.

\subsection{Rainbows with short nanotube embedded in dielectric media}

Experimentally, the main problem that prevent the observation of ion
channeling through carbon nanotubes is the problem of ordering,
straightening and holding nanotubes. The most promising method
enabling us to solve this problem is based on the growing of
nanotubes inside the holes in a dielectric medium \citep{zhu05,
berd08}. Besides, in many applications of nanotubes it is desirable
to have them embedded in SiO$_2$ \citep{tset06}, or clamped by a Ni
shield \citep{guer94,misk07}.

When a nanotube is embedded in a dielectric medium, the static part
of its interaction potential with the proton is the same as in the
case when it is in vacuum, which is given by Eq. (19). The dynamic
part of the interaction potential is obtained by a hydrodynamic
model of the nanotube atoms valence electrons based on the
jellium-like description of the ion cores on the nanotube walls, as
in the case when the nanotube is in vacuum, but extended to include
the polarization of the dielectric boundary \citep{mowb07,bork08b}.
It should be noted that this model has also been applied
successfully in the related area of ion interaction with dielectric
suported sheets of carbon atoms \citep{rado07,rado09}.

\begin{figure}[ht!]
\centering
\includegraphics[width=0.45\textwidth]{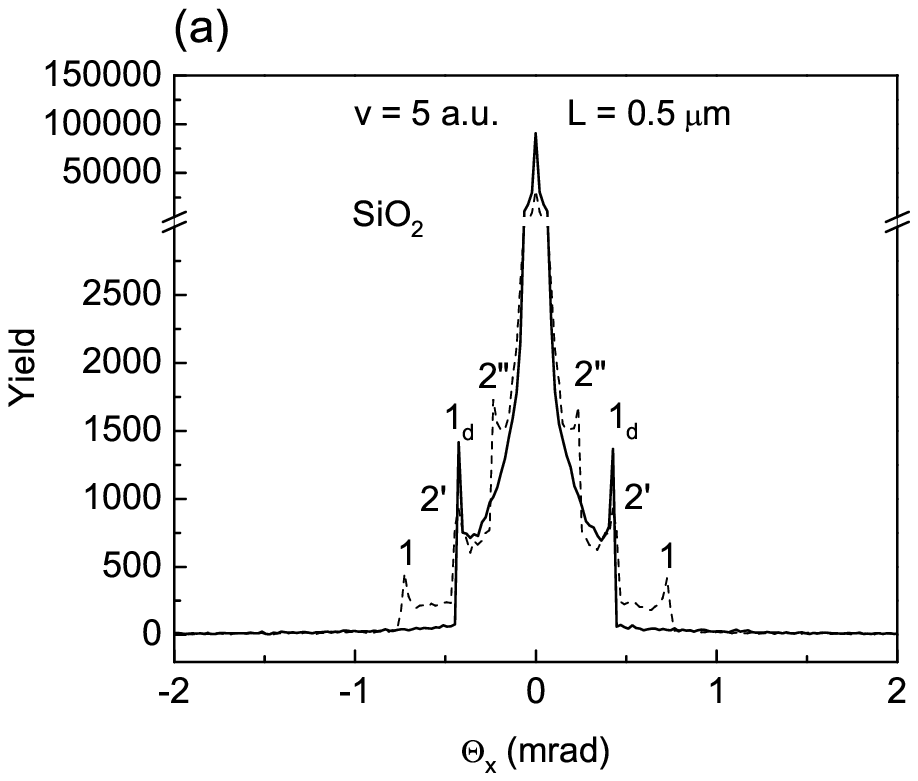}
\vspace*{0.7cm} \hspace*{1cm}
\includegraphics[width=0.45\textwidth]{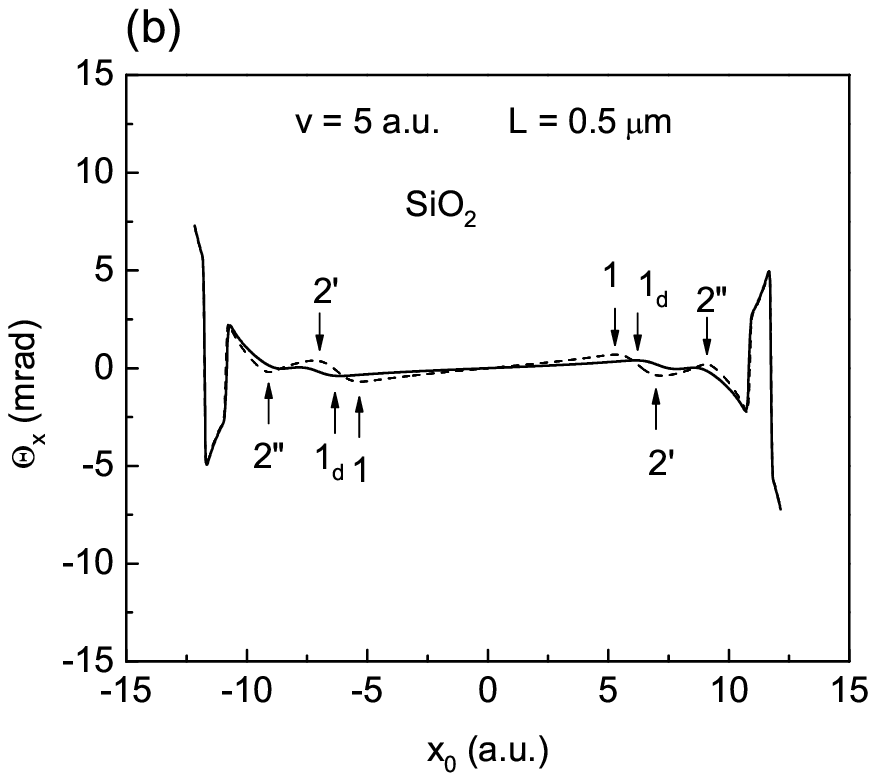}
\includegraphics[width=0.45\textwidth]{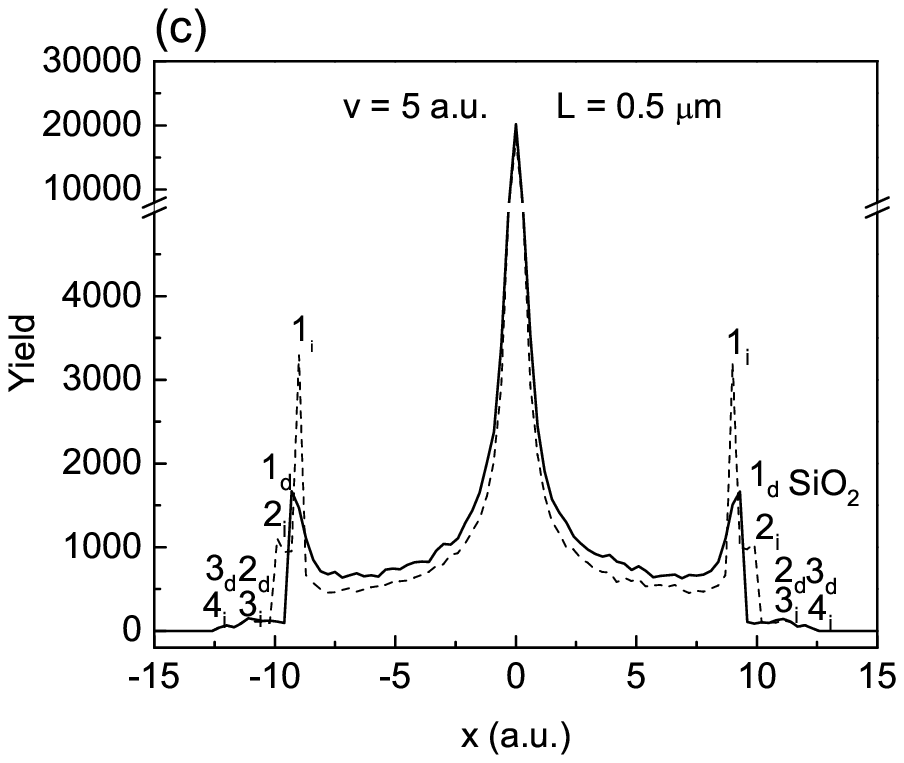}
\hspace*{1cm}
\includegraphics[width=0.45\textwidth]{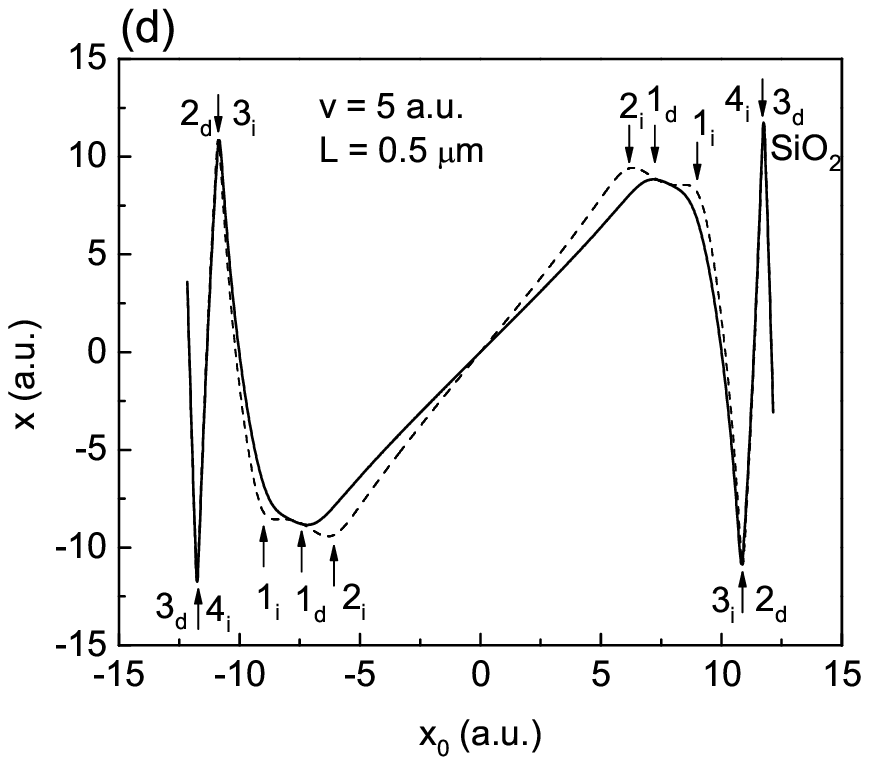}
\caption{(a) The angular distributions along the $\Theta_x$ axis of protons transmitted through a (11, 9) single-wall carbon nanotube embedded in SiO$_2$ and in vaccum for $v$ = 5 a.u. and $L$ = 0.5 $\mu$m - solid and dashed lines, respectively. (b) The corresponding   deflections functions. (c) The corresponding spatial distributions along the $x$ axis in the exit plane of the nanotube - solid and dashed lines, respectively. (b) The corresponding mappings of the $x_0$ axis in the entrance plane of the nanotube to the $x$ axis in its exit plane - solid and dashed lines, respectively.}
\label{fig6_2_1}
\end{figure}

Figure \ref{fig6_2_1}(a) shows the angular distribution along the
$\Theta_x$ axis of protons transmitted through a (11, 9) single-wall
carbon nanotube embedded in SiO$_2$ for $v$ = 5 a.u. and $L$ = 0.5
$\mu$m \citep{bork08b}; the corresponding proton kinetic energy is
0.619 MeV. This figure also shows the corresponding angular
distribution when the nanotube is in vacuum. The size of a bin along
the $\Theta_x$ axis is equal to 0.0213 mrad and the initial number
of protons is 3~141~929. One can see a central maximum and only one
pair of symmetric peripheral maxima, labeled by 1$_d$, when the
nanotube is embedded in SiO$_2$, in comparison with a central
maximum and three pairs of symmetric peripheral maxima, labeled by
1, 2' and 2'', when the nanotube is in vacuum. Figure
\ref{fig6_2_1}(b) gives the $\Theta_x(x)$ deflection functions
corresponding to the angular distributions given in Fig.
\ref{fig6_2_1}(a). We have found that each peripheral maximum of
each angular distribution is connected to an extremum of the
corresponding deflection function. Hence, all the peripheral maxima
are the rainbow maxima.

Let us know analyze the spatial distributions along the $x$ axis in
the exit plane of the nanotube of protons transmitted through a (11,
9) single-wall carbon nanotube embedded in SiO$_2$ and in vacuum for
$v$ = 5 a.u. and $L$ = 0.5 $\mu$m \citep{bork08a}. It is given in
Fig. \ref{fig6_2_1}(c). The size of a bin along the $\Theta_x$ axis
is equal to 0.3 a.u. and the initial number of protons is 3~141~929.
In the former case the spatial distribution constains a central
maximum and three pairs of symmetric peripheral maxima, labeled by
1$_d$, 2$_d$ and 3$_d$, and in the latter case the spatial
distribution includes a central maximum and four pairs of symmetric
peripheral maxima, labeled by 1$_i$, 2$_i$, 3$_i$ and 4$_i$. The
maxima designated by 2$_d$, 3$_d$, 3$_i$ and 4$_i$ are very weak.
Figure \ref{fig6_2_1}(d) gives the mappings of the $x_0$ axis in the
entrance plane of the nanotube to the $x$ axis in its exit plane
corresponding to the spatial distributions given in Fig.
\ref{fig6_2_1}(c). It is evident that when the nanotube is embedded
in SiO$_2$, the mapping has six symmetric extrema, labeled by 1$_d$,
2$_d$ and 3$_d$. When the nanotube is in vacuum, the mapping has
eight symmetric extrema, labeled by 1$_i$, 2$_i$, 3$_i$ and 4$_i$.
Extrema 2$_d$, 3$_d$, 3$_i$ and 4$_i$ are very sharp and lie near
the nanotube wall. In each case the ordinate of the extremum of the
mapping coincides with the abscissa of the corresponding peripheral
maximum of the spatial distribution. Therefore, these maxima are
attributed to the rainbow effect.

\begin{figure}[ht!]
\centering
\includegraphics[width=0.45\textwidth]{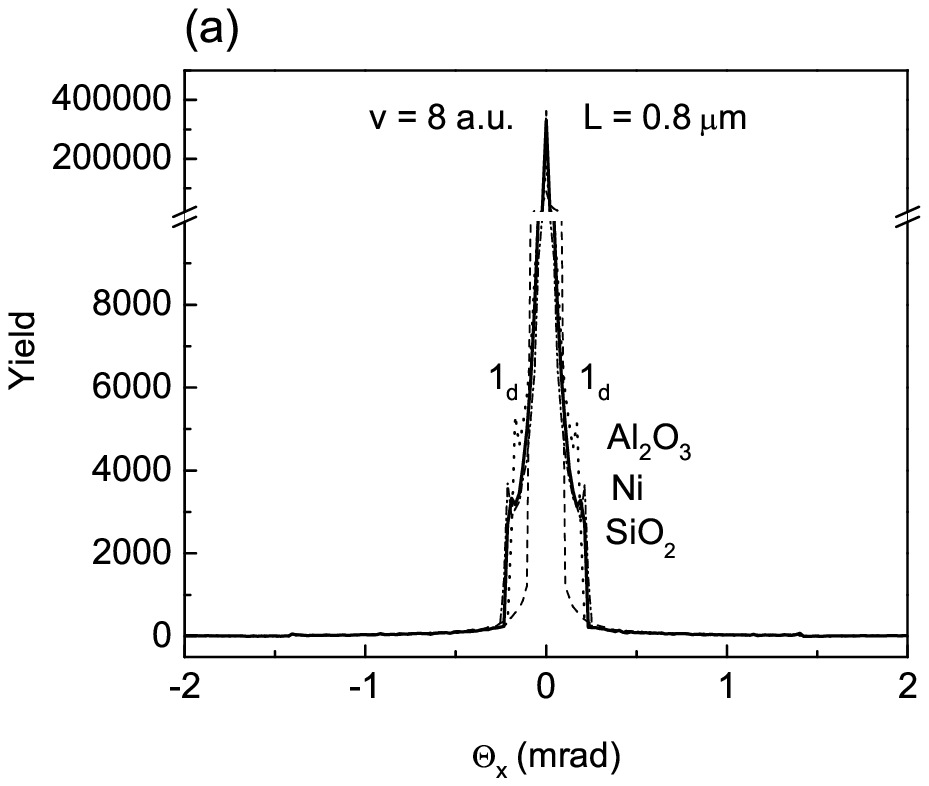}
\vspace*{0.7cm} \hspace*{1cm}
\includegraphics[width=0.45\textwidth]{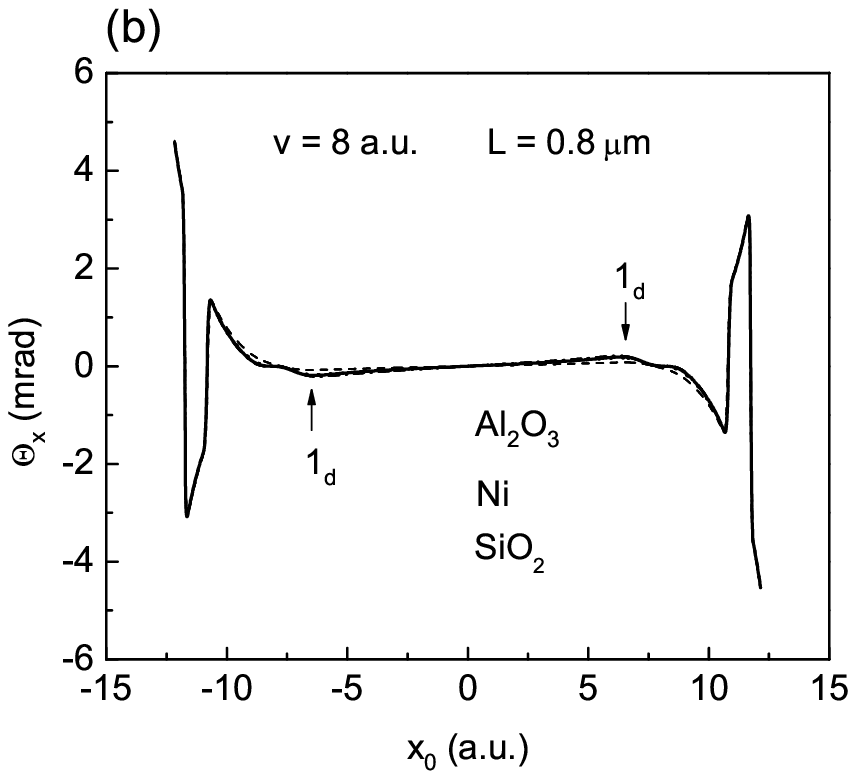}
\includegraphics[width=0.45\textwidth]{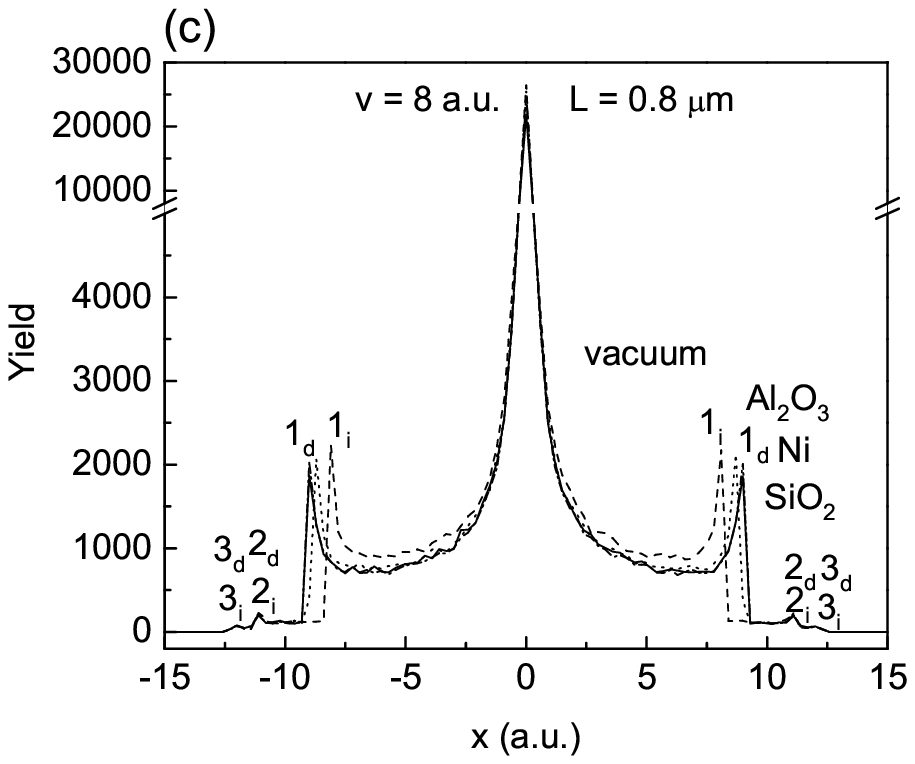}
\hspace*{1cm}
\includegraphics[width=0.45\textwidth]{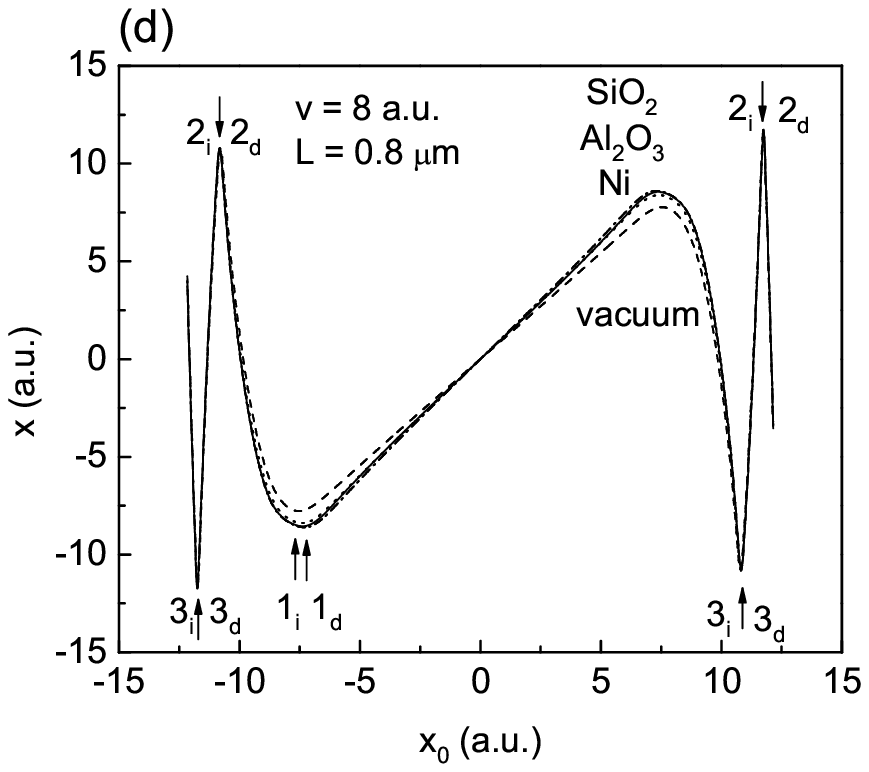}
\caption{(a) The angular distributions along the $\Theta_x$ axis of protons transmitted through a (11, 9) single-wall carbon nanotube embedded in SiO$_2$, Al$_2$O$_3$ and Ni and in vacuum for $v$ = 8 a.u. and $L$ = 0.8 $\mu$m - solid, dotted and dash dotted lines and dashed line, respectively. (b) The corresponding $\Theta_x(x)$ deflections functions. (c) The corresponding spatial distributions along the $x$ axis in the exit plane of the nanotube - solid and dashed lines, respectively. (b) The corresponding mappings of the $x_0$ axis in the entrance plane of the nanotube to the $x$ axis in its exit plane - solid and dashed lines, respectively.}
\label{fig6_2_2}
\end{figure}

Figure \ref{fig6_2_2}(a) gives the angular distribution along the
$\Theta_x$ axis of protons channeled through a (11, 9) single-wall
carbon nanotube embedded in SiO$_2$, Al$_2$O$_3$ and Ni and in
vacuum for $v$ = 8 a.u. and $L$ = 0.8 $\mu$m \citep{bork08b}; the
corresponding proton kinetic energy is 1.59 MeV. The size of a bin
along the $\Theta_x$ axis is equal to 0.0213 mrad and the initial
number of protons is 3~141~929. The angular distributions obtained
with the embedded nanotube contain one central maximum and two
symmetric peripheral maxima, labeled by 1$_d$. It is clear that
these positions do not depend much on the type of medium surrounding
the nanotube. This can be explained by the fact that in these cases
the dependences of the image force acting on the proton on $x$ are
close to each other \citep{bork08b}. When the nanotube is in vacuum,
the angular distribution contains a central maximum and no symmetric
peripheral maxima. Figure \ref{fig6_2_2}(b) shows the $\Theta_x(x)$
deflection functions corresponding to the angular distributions
shown in Fig. \ref{fig6_2_2}(a). It is clear that in the cases of
embedded nanotubes the deflection functions contain two pairs of
symmetric extrema. The analysis shows that the two peripheral maxima
appearing in the angular distributions obtained with the embedded
nanotube are connected to the former pair of extrema of the
corresponding deflection functions, also labeled by 1$_d$. This
means that these maxima are due to the rainbow effect. The angular
distributions obtained with the embedded nanotube do not include any
maxima that would be connected to the latter pair of extrema of the
corresponding deflection functions, lying close to the nanotube
wall. It is also clear that in the case of nanotube in vacuum the
deflection function includes only one pair of symmetric extrema,
lying close to the nanotube wall. However, as it has been already
said, the angular distribution obtained with the nanotube in vaccum
does not have any maxima that would be connected to the pair of
extrema of the corresponding deflection function.

Now, we are going to consider the spatial distributions along the
$x$ axis in the exit plane of the nanotube of protons channeled
through a (11, 9) single-wall carbon nanotube embedded in SiO$_2$,
Al$_2$O$_3$ and Ni and in vacuum for $v$ = 8 a.u. and $L$ = 0.8
$\mu$m \citep{bork08a}. They are shown in Fig. \ref{fig6_2_2}(c).
The size of a bin along the $\Theta_x$ axis is equal to 0.3 a.u. and
the initial number of protons is 3~141~929. All these spatial
distributions contain a central maximum and three pairs of symmetric
peripheral maxima. In the cases of embedded nanotubes the peripheral
maxima are labeled by 1$_d$, 2$_d$ and 3$_d$, and in the case of
nanotube in vacuum the peripheral maxima are labeled by 1$_i$, 2$_i$
and 3$_i$. One can see that the angular distributions obtained with
the embedded nanotubes do not differ much from each other, and that
they differ from the angular distribution obtained with the
nanotaube in vacuum. The spatial distributions obtained with the
nanotube embedded in SiO$_2$ and Ni almost coincide. Figure
\ref{fig6_2_2}(d) shows the mappings of the $x_0$ axis in the
entrance plane of the nanotube to the $x$ axis in its exit plane
corresponding to the spatial distributions shown in Figure
\ref{fig6_2_2}(c). One can observe that all these mappings include
three pairs of symmetric extrema. In the cases of embedded nanotubes
the extrema of the mappings are labeled by 1$_d$, 2$_d$ and 3$_d$,
and in the case of nanotube in vacuum the extrema are labeled by
1$_i$, 2$_i$ and 3$_i$. In each case the ordinate of the extremum of
the mapping coincides with the abscissa of the corresponding
peripheral maximum of the spatial distribution. Therefore, these
maxima are the rainbow maxima.

Let us now compare the above described angular and spatial
distributions of channeled protons, obtained for $v$ = 5 a.u. and
$L$ = 0.5 $\mu$m, and for $v$ = 8 a.u. and $L$ = 0.8 $\mu$m. All of
them are characterized by the same proton dwell time, i.e., the same
duration of the process of proton channeling. When the nanotube is
in vacuum, the numbers of rainbow maxima of the angular and spatial
distributions are larger in the former case than in the latter case.
This is attributed to the weakening of the image force acting on the
proton with the increase of $v$. On the other hand, when the
nanotube is embedded, the numbers of rainbow maxima of the angular
and spatial distributions is the same in the former and latter
cases. This persistency of the rainbow effect with the increase of
$v$ is explained by the increase of the image force due to the
polarization of the dielectric boundary.

Our findings indicate that a dialectric medium surrounding carbon
nanotubes can influence their properties. Also, the changing of the
spatial distribution of channeled protons with their kinetic energy
we have demonstrated, based on the changing of the dynamic
polarization effect, may be used to probe the atoms or molecules
intercalated in the nanotubes. Besides, we think that nanotubes
could be employed to produce nanosized ion beams for applications in
biomedicine.

\subsection{Donut effect with protons and a short nanotube in vacuum}

In ion channeling experiments the always present questions are the
questions of ion beam misalignment and divergence. Therefore, it is
important to explore how the dynamic polarization effect influences
the angular and spatial distributions of ions channeled through
carbon nanotubes when the initial ion velocity vector is not
parallel to the nanotube axis. We are going to focus on the angular
distribution of protons channeled through a (11, 9) single-wall
carbon nanotube in vacuum for $v$ = 3 a.u. and $L$ = 0.2 $\mu$m and
for the angle of the initial proton velocity vector relative to the
nanotuge axis, $\varphi$, between 0 and 10 mrad \citep{bork10}; the
corresponding proton kinetic energy is 0.223 MeV. In this case the
critical angle for channeling is $\Psi_C$ = 11 mrad. It should be
noted that the inclusion of a dielectric medium around the nanotube
will modify the results of the calculation only slightly.

It is well-known that when $\varphi$ is close to $\Psi_C$, the
angular distribution of ions transmitted through crystal channels
takes the shape of a donat \citep{chad70,rosn78,ande80}. This effect
is called the donut effect. It has been explained afterwards by the
theory of crystal rainbows \citep{nesk02,bork03}. It must be noted
that the donut effect has been observed in computer simulations of
ion propagation through nanotubes \citep{zhev98,zhev00}. However,
the authors did not connect the obtained results to the rainbow
effect.

\begin{figure}[ht!]
\centering
\includegraphics[width=0.46\textwidth]{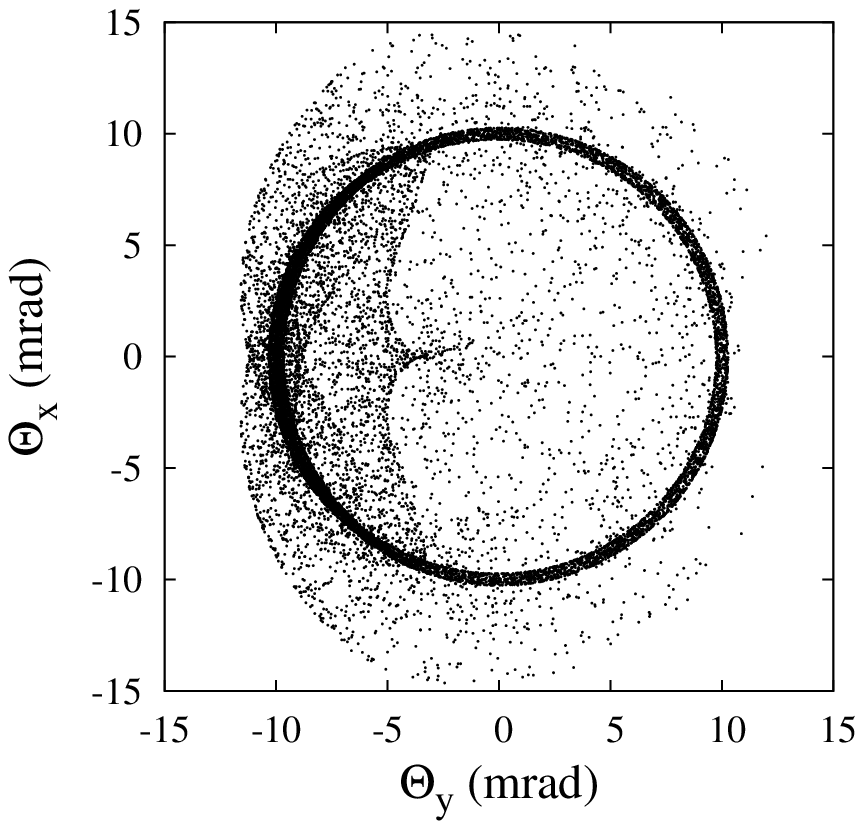}
\hspace*{1cm}
\includegraphics[width=0.42\textwidth]{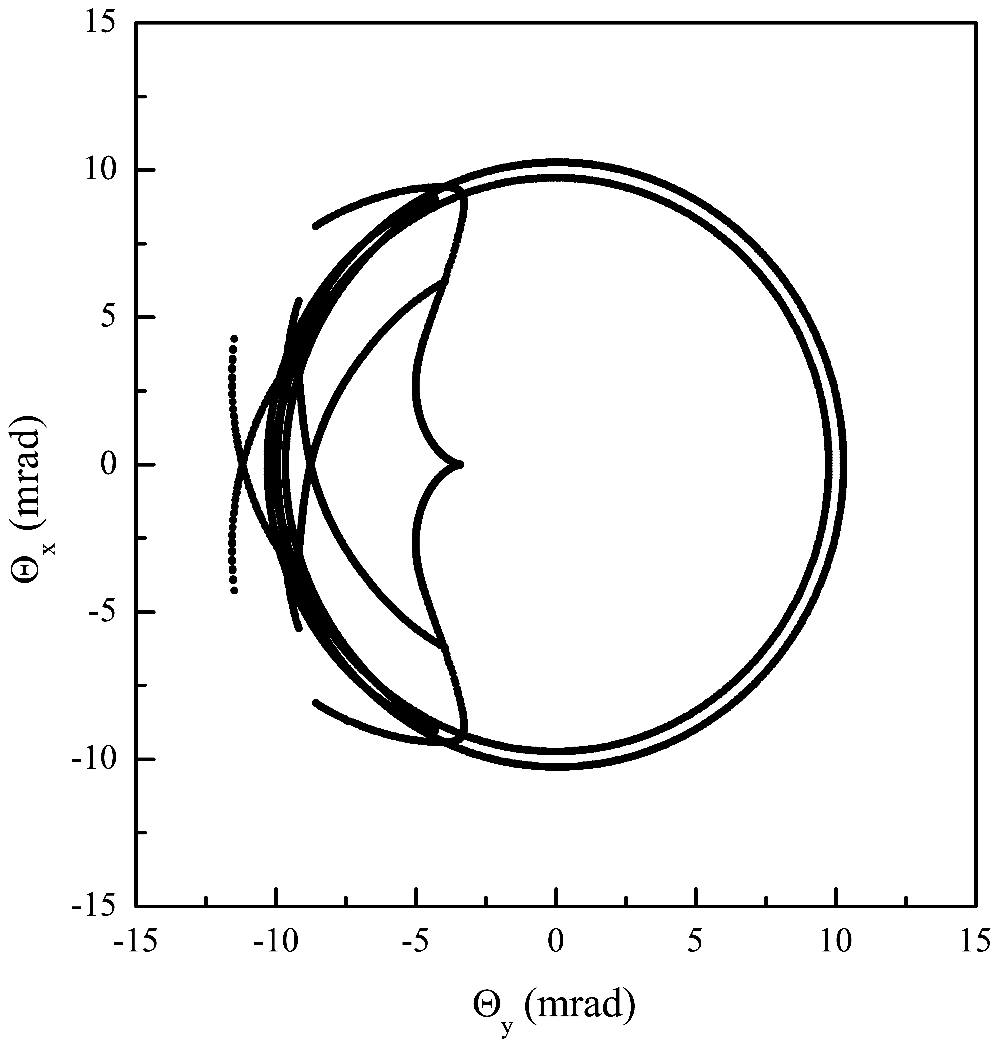}
\caption{(a) The angular distribution of protons transmitted through a (11, 9) single-wall carbon nanotube for $v$ = 3 a.u., $L$ = 0.2 $\mu$m and $\varphi$ = 10 mrad. The image force acting on the proton is taken into account. (b) The corresponding rainbow pattern in the transmission angle plane.}
\label{fig6_3}
\end{figure}

Figure \ref{fig6_3}(a) shows the angular distribution of transmitted
protons for $\varphi$ = 10 mrad, being close to $\Psi_C$
\citep{bork10}. The image force acting on the proton is taken into
account. One can see clearly that its has a ring-like structure, and
that the yield of protons along the ring is very high. This is the
fully developed donut effect. The corresponding rainbow lines in the
transmission angle plane are shown in Fig. \ref{fig6_3}(b). These
lines demonstrate that the structure of the angular distribution,
including the donut effect, is due to the rainbow effect. We have
established that the influence of the dynamic polarization effects
on the angular distribution is significant for $\varphi$ between 0
and $\Psi_C/2$. However, when $\varphi$ becomes close to $\Psi_C$,
this influence is reduced strongly \citep{bork10}. This means that
the contribution of the dynamic polarization effect to the donut
effect is minor.

\section{Dynamic polarization effect in proton channeling through short nanotubes}
\markright{Dynamic polarization effect...}

The channeling star effect was observed for the first time in the
early experiments on ion transmission through axial crystal channels
\citep{gemm74}. The characteristic "arms" of a channeling star were
attributed to the ions moving along the planar channels that cross
the axial channel. It must be noted that such a motion can occur
only when the ion beam divergence is larger than the critical angle
for channeling \citep{petr02}. Usually, the effect has been used for
precise alignment of the ion beam with the channel axis.

We are going to describe here the channeling star effect in the
transmission of protons through the bundles of (10, 10) single-wall
carbon nanotubes \citep{bork06b}. The initial proton energy is 1 GeV
and the bundle length is varied between 2.2 to 14.3 $\mu$m,
corresponding to the reduced bundle length associated with the
protons moving close to the centers of the triangular channels of
the bundle, $\Lambda_2$, between 0.15 and 1, respectively (see part
\S 4 of this review). The study is focused on the problem of mutual
orientation of the neighboring nanotubes within the bundle, which
has not been resolved yet \citep{kwon00}. The nanotubes in question
are achiral. It is assumed that they form a bundle whose transverse
cross-section can be descirbed via a hexagonal or rhombic
superlattice with one nanotube per primitive cell \citep{thes96}. We
choose the $y$ axis of the reference frame to coincide with the
bundle axis and its origin to lie in the entrance plane of the
bundle. The arrangement of the nanotubes is such that their axes
intersect the $x$ and $y$ axes of the reference frame, which are the
vertical and horizontal axes, respectively \citep{petr05c}. We take
into account the contributions of the nanotubes lying on the two
nearest rhombic coordination lines, relative to the center of the
(rhombic) primitive cell of the superlattice. The initial proton
beam axis coincides with the bundle axis. The proton beam divergence
angle is chosen to be $\Omega_d = 6 \psi_c$, where $\psi_c$ = 0.314
mrad is the critical angle for channeling. The calculations are
performed using the theory of crystal rainbows, which is described
in part \S 3.2 of this review. The interaction of the proton and a
nanotube atom is described by the Moli\`{e}re's approximation
\citep{moli47} of the Thomas-Fermi interaction potential, which is
given by Eq. (\ref{equ13}). The application of the continuum
approximation gives the Moli\`{e}re's proton-bundle continuum
interaction potential, which is defined by Eq. (\ref{equ14}). The
number of nanotubes within the bundle $I$ = 16, the number of atomic
strings of a nanotube $J$ = 40, and the distance between the atoms
of an atomic string $d$ = 0.24 nm. The thermal vibrations of the
nanotube atoms are taken into account. This is done by Eq.
(\ref{equ17}). The one-dimensional thermal vibration amplitude of
the nanotube atoms $\sigma_{th}$ = 5.3 pm \citep{hone00}.

The electronic proton energy loss and dispersion of its proton
transmission angle, caused by its collisions with the nanotube
electrons, are taken into account. For the specific proton energy
loss we use expression

\begin{equation}
- \frac{dE_e}{dz} = \frac{4 \pi Z_1^2 e^4}{m_e v^2(z)} n_e \left[ \ln \frac{2m_e \gamma^2 v^2(z)}{\hbar \omega_e} - \beta^2 \right],
\label{equ23}
\end{equation}

\noindent where $m_e$ is the electron mass, $v(z)$ the magnitude of
the proton velocity vector, $\beta = v(z)/c$, $c$ the speed of
light, $\gamma^2 = (1 - \beta)^{-1}$, $n_e = (\partial_{xx} +
\partial_{yy})U^{th}(x,y) / 4 \pi$ the average along the $z$ axis of
the density of the nanotube electron gas, $U^{th}(x,y)$ the
Moli\`{e}re's proton-bundle continuum interaction potential with the
effect of thermal vibrations of the nanotube atoms taken into
account, $\hbar$ the reduced Planck constant, and $\omega_e =
\left(4 \pi e^2 n_e / m_e \right)^{\frac{1}{2}}$ the angular
frequency of the proton induced oscillations of the nanotube
electron gas. For the specific change of the dispersion of the
proton transmission angle we use expression

\begin{equation}
\frac{d\Omega_e^2}{dz} = \frac{m_e}{m_p^2 v^2(z)} \left( - \frac{dE_e}{dz} \right),
\label{equ24}
\end{equation}

\noindent where $m_p$ is the (relativistic) proton mass. The
corresponding standard deviations of the components of the proton
transmission angle, $\Theta_x$ and $\Theta_y$, are $\Omega_{ex} =
\Omega_{ey} = \Omega_e / \sqrt 2$.

The nanotube diameter is 1.34 nm  \citep{sait01}, and the distance
between the axes of two neighboring nanotubes is 1.70 nm
\citep{thes96}. The diameter of the bundle is chosen to be 169.64
nm, corresponding to 100 nanotubes lying on the $y$ axis. The
components of the proton impact parameter vector are chosen
uniformly within the region of the bundle. For each proton impact
parameter the $x$ and $y$ components of its initial velocity,
$v_{0x}$ and $v_{0y}$, respectively, are chosen within the Gaussian
distributions with the standard deviations $\Omega_{bx} =
\Omega_{by} = \Omega_b / \sqrt 2$, where $\Omega_b$ is the proton
beam divergence angle. The initial number of protons is 859~144.

\begin{figure}[ht!]
\centering
\includegraphics[width=0.45\textwidth]{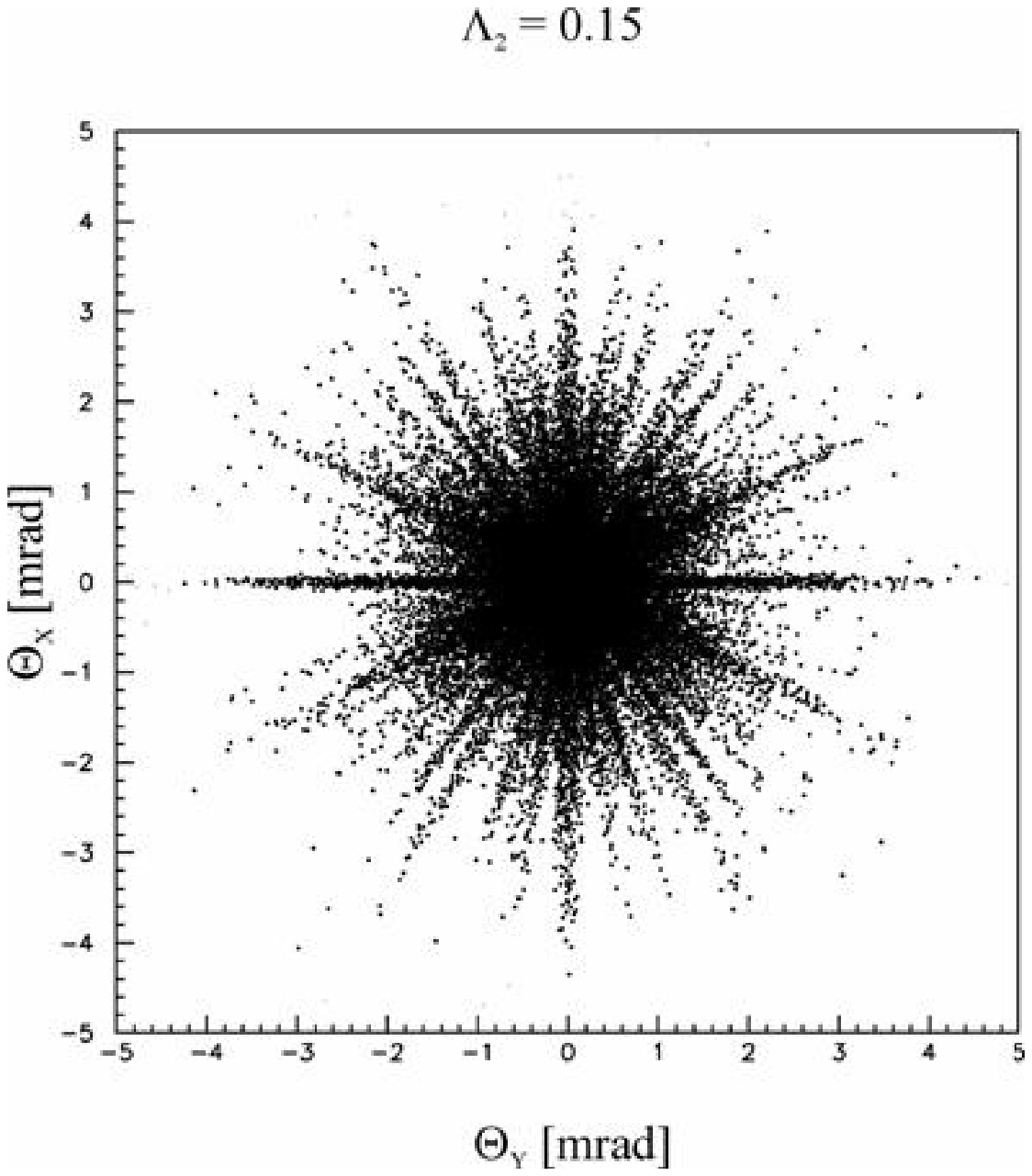}
\vspace*{0.5cm} \hspace*{0.8cm}
\includegraphics[width=0.45\textwidth]{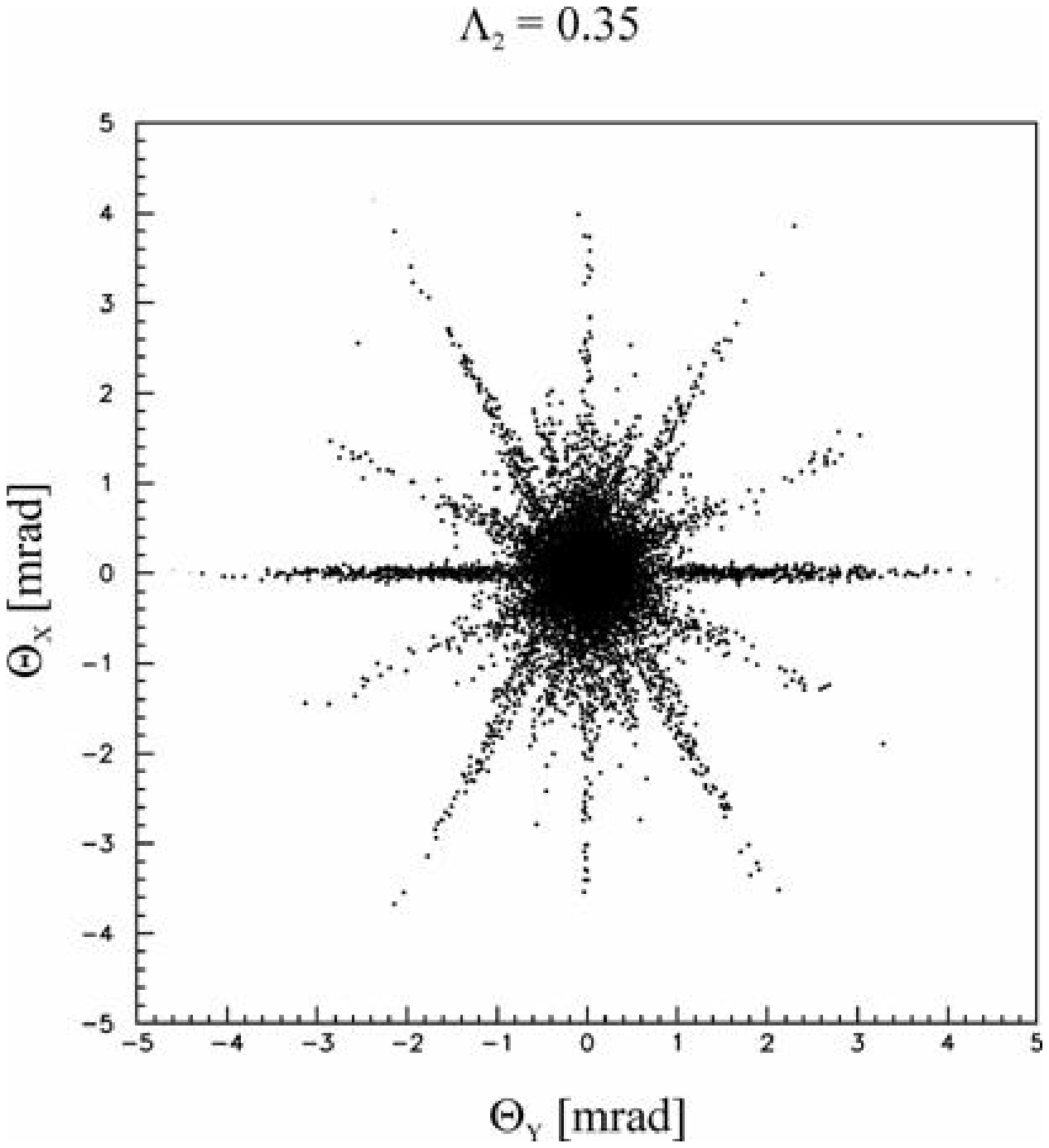}
\includegraphics[width=0.45\textwidth]{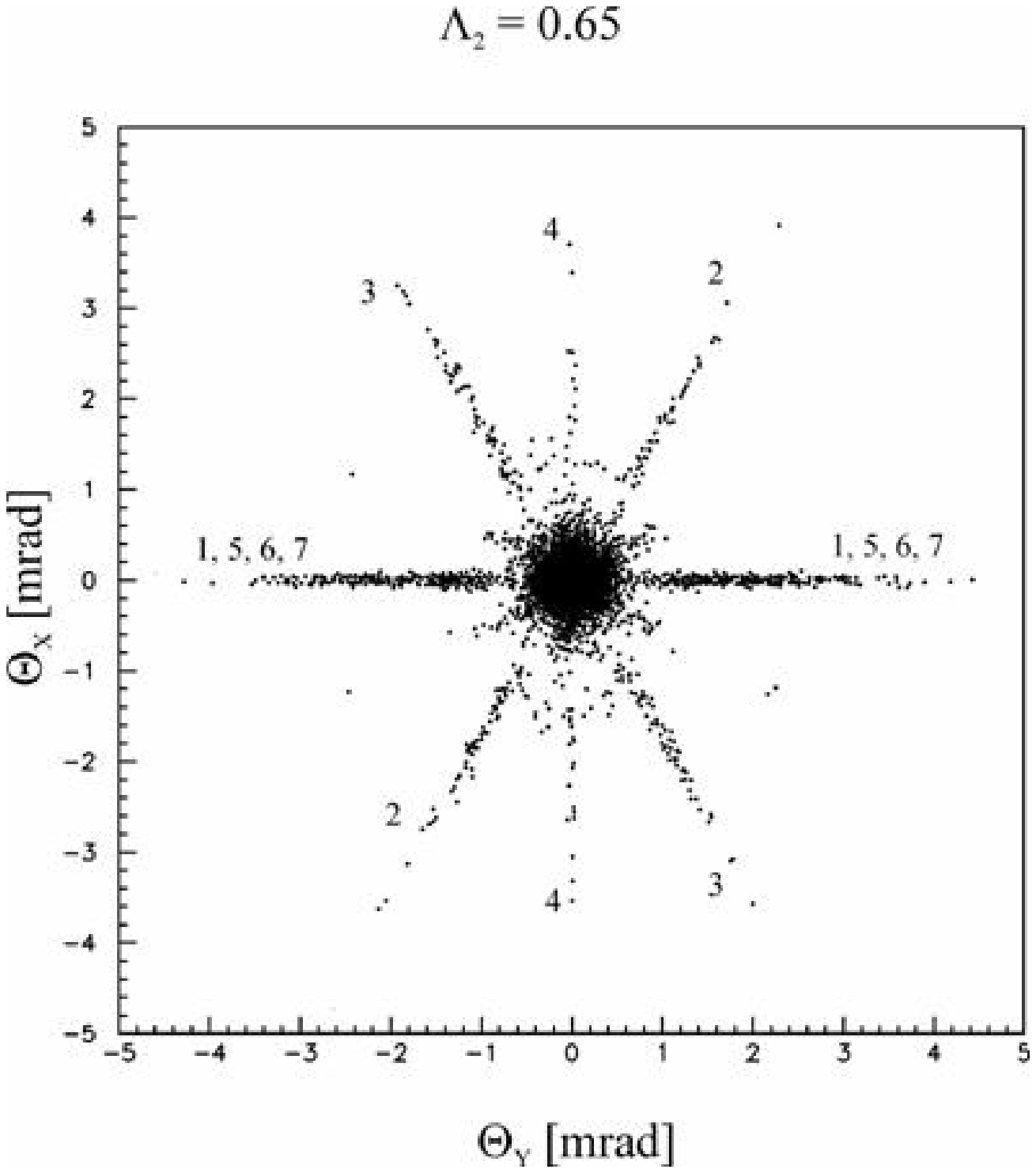}
\hspace*{1cm}
\includegraphics[width=0.45\textwidth]{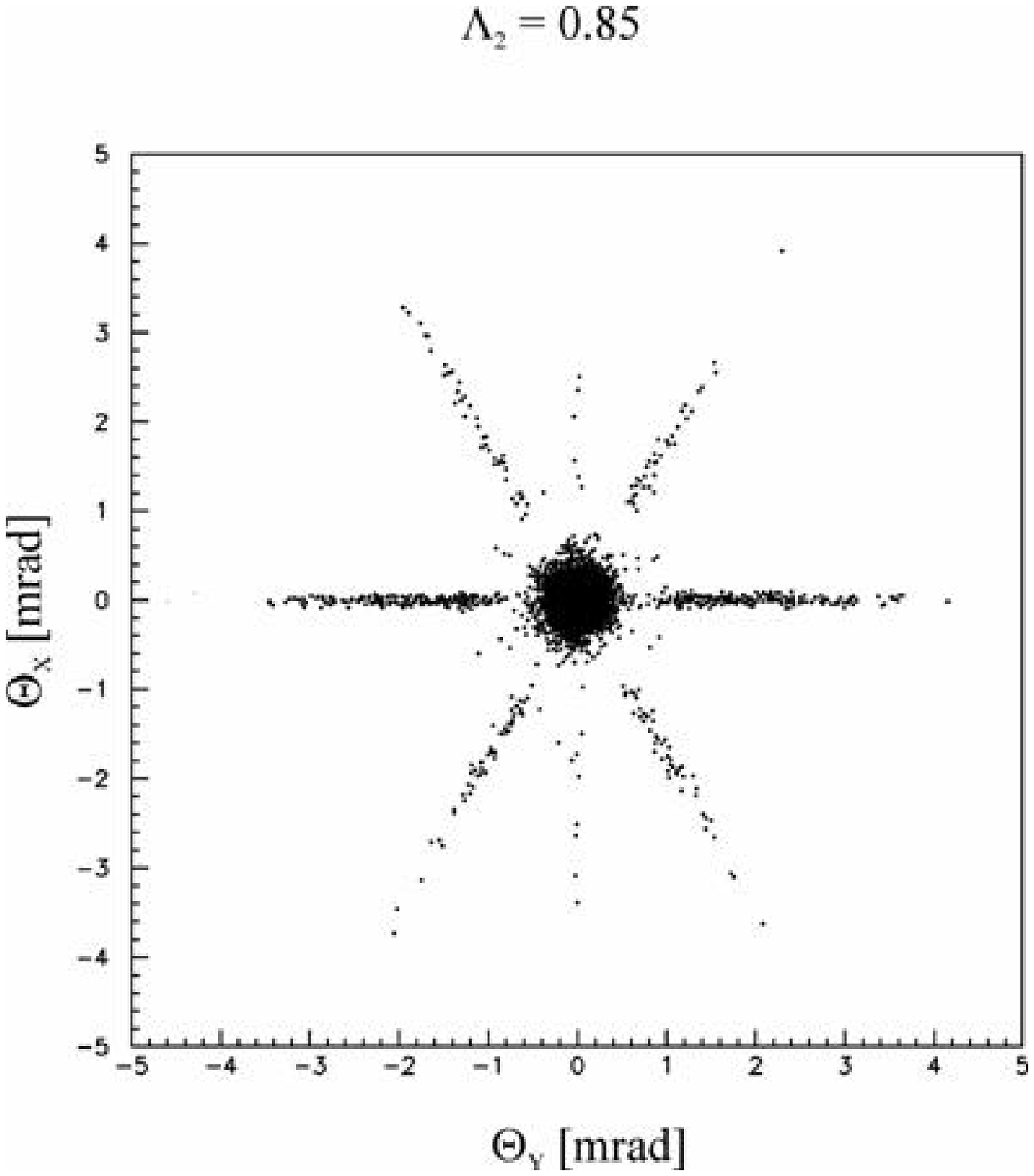}
\caption{The angular distributions of 1 GeV protons transmitted through the bundles of (10, 10) single-wall carbon nanotubes of the reduced lengths (a) $\Lambda_2$ = 0.15, (b) $\Lambda_2$ = 0.35, (c) $\Lambda_2$ = 0.65, and (d) $\Lambda_2$ = 0.85. The proton beam divergence angle is $\Omega_d$ = 1.884 mrad.}
\label{fig7_1}
\end{figure}

Figure \ref{fig7_1} show the angular distributions of transmitted
protons for $\Lambda_2$ = 0.15, 0.35, 0.65 and 0.85. For $\Lambda_2$
= 0.15 a channeling star is clearly visible in the angular
distribution, and one can distinguish its 40 arms. One can connect
this to the fact that each nanotube has 40 atomic strings. This
means that the channeling star effect can be used to learn about the
transverse structure of a nanotube. For $\Lambda_2$ = 0.35, 12 arms
of the channeling star are still visible, while for $\Lambda_2$ =
0.65 and 0.85, eight arms remain, with the intensity being lower in
the latter case. The analysis shows that the observed intensity
lowering and disappearance of the arms as $\Lambda_2$ increases can
be attributed to the process of proton dechanneling, which is very
pronounced since $\Omega_d \gg \psi_c$. For $\Lambda_2 >$ 1 the
channeling star does not exist any more.

\begin{figure}[ht!]
\centering
\includegraphics[width=0.45\textwidth]{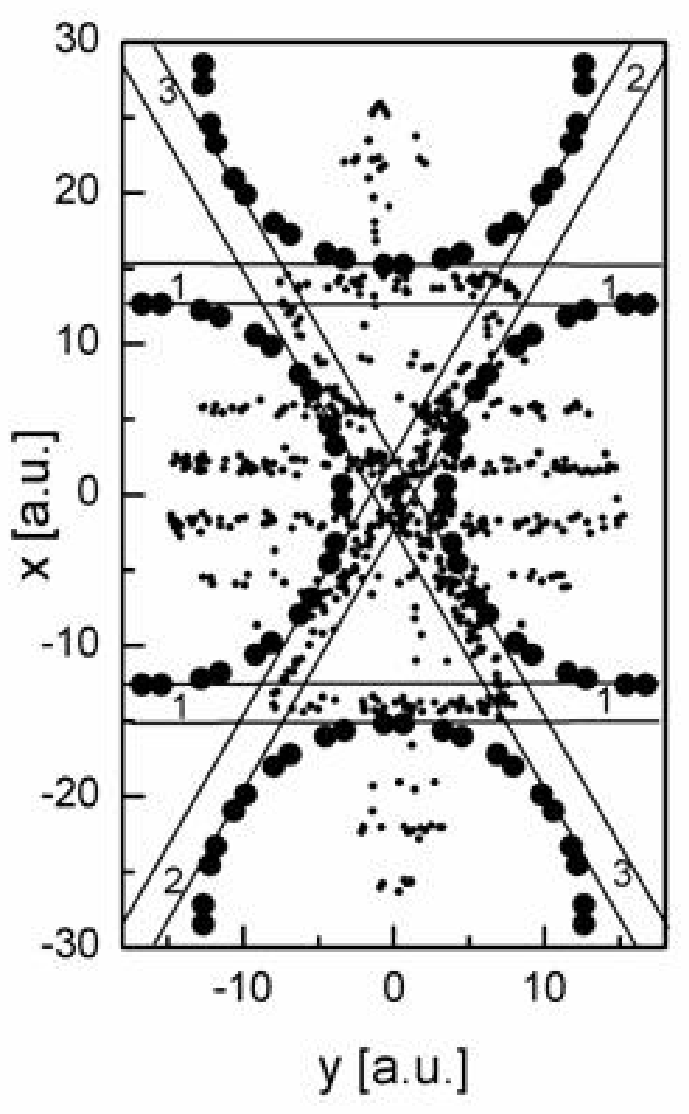}
\hspace*{1cm}
\includegraphics[width=0.45\textwidth]{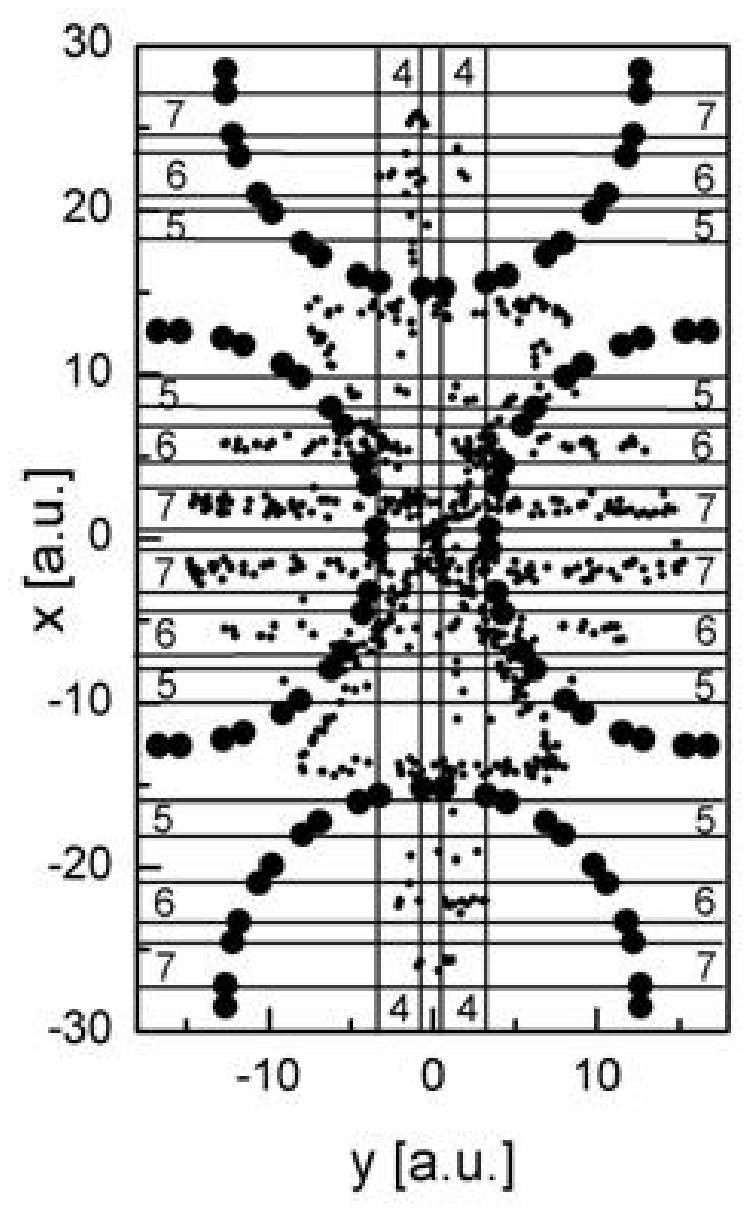}
\caption{The components of the proton impact parameter vectors, within the primitive cell of the superlattice, that correspond to the arms of the channeling star for the reduced bundle length $\Lambda_2$ = 0.65. (a) The solid lines represent four pairs of planes defined by the atomic strings of the nanotubes, designated by 1, 2 and 3, that enable the planar channeling of the protons through the bundle. (b) The solid lines represent 14 additional pairs of planes defined by the atomic strings of the nanotubes, designated by 4, 5, 6 and 7 that enable the planar channeling of the protons through the bundle.}
\label{fig7_2}
\end{figure}

\begin{figure}[ht!]
\centering
\includegraphics[width=0.7\textwidth]{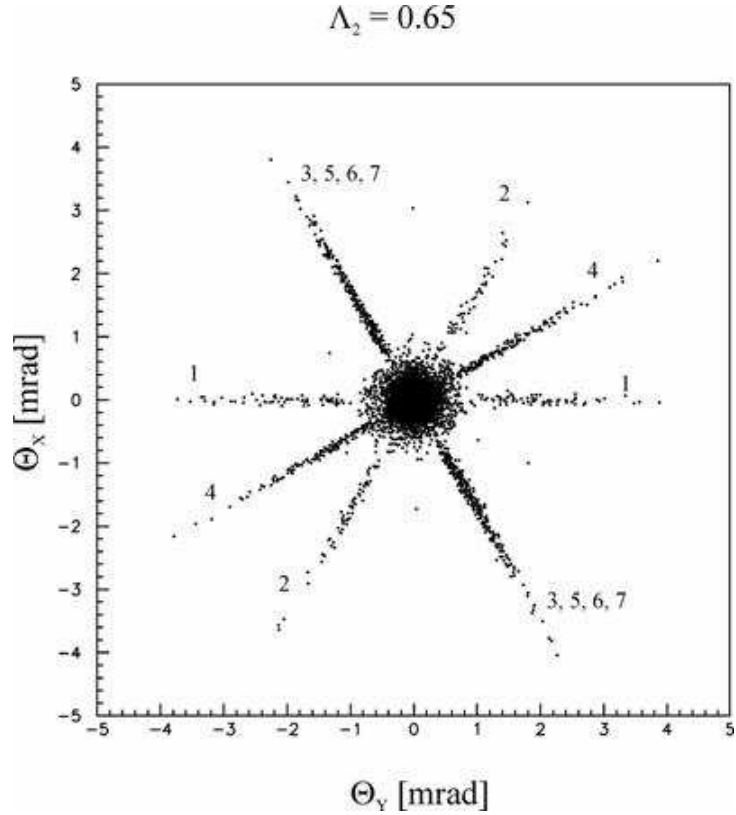}
\caption{The angular distributions of 1 GeV protons transmitted through the bundle of (10, 10) single-wall carbon nanotubes of the reduced length $\Lambda_2$ = 0.65 when the nanotubes are rotated about their axes by $\pi / 60$ counterclockwise relative to their assumed positions. The proton beam divergence angle is $\Omega_d$ = 1.884 mrad.}
\label{fig7_3}
\end{figure}

\begin{figure}[ht!]
\centering
\includegraphics[width=0.45\textwidth]{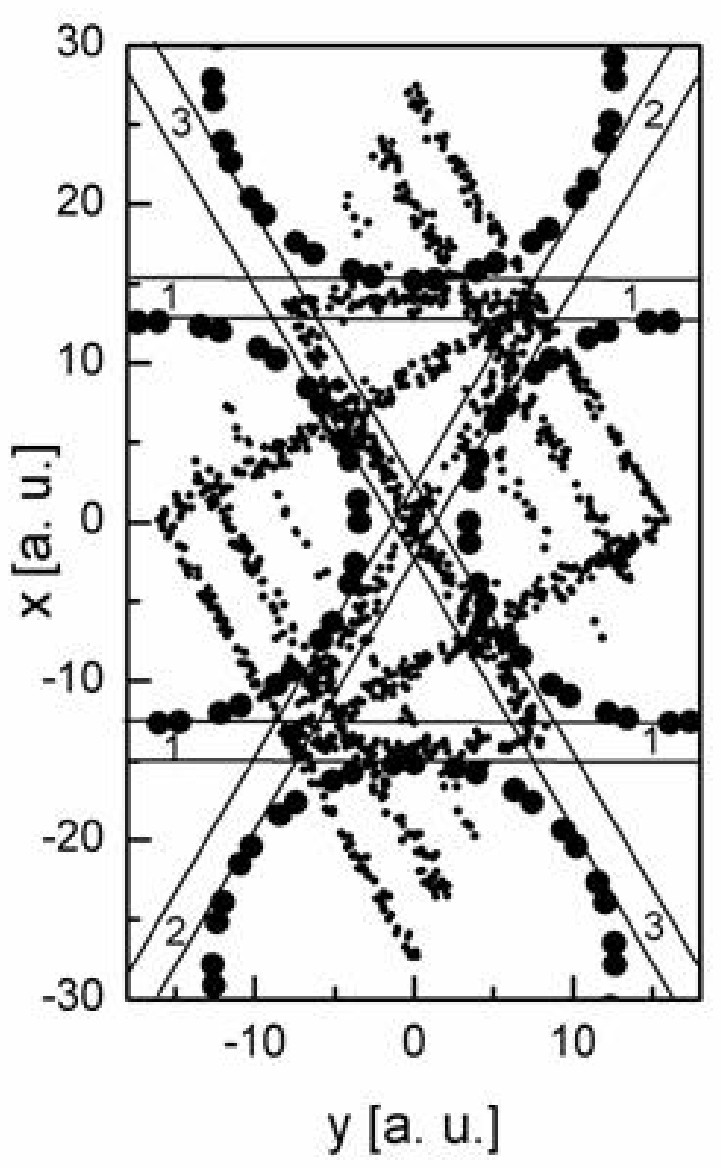}
\hspace*{1cm}
\includegraphics[width=0.45\textwidth]{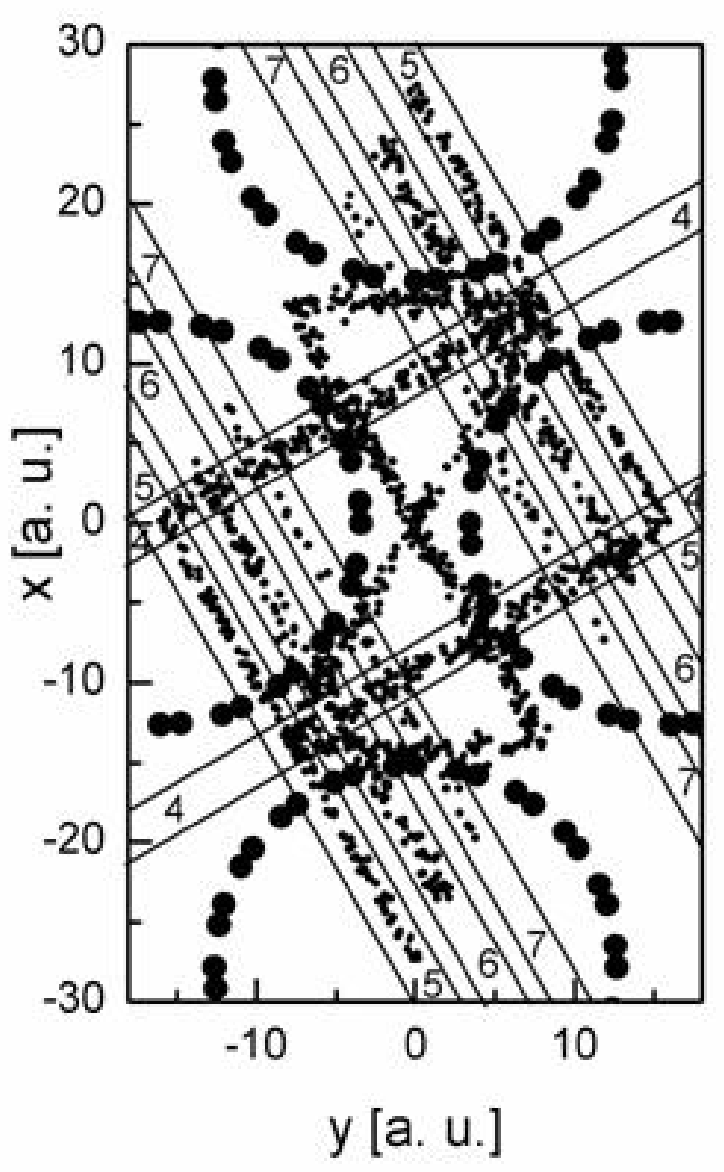}
\caption{The components of the proton impact parameter vectors, within the primitive cell of the superlattice, that correspond to the arms of the channeling star for the reduced bundle length $\Lambda_2$ = 0.65 when the nanotubes are rotated about their axes by $\pi / 60$ counterclockwise relative to their assumed positions. (a) The solid lines represent four pairs of planes defined by the atomic strings of the nanotubes, designated by 1, 2 and 3, that enable the planar channeling of the protons through the bundle. (b) The solid lines represent eight additional pairs of planes defined by the atomic strings of the nanotubes, designated by 4, 5, 6 and 7 that enable the planar channeling of the protons through the bundle.}
\label{fig7_4}
\end{figure}

Let us now concentrate on the channeling star for $\Lambda_2$ =
0.65. It has two arms lying along the $\Theta_x$ axis, two arms
lying along the $\Theta_y$ axis, two arms lying along the line
defined by $\varphi = \tan^{-1}(\Theta_y / \Theta_x) = n \pi / 3$
and $7 \pi / 6$, and two arms lying along the line defined by
$\varphi = 5 \pi / 6$ and $11 \pi / 6$. Figure \ref{fig7_2} show the
components of the proton impact parameter vectors, within the
(rhombic) primitive cell of the superlattice, that correspond to the
arms of this channeling star. They are selected taking that the
proton transmission angle $\Theta = (\Theta_x^2 +
\Theta_y^2)^{\frac{1}{2}} \ge$ 1.6 mrad. The analysis shows that
these impact parameter vectors are located between seven classes of
pairs of parallel lines, representing seven classes of pairs of
planes defined by the atomic strings of the nanotubes, which are
parallel to the bundle axis. There are two equivalent pairs of
planes designated by 1, one pair designated by 2, and one pair
designated by 3, which are shown in Fig. \ref{fig7_2}(a), and two
equivalent pairs of planes designated by 4, four equivalent pairs
designated by 5, four equivalent pairs designated by 6, and four
equivalent pairs designated by 7, which are shown in Fig.
\ref{fig7_2}(b). Each pair of planes defines a planar channel, which
corresponds to two opposite arms of the channeling star. The pairs
of planes that are parallel to each other correspond to the same two
opposite arms. In accordance with this, the arms shown in Fig.
\ref{fig7_1}(c) are designated by 4, by 1, 5, 6 and 7, by 2, and by
3.

In order to analyze the sensitivity of the channeling star effect to
the mutual orientation of the neighboring nanotubes within the
bundle, we rotated each nanotube about its axis by $\pi / 60$
counterclockwise relative to its assumed position  4.2 . The
resulting angular distribution of transmitted protons, for
$\Lambda_2$ = 0.65, is given in Fig. \ref{fig7_3}. In this case the
channeling star also has eight arms, as the one given in Fig.
\ref{fig7_1}(c), but the arrangement of its arms is different. Two
of the arms lie along the $\Theta_y$ axis, two arms lie along the
line defined by $\varphi = \pi / 6$ and $7 \pi / 6$, two arms lie
along the line defined by $\varphi = 2 \pi / 6$ and $ = 8 \pi / 6$,
and two arms lie along the line defined by $\varphi = 5 \pi / 6$ and
$11 \pi / 6$. Figure \ref{fig7_4} show the components of the proton
impact parameter vectors, within the (rhombic) primitive cell of the
superlattice, that correspond to the arms of the channeling star.
They were selected taking that $\Theta \ge$ 1.0 mrad. As in the
above considered case, the impact parameter vectors are located
between seven classes of pairs of parallel lines, representing seven
classes of pairs of planes defined by the atomic strings of the
nanotubes. There are two equivalent pairs of planes designated by 1,
one pair designated by 2 and one pair designated by 3, which are
shown in Fig. \ref{fig7_4}(a), and two equivalent pairs of planes
designated by 4, two equivalent pairs designated by 5, two
equivalent pairs designated by 6, and two equivalent pairs
designated by 7, which are shown in Fig. \ref{fig7_4}(b). In
accordance with this, the arms shown in Fig. \ref{fig7_3} are
designated by 1, by 2, by 4, and by 3, 5, 6 and 7.

It should be noted that in both of the above considered cases the
pairs of planes designated by 1, 2 and 3 do not cross the nanotubes,
and that the pairs of planes designated by 4, 5, 6 and 7 cross the
nanotubes. Namely, the protons channeled between planes 1, 2 and 3
move only in between the nanotubes while the protons channeled
between planes 4, 5, 6 and 7 move in between as well as through the
nanotubes. Also, it is clear that planar channels 1, 2 and 3 in the
former case practically coincide with planar channels 1, 2 and 3 in
the latter case. However, planar channels 4, 5, 6 and 7 in the
former case considerably differ from the planar channels 4, 5, 6 and
7 in the latter case. This explains the difference between the
arrangements of the arms in these two cases. Thus, one can say that
the channeling star effect is sensitive to the mutual orientation of
the neighboring nanotubes within a bundle. This means that it is
possible to measure the effect and employ the obtained results to
determine this orientation.

\markright{References}

\vfil

\end{document}